\documentclass{emulateapj}

\lefthead{Nidever, Zasowski, \& Majewski}
\righthead{Lifting the Dust Veil}

\usepackage{hyperref}

\shorttitle{RJCE Extinction Maps}

\shortauthors{Nidever, Zasowski, \& Majewski}

\begin{document}

\title{Lifting the Dusty Veil With Near- and Mid-Infrared Photometry: \\
III. Two-Dimensional Extinction Maps of the Galactic Midplane Using the Rayleigh-Jeans
Color Excess Method}

\author{
David L. Nidever, Gail Zasowski, \& Steven R. Majewski}
\affil{Department of Astronomy, University of Virginia,
    Charlottesville, VA 22904}
\email{dln5q, gz2n, srm4n@virginia.edu}

\begin{abstract}
We provide new, high-resolution $A(K_s)$
extinction maps of the heavily reddened Galactic midplane
based on the Rayleigh-Jeans Color Excess (``RJCE'') method.  RJCE determines star-by-star reddening based on a combination
of near- and mid-infrared photometry.
The new RJCE-generated maps 
have $2^\prime \times 2^\prime$ pixels and
span some of the most severely extinguished regions of the Galaxy --- those covered with {\it Spitzer}+IRAC
imaging by the GLIMPSE-I, -II, -3D, and Vela-Carina surveys, from 
$256^{\circ}<l<65^{\circ}$ and, in general, for
$|b| \le 1^{\circ}$--$1.5^{\circ}$ (extending up to $|b|\leq4^{\circ}$ in the bulge).
Using RJCE extinction measurements, we generate dereddened color-magnitude diagrams and, in
turn, create maps based on main sequence, red
clump, and red giant star tracers, each probing different distances
and thereby providing coarse three-dimensional information on the relative
placement of dust cloud structures.  The maps generated from red giant stars, which reach to
$\sim$18--20 kpc, probe beyond most of the Milky Way extinction in
most directions and provide close to a ``total Galactic extinction'' map --- at minimum they provide 
high angular resolution maps of lower limits on $A(K_s)$.
Because these maps are generated directly from measurements of reddening by the very dust being
mapped, rather than inferred on the basis of some less direct means, they are likely the
most accurate to date for charting in detail
the highly patchy differential extinction in the
Galactic midplane.
We provide downloadable FITS files and an IDL tool
for retrieving extinction values for any line of sight within our mapped regions.
\end{abstract}


\section{Introduction} \label{sec:intro}

The existence of interstellar dust and its light-obscuring effects, 
though postulated more than a century and a half ago by Struve (1847)
as one (incorrect) solution to Olber's Paradox, was not definitively proven
until Trumpler's (1930) insightful experiment to compare distances 
of open clusters derived in separate extinction-dependent and extinction-independent methods.
Though the amount of extinction resulting from
this dust was historically often approximated with a standard accumulation
by distance under an assumed plane-parallel distribution (i.e., the cosecant law; e.g.,
Shane \& Wirtanen 1967, Peterson 1970, Noonan 1971, Sandage 1973, 
de Vaucouleurs, de Vaucouleurs \& Corwin 1976, 
de Vaucouleurs \& Buta 1983, Fabbri et al. 1986; but see Kron \& Guetter 1973, 
Holmberg 1974, Knude 1996 for examples of discussions
of simple variations about the basic cosecant law), 
it was always well-understood that the distribution of Galactic dust is in fact quite
irregular.\footnote{Even relatively recently, Knude (1996)
went through great pains to determine whether the cosecant law can at least provide a reasonable 
approximation in the North Galactic polar cap and concluded that only a piecewise application
of the cosecant law with varying coefficients can work, and only 
in certain ranges of Galactic latitude and longitude.}
Once it is appreciated that an irregular selective obscuration of a more regular distribution of stars, 
and not an irregular distribution of the stars themselves, is shaping the net configuration of observed starlight,
then any observation of the density of --- or light from --- stars in the Galactic plane of the Milky Way
clearly illustrates this patchiness (e.g., Herschel's [1785] drawings of the distribution of stars
in the Milky Way, or the early photographs of the Galactic plane by Barnard et al.\ 1927).

For extragalactic studies, a long-standing default strategy has been simply to
steer clear of the ``Zone of Avoidance" at low Galactic latitudes altogether, rather than contend
with assessing the amount of foreground extinction.  Unfortunately, Nature is
not always so accommodating with the placement of important and interesting
sources, and corrections for foreground dust --- generally attributable to 
dust in our own Milky Way --- are almost always needed; even at the highest Galactic
latitudes, one can encounter appreciable amounts of dust (e.g., the North Galactic
Pole has $A(B)$ as high as $0.05$ to $0.1$ mags, depending on the reddening calibration and extinction law used;
e.g., Hilditch et al.\ 1983, Knude 1996, Peek \& Graves 2010).

Early attempts to map the variation of total line-of-sight Galactic extinction explored a 
number of techniques.  For example, under an assumed constancy of dust-to-gas
ratio, the net column density of HI, easily measured via 21-cm observations, could
be converted into a dust column density (e.g., Lilley 1955, Sturch 1969, Savage \& Jenkins 1972, Knapp \& Kerr 1974, Burstein \& Heiles 1982).
Under the assumption of a 
homogeneous spatial distribution of galaxies, the counts of galaxies to a given magnitude
limit were also used to infer the degree of extinction along a line of sight
(Burstein \& Heiles 1978, 1982).
Similarly, the assumption of spatial constancy of the {\it stellar} distribution (at least on relatively small angular scales) has been
employed to derive extinction using star counts --- comparing stellar densities across a clearly extinguished region 
(often on the edges of a dark cloud) to those in
nearby ``control'' regions (e.g., Wolf 1923, Bok 1956, Froebrich et al.\ 2005).

Directly {\em measuring} the effects of dust via such affected tracers
is the most reliable method to map the Galactic extinction.
Extracting the reddening and extinction effects
of foreground dust from the intrinsic spectral energy distributions (SEDs) of observed
sources has traditionally required direct, detailed spectroscopic 
information of each source to classify it explicitly and thereby infer its intrinsic properties 
independent of observed colors --- quite a demanding prospect in most situations.  
In recent decades, a large body of work has amassed that attempts to
determine the total amount of extinction
foreground to heavily-reddened sources by {\it photometric} means; under the
assumption of a homogeneous color distribution of 
either stars or galaxies (or particular classes of either), 
deviations from the mean observed colors are attributed to interstellar reddening.
The decreased temperature dependencies of infrared (IR) colors, combined with the decreased extinction
effects at IR wavelengths, has led to the wide adoption of IR and optical-IR color excesses as high-quality gauges of 
interstellar reddening, especially in highly extinguished regions (since e.g., Jones et al. 1980, 1984, Smith 1987).
Lada et al.\ (1994) introduced a powerful and efficient method of
extinction mapping with the NICE (Near Infrared Color Excess) technique,
which applies elements of both IR excess measurements and star-counting techniques to large sets of IR
photometry; this method has undergone significant evolution (NICER, Lombardi \& Alves 2001; 
V-NICE, Gosling et al.\ 2009; NICEST, Lombardi 2009) and has been applied
to a series of dense clouds and dark nebulae (e.g., Alves et al.\ 1998, Lada et al.\ 2009, Lombardi et al.\ 2010). 
The color excess method has also been applied to galaxy colors (e.g., Peek \& Graves 2010), 
though galaxy ensembles are obviously best-suited to gauging {\it total} line-of-sight Galactic reddening.
Greater accuracy may be attained through use of objects with well-defined (and limited ranges of) colors
like RR Lyrae stars, globular clusters,
or red clump stars.
Another method advocated by Burstein and collaborators
(Burstein et al.\ 1988, Faber et al.\ 1989)
is based on the observation that the integrated spectra of the stellar populations
of giant elliptical galaxies are strikingly uniform, so that the intrinsic $(B-V)_0$ colors
of these systems can be inferred on the basis of the spectroscopic measurement of 
their Mg$_2$ line index.

Early investigations into the consistency of extinction derived with these various methods
generally led to disappointing results, with no simple relationships found amongst the 
different dust proxies (Heiles 1976).  Partly this is because of 
the failure of the very assumptions on which each method is based --- e.g., 
the  high ``noise" in galaxy counts due to large scale structure, or variations
in the dust-to-gas ratio (Seki 1973, Jenkins \& Savage 1974, Bohlin 1975, Savage et al. 1977, 
Burstein \& Heiles 1978, Fong et al. 1987, Burstein 2003).

Currently the most commonly used large-scale reddening maps are those by Schlegel, Finkbeiner \& Davis (1998,
``SFD" hereafter), which have been derived from 100$\mu$m emission maps
using the COBE/DIRBE and IRAS/ISSA data.  These maps have proven reliable for 
differential line-of-sight reddening in low extinction regions, albeit with some 
uncertainties in zero-points (see discussion by Burstein 2003).
Unfortunately, these maps have a resolution of only $\sim$6$^\prime$, which is particularly
problematical where the reddening varies dramatically on much smaller angular scales.  In addition, 
a number of studies have shown that 
the SFD maps overestimate the extinction at low latitudes by as much
as 50\% in $A(V)$ (Stanek 1998a,b, Arce \& Goodman 1999a, Chen et al.\ 1999, 
Ivans et al.\ 1999, von Braun \& Mateo 2001, 
Choloniewski \& Valentijn 2003, Rocha-Pinto et al.\ 2004, Dutra et al.\ 2003a,b, 
Cambr\'esy et al.\ 2005, Am\^ores \& L\'epine 2005, 2007, Yasuda et al.\ 2007, Peek \& Graves 2010).  
A physical reason for the complications of the SFD maps at low latitudes 
is explored in the analysis in Paper I --- namely that far-infrared-{\it emitting} dust, 
which provides the basis of the SFD extinction maps,
is simply not a good proxy for the dust causing {\it extinction} at shorter wavelengths.
Comparison to other extinction tracers reveals that many of the SFD ``high extinction'' knots
actually trace supernova remnant or star-forming environments, and a number of cold, dense clouds
can be identified in extinction that lack the far-IR emission actually traced by the SFD maps.

Thus,
the midplane extinction in the Milky Way is far from being unambiguously mapped,
and for studies of the Milky Way itself, the patchy intermingling of dust and
stars remains a particularly troublesome obstacle.
Both photometric and spectroscopic parallax determinations must include
an accurate assessment of the extinction {\it foreground} to a star to gauge its distance.
While studies of specific dark clouds using
stellar IR color excess methods (particularly the extensive NICE family of techniques) have achieved
remarkable success at mapping extinction,
reliable assessments of extinction foreground to stars and astronomical sources 
{\it not} associated with these dark clouds have proven more elusive.

Now, more accurate measurements of individual stellar extinction levels,
including full three-dimensional spatial distribution information, 
is becoming possible through combined near- and mid-IR observations of the 
Rayleigh-Jeans flux distributions of stars (Majewski et al. 2011, Paper I hereafter);  
because the long wavelength SEDs of normal
stars have approximately the same shape (i.e., colors) on their Rayleigh-Jeans tails, the
{\it observed} Rayleigh-Jeans colors contain information on the foreground reddening to
a star explicitly. 
We show in Paper I that this Rayleigh-Jeans Color Excess (RJCE) method is
superior to the SFD method for mapping
the extinction by midplane dust and is an improvement over the other
IR color-excess methods due to the improved homogeneity of RJCE's adopted
color indices.
An additional advantage of the RJCE method over its predecessors is that
the former cleanly separates reddening from intrinsic
stellar colors and preserves the latter for use in the identification of stellar 
types (e.g., main sequence versus red clump versus red giant); thus, we can use the RJCE 
method to fairly reliably isolate the most distant
stellar proxies (i.e., red giants) to create the best assessment
of the total reddening along a given line of sight.
Alternatively, one can isolate a rather reliable standard candle (in the form of red clump stars) to bring
greater accuracy to distance estimates.  In combination, stellar types of different intrinsic luminosity can yield the
distance distribution of the dust along the line of sight.

In this contribution to the present series of papers on the RJCE method, we make new, two-dimensional 
extinction maps of the Galactic midplane across the entire Galactic Legacy Infrared Mid-Plane Survey Extraordinaire
(GLIMPSE-I; Benjamin et al.\ 2003; GLIMPSE-II/-3D Churchwell et al.\ 2009),
and Vela-Carina ({\it Spitzer} PID 40791) IRAC survey areas.  
Section~\ref{sec:overview} summarizes how the maps were created; 
the reader is directed to Paper I for a fuller description and analysis of the RJCE method and
its application to different color index combinations and different stellar types.
In Section~\ref{sec:atlas} we present a full atlas of these maps based on use of main sequence, red clump, and red giant tracers, and
in Section~\ref{sec:caveats} we alert the reader to various caveats for using the maps.
Electronic versions of these maps have been made publicly available, and we describe how to access them in Section~\ref{sec:publicmaps}.

\section{Creation of the Maps} \label{sec:overview}

\subsection{Summary of the RJCE Method} \label{sec:rjce}

IR imaging, like that in 2MASS (Skrutskie et al.\ 2006) and UKIDSS (Lucas et al.\ 2008),
can more efficiently penetrate the obscuration in the dusty midplane and Galactic Center 
than observing at shorter wavelengths, and much work
has been done using IR observations of low-latitude stars as probes of Galactic structure and the interstellar medium (ISM) there.
Measuring ``color excesses'' is one common use of this NIR photometry --- 
comparing the observed colors of a star (or the average of a group of stars) to the
expected intrinsic colors to determine the amount of line-of-sight reddening,
which can then be converted into an extinction under an assumed extinction law.
Figure 1 in Paper I
demonstrates some of the primary complications when this method is applied to broadband NIR colors in isolation.  
First, the spread in intrinsic NIR color is significant: A, F, G, and K stars span about a magnitude in
$(J-H)_0$ and $(J-K_s)_0$ and almost half of a magnitude in $(H-K_s)_0$.  In addition, the NIR reddening
vector very closely parallels the intrinsic stellar locus, leading to high degeneracy in derived reddening
estimates based on observed stellar colors.

\begin{figure*}[ht!]
\begin{center}
\includegraphics[angle=0,width=0.9\textwidth]{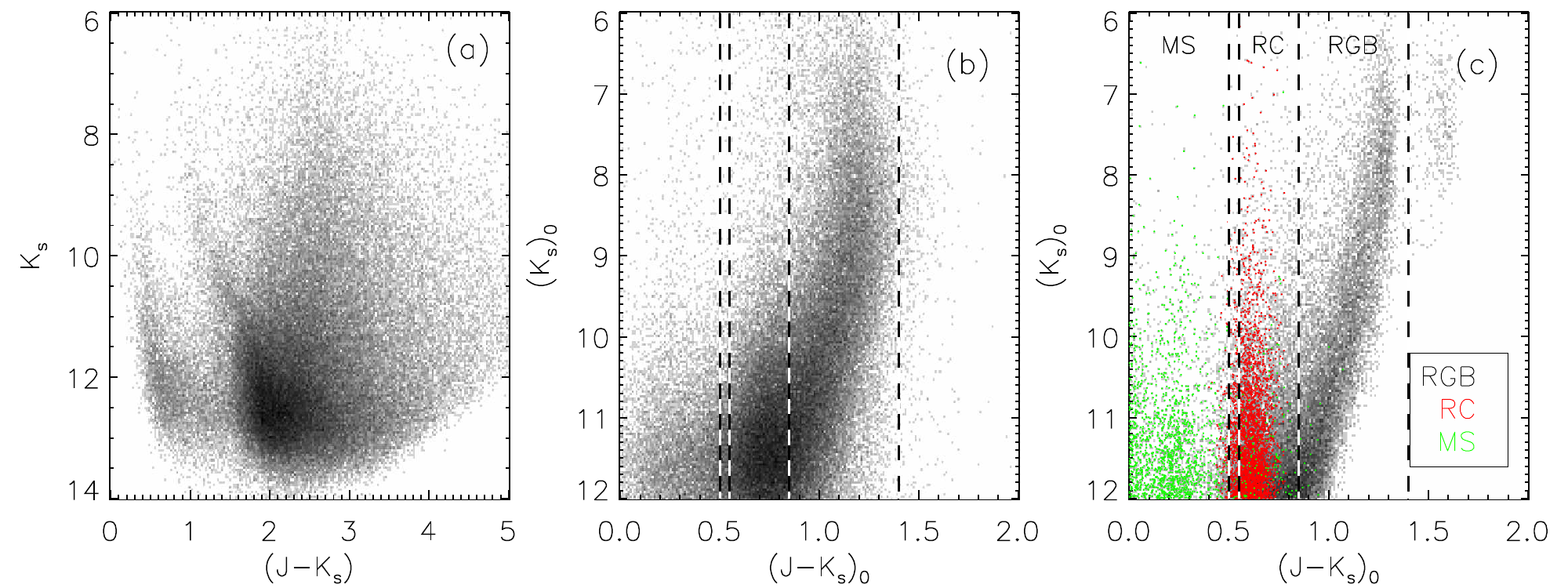}
\end{center}
\caption{Three color-magnitude diagrams (CMDs) of a 4 deg$^2$ field at $(l,b)$ = (42.0$^\circ$,0$^\circ$) for all sources
with magnitude errors less than 0.5 mag in the three 2MASS and four {\it Spitzer}-IRAC bands.
Sources are binned in bins of 0.025 in $J-K_s$ and 0.04 in $K_s$.  $(a)$ The raw
2MASS data.  $(b)$ The RJCE-corrected CMD. $(c)$ An extinction-free model of the same field from TRILEGAL
(Girardi et al.\ 2005) with photometric errors; color-coded to highlight the different stellar populations
(MS--main sequence, RC--red clump, and RGB--red giant branch).  The vertical dashed lines in $(b)$ and
$(c)$ indicate the color divisions between MS, RC and RGB stars.
}
\label{fig:cmds}
\end{figure*}

Both of these critical problems are greatly ameliorated by using longer-wavelength, longer-baseline colors
including MIR photometry from, e.g., the {\it Spitzer Space Telescope}'s IRAC instrument.  These longer-wavelength
filters sample the Rayleigh-Jeans tail of a normal star's SED; the slope of this tail (and thus the colors measured from
it) has very little dependence on stellar temperature, so the intrinsic variation in MIR and NIR-MIR colors is
significantly smaller than in NIR colors.  For example, the A--K stars mentioned earlier differ by only $\sim$0.1 mag
in $(H-[4.5\mu])_0$.  Furthermore, the reddening vector in these colors is very nearly perpendicular to the stellar
RGB locus\footnote{It is true that the reddening vector more closely parallels the locus of cool main-sequence stars;
however, these are very sparse in the 2MASS and GLIMPSE surveys}, facilitating more accurate estimates of the stellar reddening.
The RJCE method takes advantage of these NIR-MIR color properties to produce reliable estimates of the line-of-sight
reddening to each individual star.  Analysis of many factors, including intrinsic color variation and typical
photometric quality (see Paper I), has led us to
adopt $(H-[4.5\mu])$ as the best single NIR-MIR color for stellar dereddening.  

Also shown in Paper I are comparisons of RJCE-derived  extinction maps to those of SFD
as well as to maps of $^{13}$CO emission, 8$\mu$m dust and PAH emission, and NIR star counts.  The RJCE maps
bear a striking structural similarity to the $^{13}$CO and star count maps, both expected to be good tracers
of cold, dense extinction regions, while the SFD maps more closely resemble the [8.0$\mu$] images.

Moreover, the RJCE method {\it directly} measures the reddening on a star-by-star basis, and is therefore
superior to methods that use proxies to calculate extinction
(e.g., dust emission, HI, or galaxy counts).  The RJCE method relies only on the availability
of stars, which are, fortunately, very abundant in those parts of the Galaxy where the effects of dust are strongest.

\subsection{Defining the Stellar Samples} \label{sec:starsamples}

Arguably, the RJCE method's most powerful and unique feature is the ability to recover a star's approximate
stellar type after the reddening has been estimated.  The strong temperature dependence of the {\it corrected}
NIR colors (especially those involving $J$-band) allows us to derive an approximate photometric temperature
(hence spectral type and, based on the apparent magnitude, the luminosity class as well\footnote{There is
certainly dwarf/giant degeneracy at redder NIR colors, but Galactic stellar population models (e.g., TRILEGAL;
Girardi et al.\ 2005) guide us as to which magnitudes, for a given color, we may expect a low fraction of
late-type dwarf contaminants.  Fortunately, these models show that late-type dwarfs are in any case very rare
in the 2MASS+GLIMPSE catalog.}).  For the initial publicly-available maps presented here, we have used this
powerful feature to split the 2MASS+{\it Spitzer} stellar sample into three subsamples: main sequence (MS)
dwarfs, which are dominant for 0.0 $\leq$ $(J-K_s)_0$ $\leq$ 0.5, red clump (RC) giants for
0.55 $\leq$ $(J-K_s)_0$ $\leq$ 0.85, and red giant branch (RGB) giants for 0.85 $<$ $(J-K_s)_0$ $\leq$ 1.4, 
where $(J-K_s)_0$ for each star is calculated using the RJCE method described in Section \ref{sec:rjce} and Paper I.  
See Figure~\ref{fig:cmds} for an example of a corrected color-magnitude diagram (CMD) and the rationale for the
adopted stellar color divisions.  Each of these stellar subsamples probes a different range of heliocentric
distance (see \S \ref{sec:dists} and Figure \ref{fig:tracerdist}), so maps using each tracer, taken together,
provide coarse {\it three-dimensional} information about the midplane dust distribution, based solely on which
of the actual stellar tracers are being reddened.

\section{Extinction Maps of the Milky Way Midplane} \label{sec:atlas}

Figure \ref{fig:extmaps6065} contains the ``Version 1'' RJCE extinction maps, 
in units of $A(K_s)$, over $\sim$170$^\circ$
of longitude in the Galactic midplane.
For $|l|$ $\leq$ 65$^\circ$, the latitude range is centered on the Galactic equator (and includes
$|b|$ $\leq$ 1$^\circ$ for most longitudes but up to $|b|$ $\leq$ 4$^{\circ}$ in the bulge).
In the longitude range covered by the Vela-Carina survey (295$^\circ$ $\leq$ $l$ $\leq$ 255$^\circ$), the map
center varies between $b$ = +0.5$^\circ$ and $b$ = $-0.5^\circ$, but the latitude coverage is consistently 2$^\circ$ tall.

The basic procedure used to make the maps is as follows: Using the reliable band-merged 2MASS+IRAC 
Catalog\footnote{\url{http://irsa.ipac.caltech.edu/data/SPITZER/GLIMPSE/doc/glimpse1\_dataprod\_v2.0.pdf}}, 
we selected stars having detections and photometric errors $<$0.2 mag in all four of the $J$, $H$, $K_s$ and [4.5$\mu$] bands.
The extinction was calculated for each star separately, using
\begin{equation}
A(K_s) = 0.918 \times (H - [4.5\mu] - 0.08)
\end{equation}
from Paper I.
Each star was dereddened and classified as MS, RC, or RGB using the RJCE method and the stellar color bins
described in Section~\ref{sec:overview}.  Each of these three subsamples was binned separately into a
spatial map with 2$^\prime$ resolution, and two extinction maps were created for each subsample
(for a total of six distinct maps) using the median and 90th percentile values of $A(K_s)$ measured within each
spatial pixel.  Similar maps were created for the larger sample of all stars, with no color cuts.  All of
these extinction maps were Gaussian smoothed
(FWHM=1.4$^\prime$)
to reduce noise.  Not only do the 90th percentile maps represent the majority
of the extinction affecting each stellar subsample, they are also the least likely to be biased by contaminants
from, e.g., foreground cool dwarfs (in the case of the RGB maps) or photometric errors.
All extinction values are given in magnitudes of $A(K_s)$.  We also provide star count
and $A(K_s)$ dispersion (not to be confused with the uncertainty) maps for each subsample.
Pixels with less than two stars in them are set to white in the $A(K_s)$ quick-look images (black in the starcount quick-look
images) and have a value of 999999.0 in the queriable FITS tables.

\begin{figure}[ht!]
\begin{center}
\includegraphics[scale=0.5]{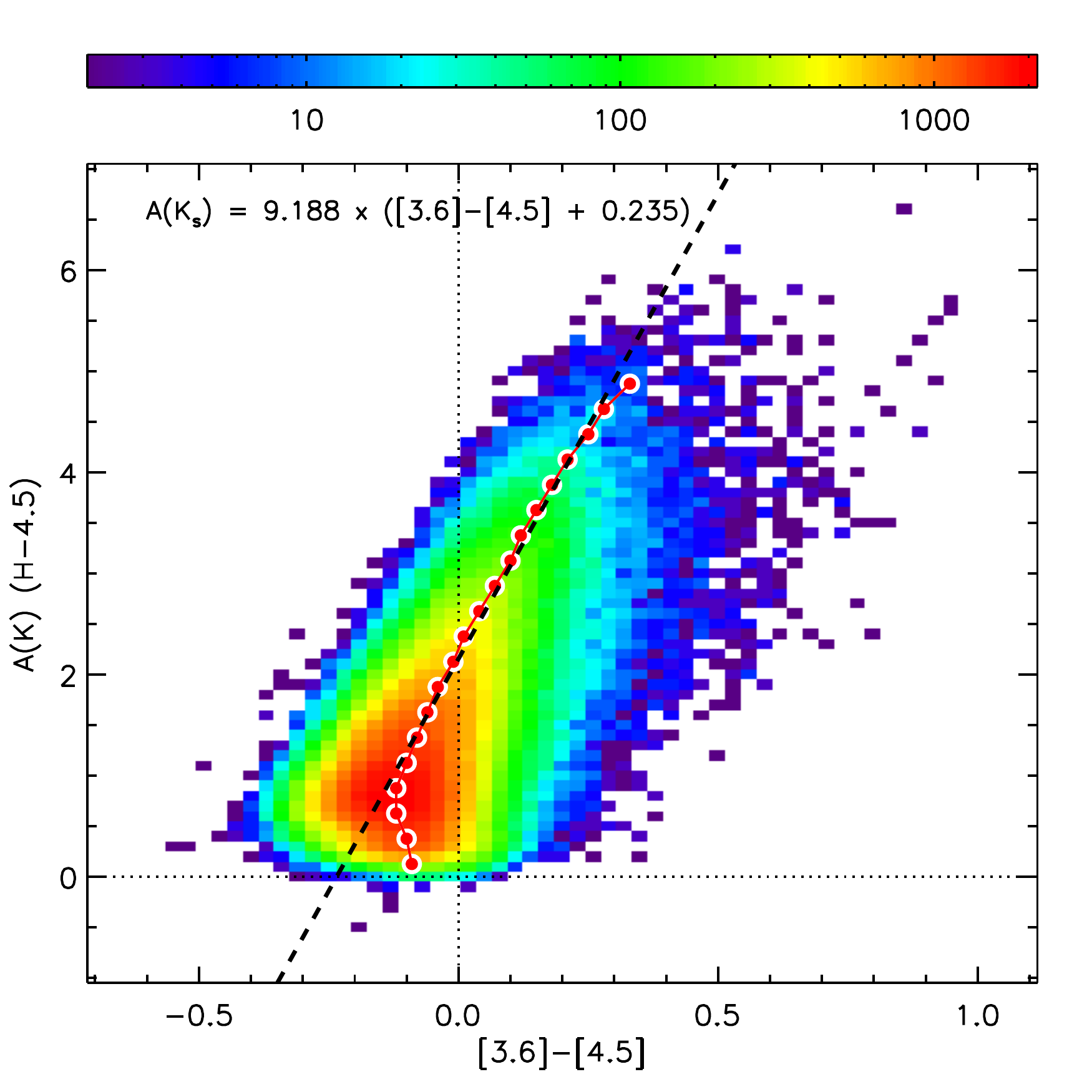}
\end{center}
\caption{Linear fitting of $E(H-[4.5\mu])$-derived $A(K_s)$ values as a function
of $([3.6\mu]-[4.5\mu])$ color for all stars with photometric errors in the $H$,$[3.6\mu]$,$[4.5\mu]$
bands less than 0.07 mag.  The values were binned in bins of 0.25 mag in $A(K_s)$ (red dots) and the
robust linear-fit perfomed with the binned values.  The best-fit line is shown in black.}
\label{fig:aklm}
\end{figure}

Finally, we also include ``lower limit total extinction'' maps produced using only the IRAC [3.6$\mu$] and [4.5$\mu$]
bands; these are 90th percentile extinction maps made using {\it all} stars with detections and photometric
errors $<$0.5 mag in both of these bands (see Section~\ref{sec:limits}).  To derive extinctions from
([3.6$\mu$]-[4.5$\mu$]) we use,
\begin{equation}
A(K_s) = 9.188 \times ([3.6\mu] - [4.5\mu] + 0.235),
\end{equation}
which was found by linear fitting the $E$($H$$-$[4.5$\mu$])--derived $A(K_S)$ values and the ([3.6$\mu$]$-$[4.5$\mu$])
colors for all stars in the longitude range 20$^{\circ} \leq l \leq 30^{\circ}$ and photometric errors in H, [3.6$\mu$],
and [4.5$\mu$] less than 0.07 mag (see Fig.\ \ref{fig:aklm}).  The intrinsic scatter and photometric errors in
([3.6$\mu$]$-$[4.5$\mu$]) are generally higher than for ($H$$-$[4.5$\mu$]), and, therefore, the derived extinctions
are generally less reliable and mainly useful for high extinction regions (where the [$H$$-$4.5$\mu$] maps often drop out).
Including the ([3.6$\mu$]$-$[4.5$\mu$]) extinction maps, there are a total of 18 maps in the downloadable FITS files,
as listed in Table \ref{table_maps}.  Examples of the extinction maps are also shown in Figure \ref{fig:extmaps6065}.

\begin{deluxetable*}{cccc}
\tablecaption{Maps in Downloadable FITS Files}
\tablecolumns{10}
\tablewidth{0pt}
\tablehead{
\colhead{HDU\tablenotemark{a}} & \colhead{Pop. Type} & \colhead{Map Type\tablenotemark{b}} & \colhead{Statistic}
}
\startdata
1 & MSTO & RJCE $(H-[4.5\mu])$ Extinction & Median \\
2 & MSTO & RJCE $(H-[4.5\mu])$ Extinction & 90th percentile \\
3 & MSTO & RJCE $(H-[4.5\mu])$ Extinction & Dispersion\tablenotemark{c} \\
4 & MSTO & Starcounts & \nodata \\
5 & RC & RJCE $(H-[4.5\mu])$ Extinction & Median \\
6 & RC & RJCE $(H-[4.5\mu])$ Extinction & 90th percentile \\
7 & RC & RJCE $(H-[4.5\mu])$ Extinction & Dispersion \\
8 & RC & Starcounts & \nodata \\
9 &  RGB & RJCE $(H-[4.5\mu])$ Extinction & Median \\
10 & RGB & RJCE $(H-[4.5\mu])$ Extinction & 90th percentile \\
11 & RGB & RJCE $(H-[4.5\mu])$ Extinction & Dispersion \\
12 & RGB & Starcounts & \nodata \\
13 & ALL & RJCE $(H-[4.5\mu])$ Extinction & Median \\
14 & ALL & RJCE $(H-[4.5\mu])$ Extinction & 90th percentile \\
15 & ALL & RJCE $(H-[4.5\mu])$ Extinction & Dispersion \\
16 & ALL & Starcounts & \nodata \\
17 & ALL & RJCE [3.6$\mu$]$-$[4.5$\mu$] Extinction & Median \\
18 & ALL & RJCE [3.6$\mu$]$-$[4.5$\mu$] Extinction & 90th percentile \\
\enddata
\tablenotetext{1}{Extension in the FITS file.}
\tablenotetext{2}{All extinction maps are in magnitudes of $A(K_s)$.}
\tablenotetext{3}{Not to be confused with an uncertainty estimate.}
\label{table_maps}
\end{deluxetable*}

\begin{figure*}[htpb]
\begin{center}
$\begin{array}{c}
\includegraphics[height=37mm]{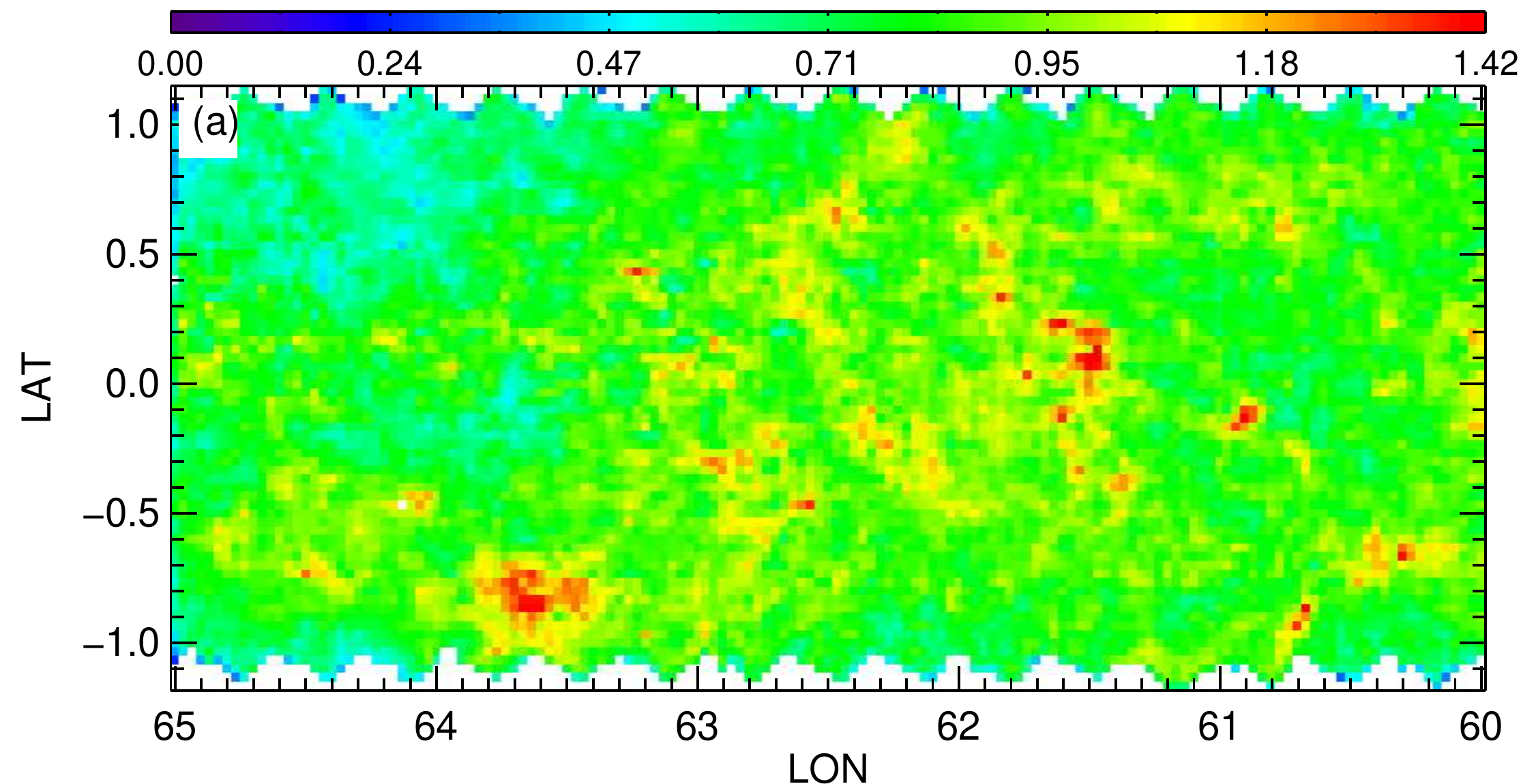} \\
\includegraphics[height=37mm]{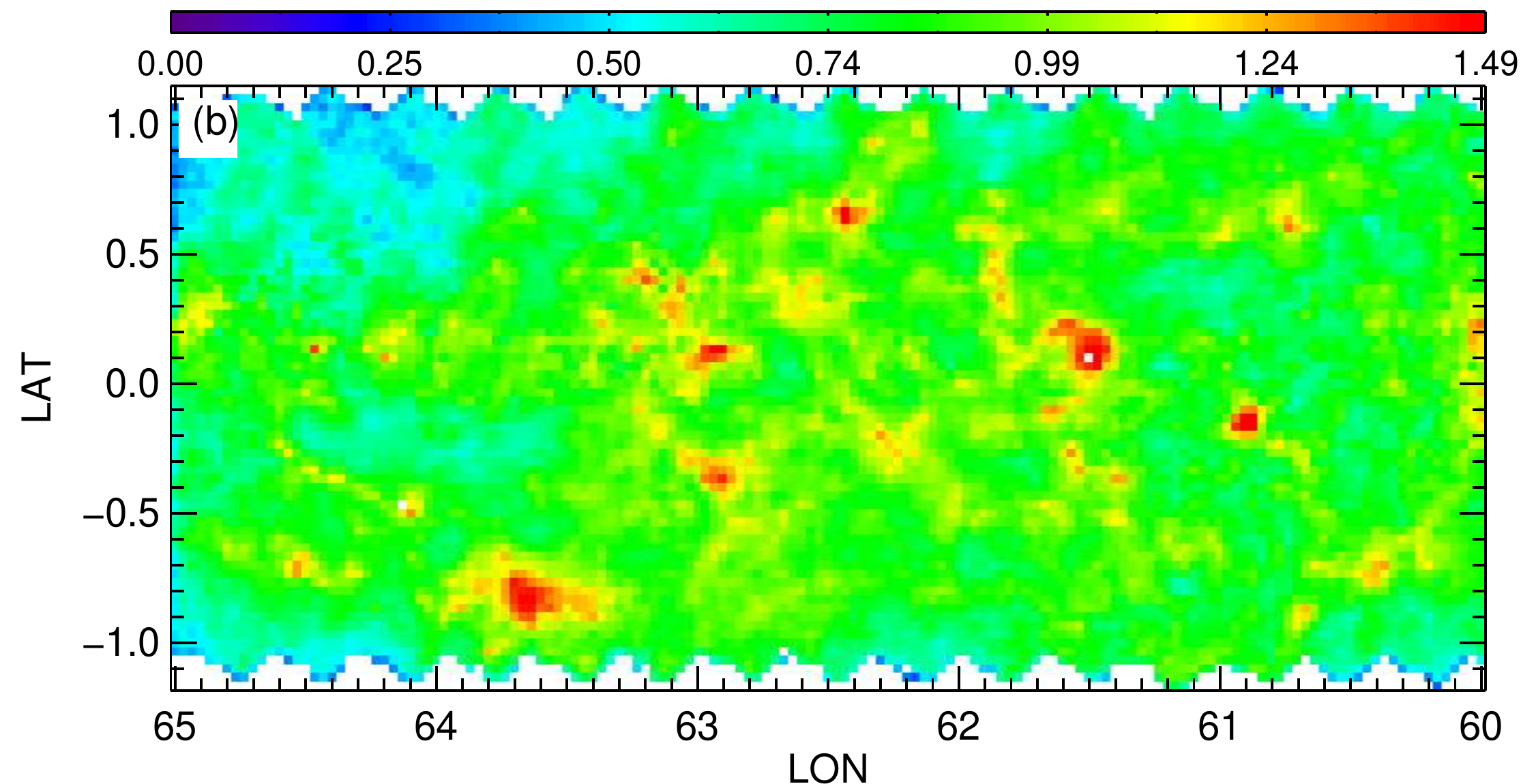} \\
\includegraphics[height=37mm]{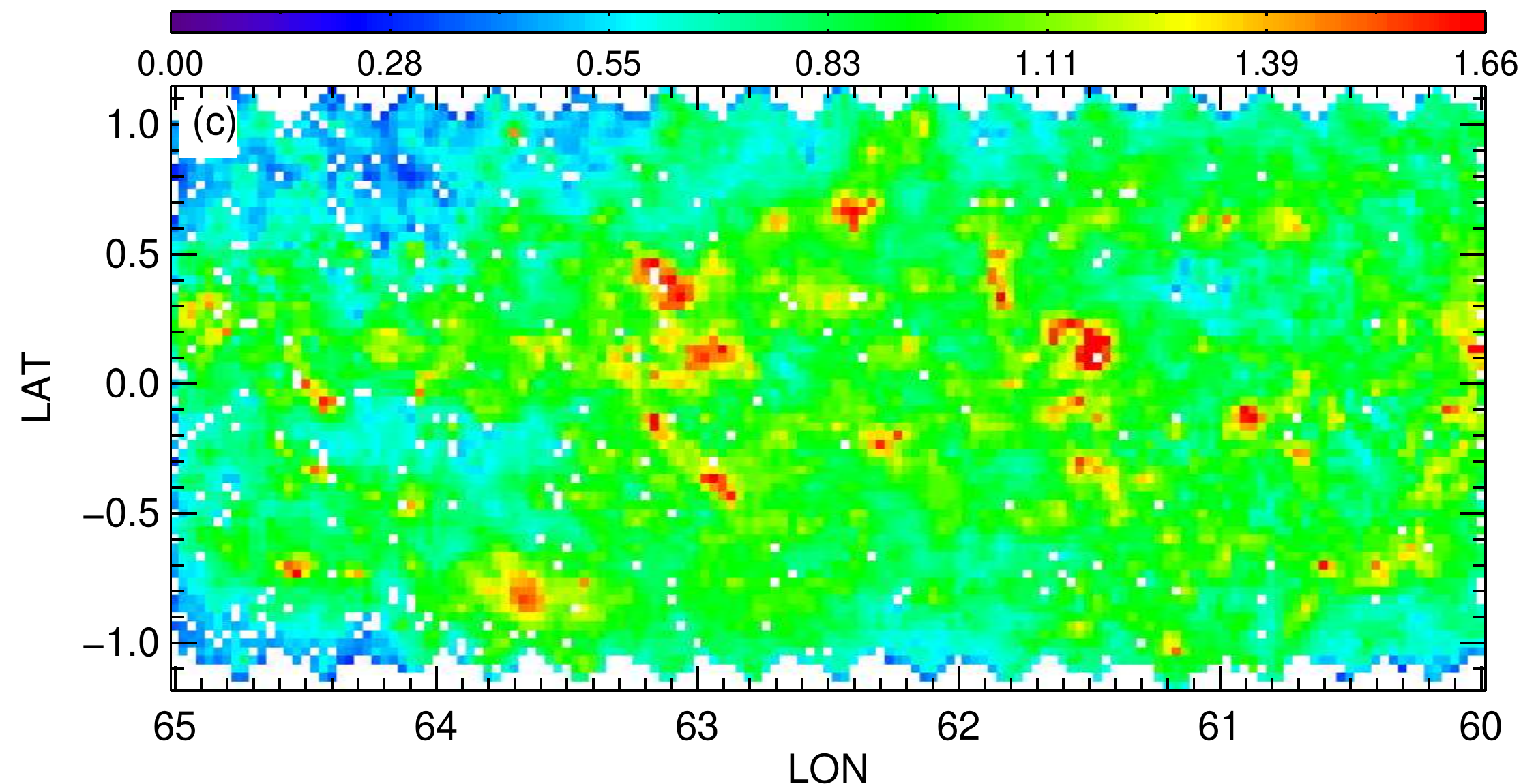} \\
\includegraphics[height=37mm]{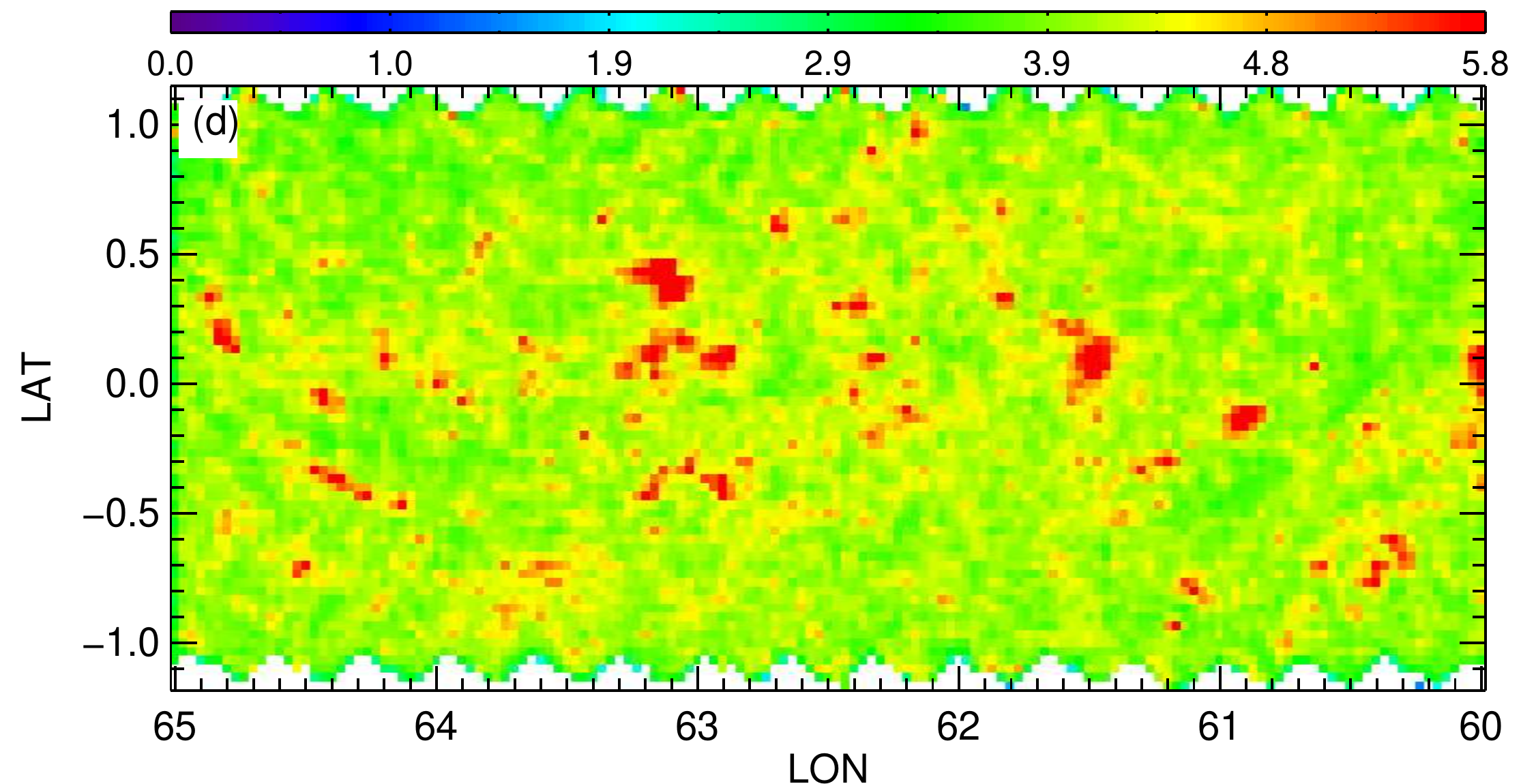} \\
\includegraphics[height=37mm]{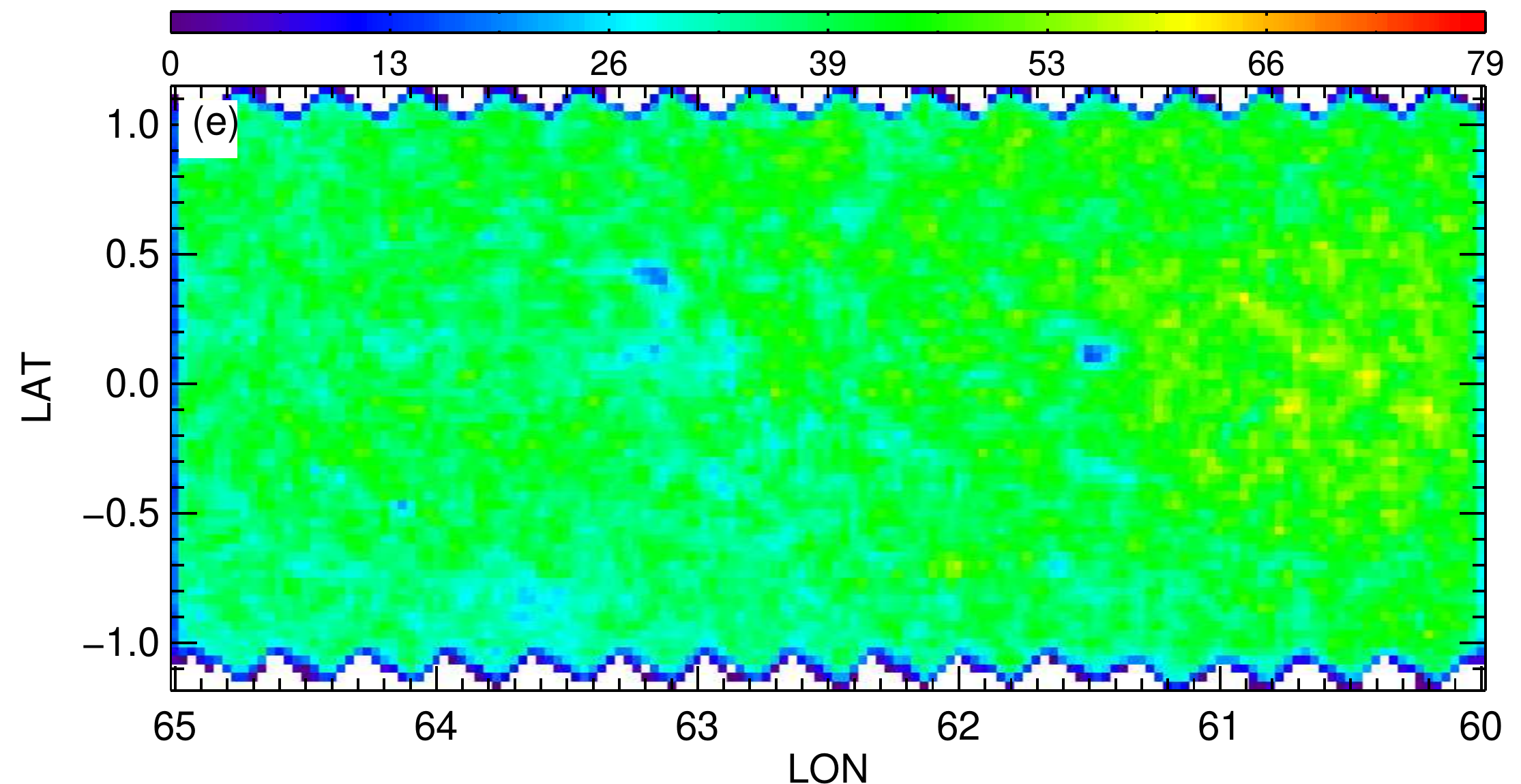}
\end{array}$
\end{center}
\caption{A subset of the publicly available maps.  The first three maps are 90th percentile $A(K_s)$ extinction maps as
derived with the RJCE method using $(H-[4.5\mu])$ colors and (a) MSTO, (b) RC, and (c) RGB stars.
Panel (d) shows the 90th percentile $A(K_s)$ extinction map derived using [3.6$\mu$]--[4.5$\mu$] colors of all stars
(note the different colorscale ranges compared to panels a--c).  Finally, panel (e) shows the starcounts map of all stars.
The GLIMPSE--I data cover $10.0<l<65.4^{\circ}$ and $295.0<l<350.3^{\circ}$ and $-1<b<+1^{\circ}$, while the
Vela-Carina data cover $255.3<l<295.0^{\circ}$ and $-1.5<b<+0.5^{\circ}$. Pixels with no stars are shown as white in
the extinction maps and black in the starcounts map.}
\label{fig:extmaps6065}
\end{figure*}

\newpage

\addtocounter{figure}{-1}
\begin{figure*}[htpb]
\begin{center}
$\begin{array}{c}
\includegraphics[height=42mm]{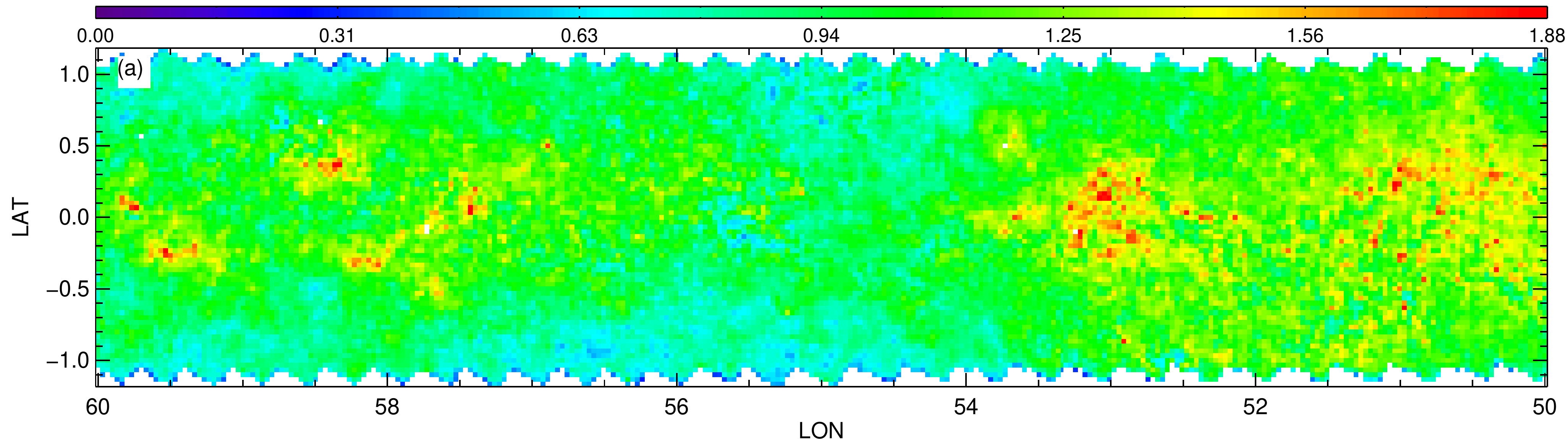} \\
\includegraphics[height=42mm]{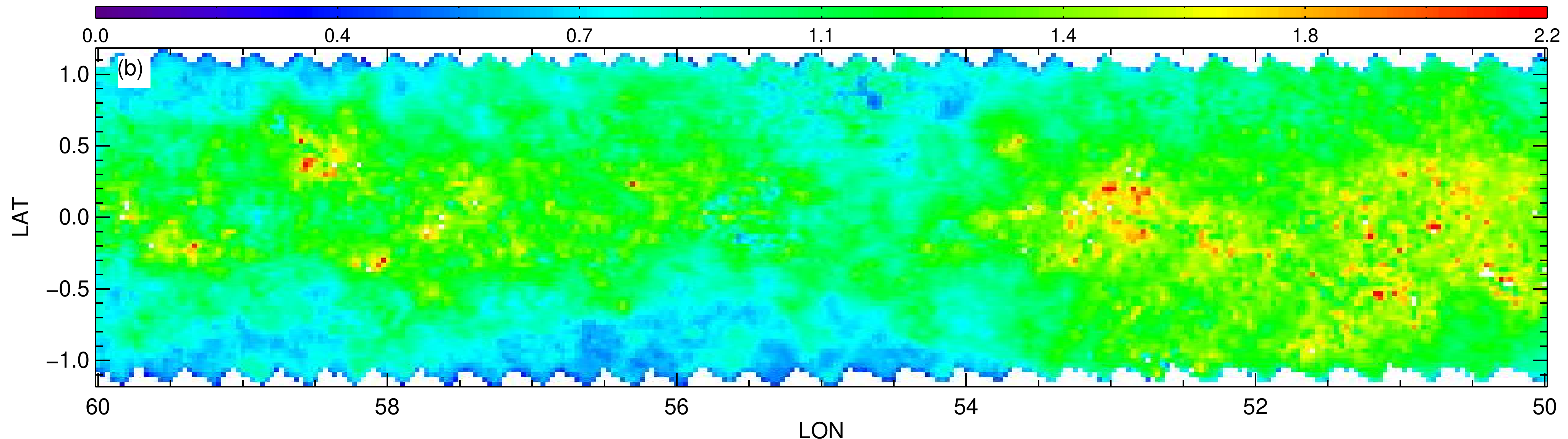} \\
\includegraphics[height=42mm]{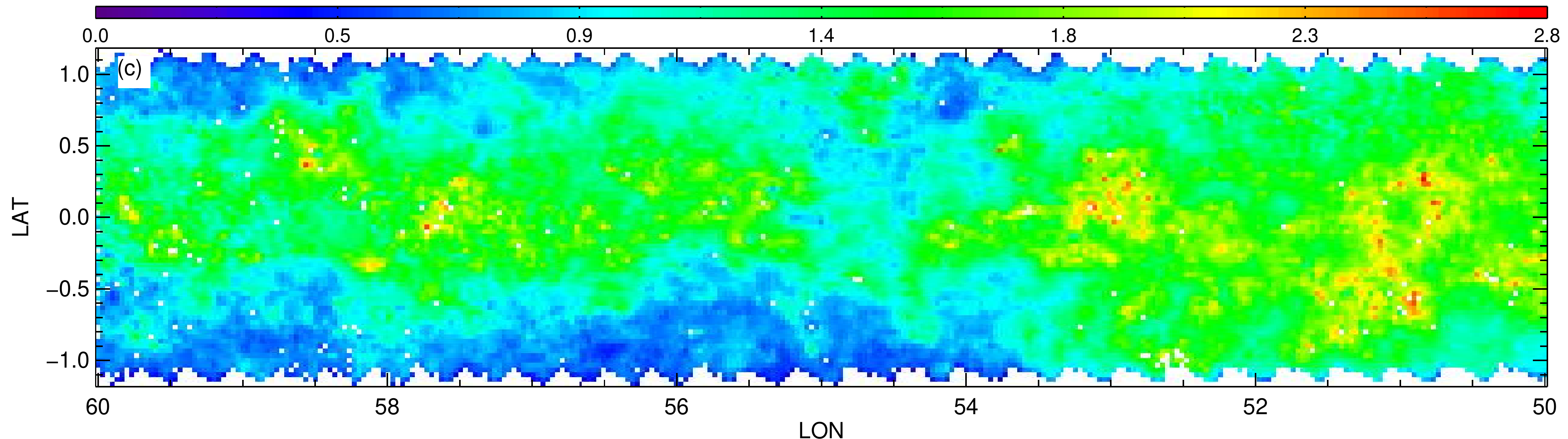} \\
\includegraphics[height=42mm]{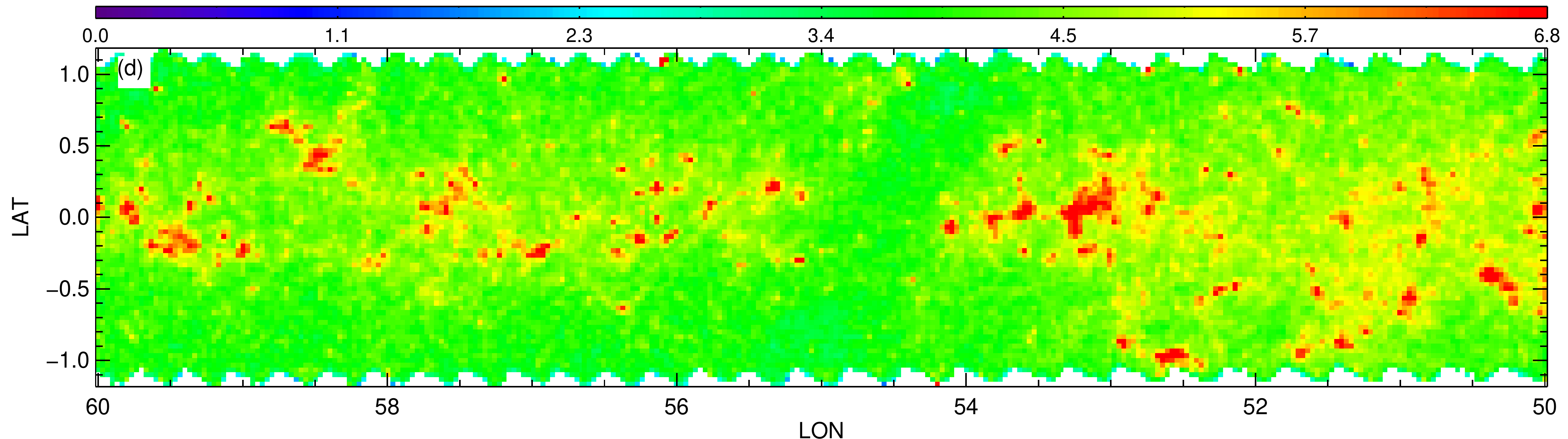} \\
\includegraphics[height=42mm]{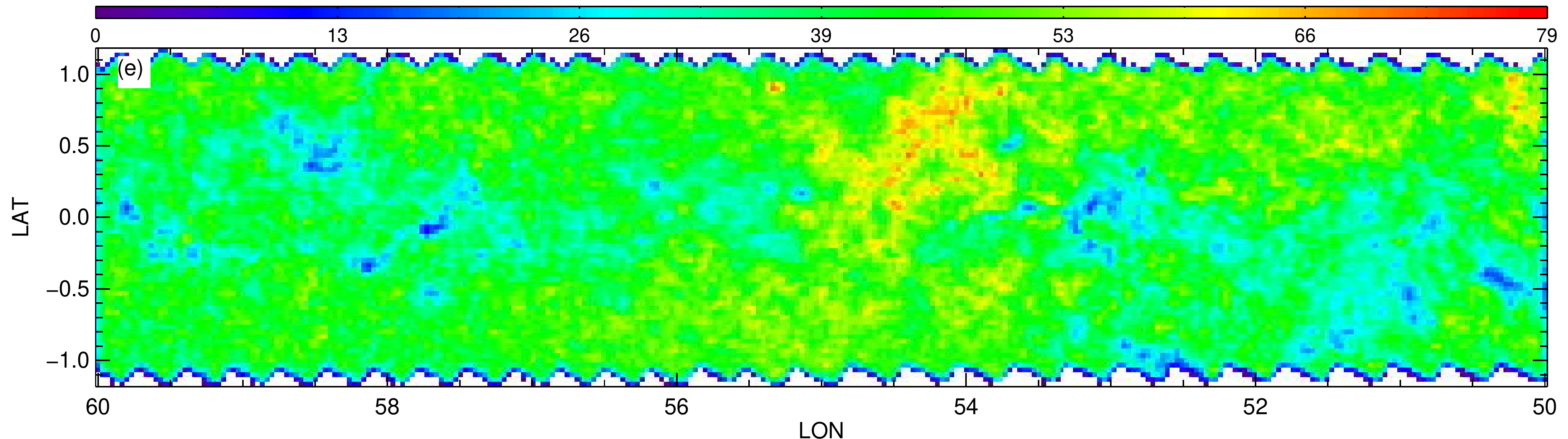}
\end{array}$
\end{center}
\caption{(cont.)}
\label{fig:extmaps5060}
\end{figure*}

\newpage

\addtocounter{figure}{-1}
\begin{figure*}[htpb]
\begin{center}
$\begin{array}{c}
\includegraphics[height=42mm]{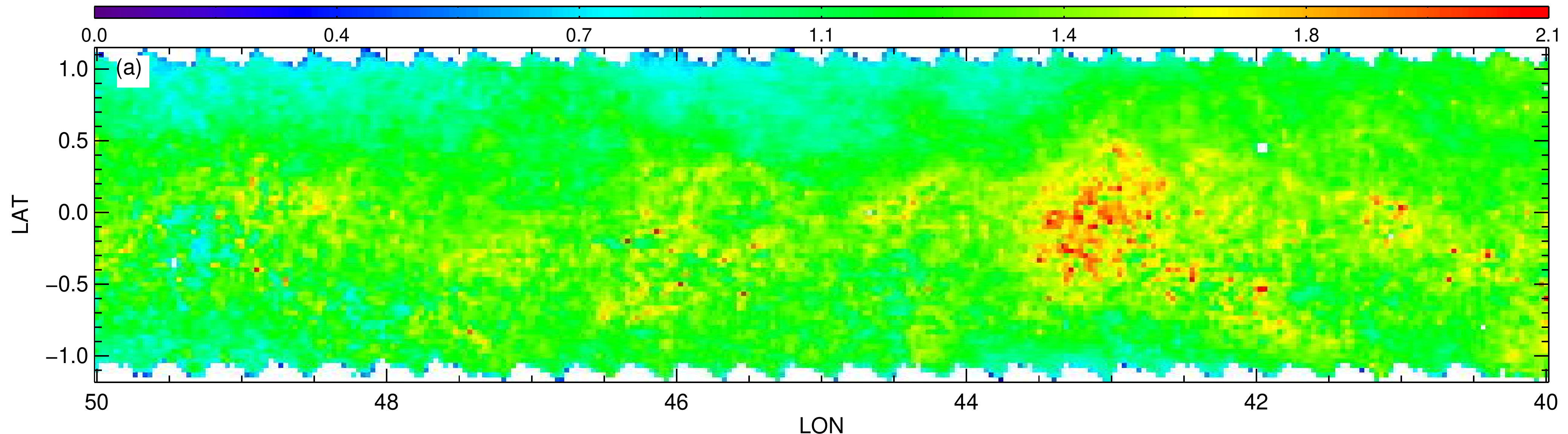} \\
\includegraphics[height=42mm]{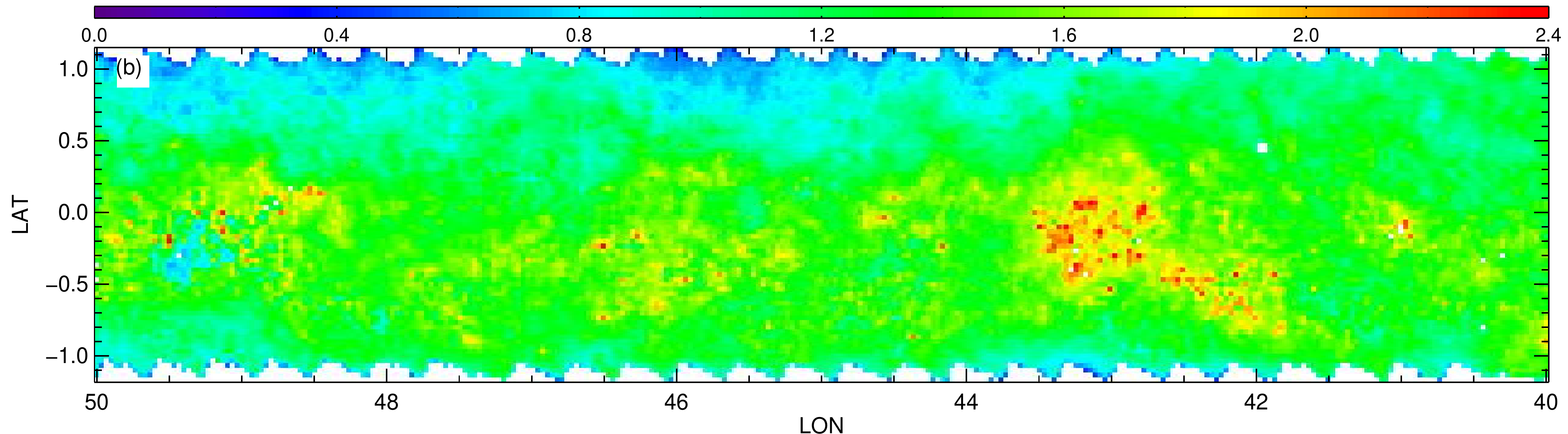} \\
\includegraphics[height=42mm]{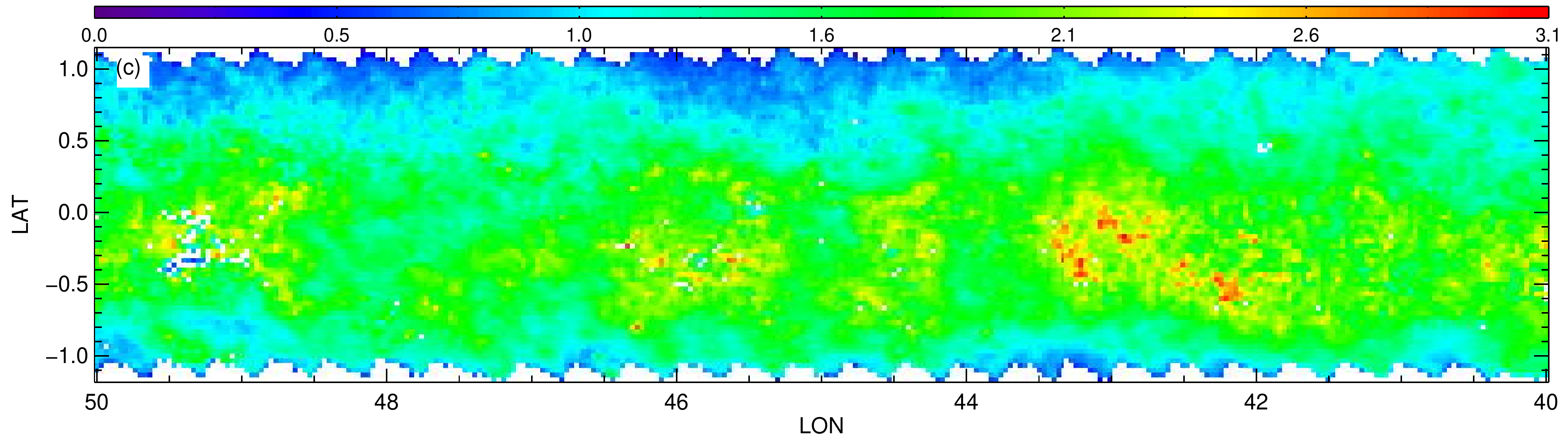} \\
\includegraphics[height=42mm]{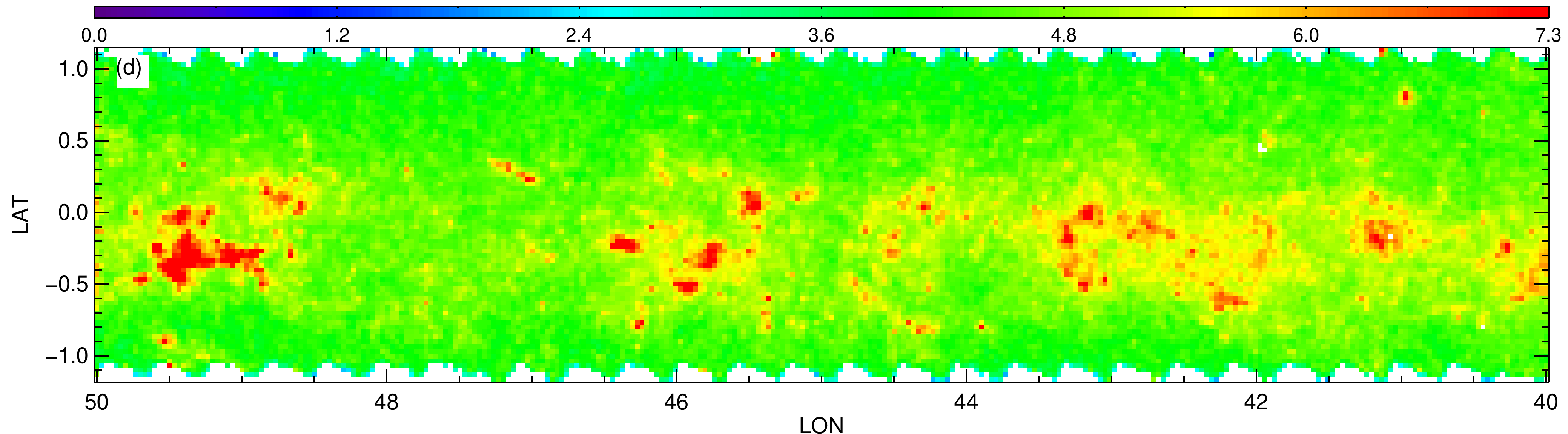} \\
\includegraphics[height=42mm]{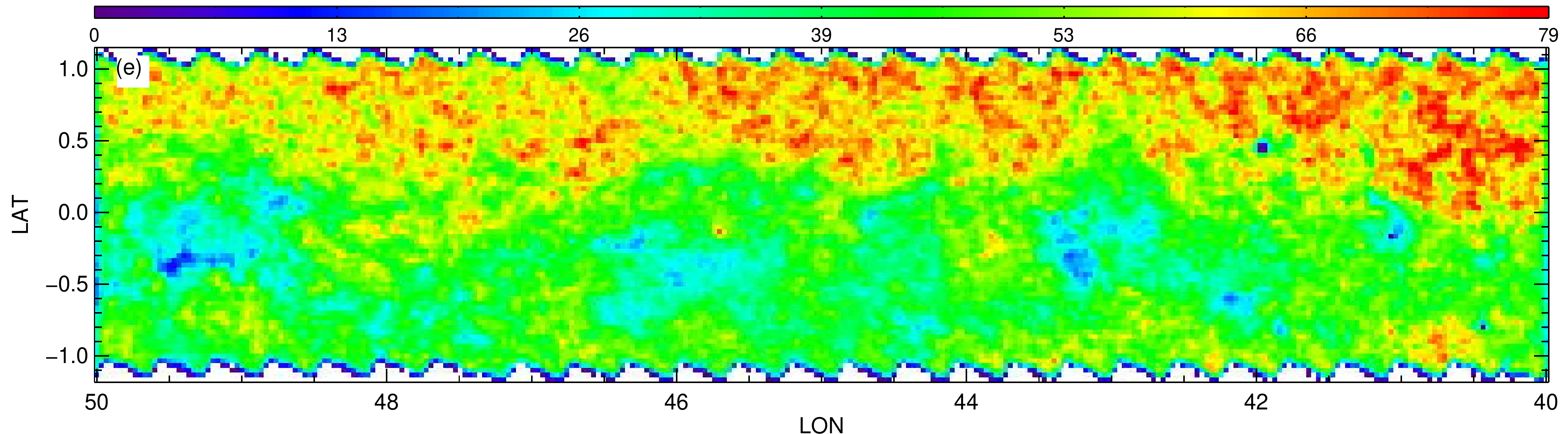}
\end{array}$
\end{center}
\caption{(cont.)}
\label{fig:extmaps4050}
\end{figure*}

\newpage

\addtocounter{figure}{-1}
\begin{figure*}[htpb]
\begin{center}
$\begin{array}{c}
\includegraphics[height=42mm]{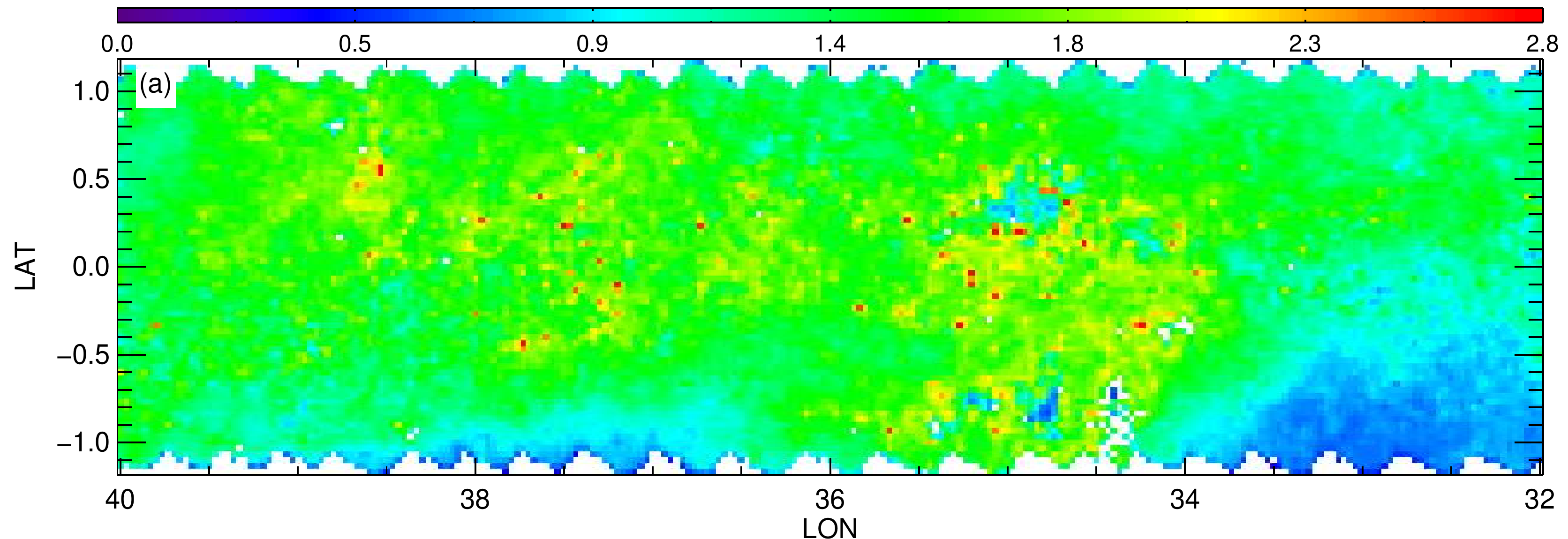} \\
\includegraphics[height=42mm]{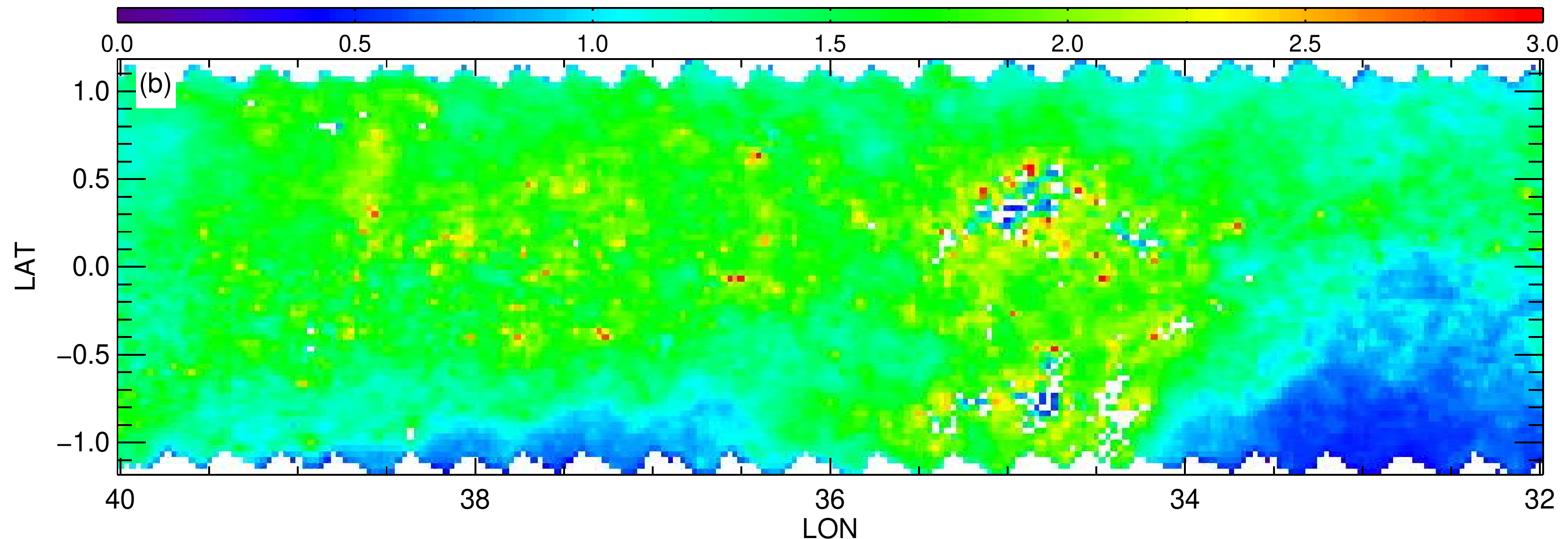} \\
\includegraphics[height=42mm]{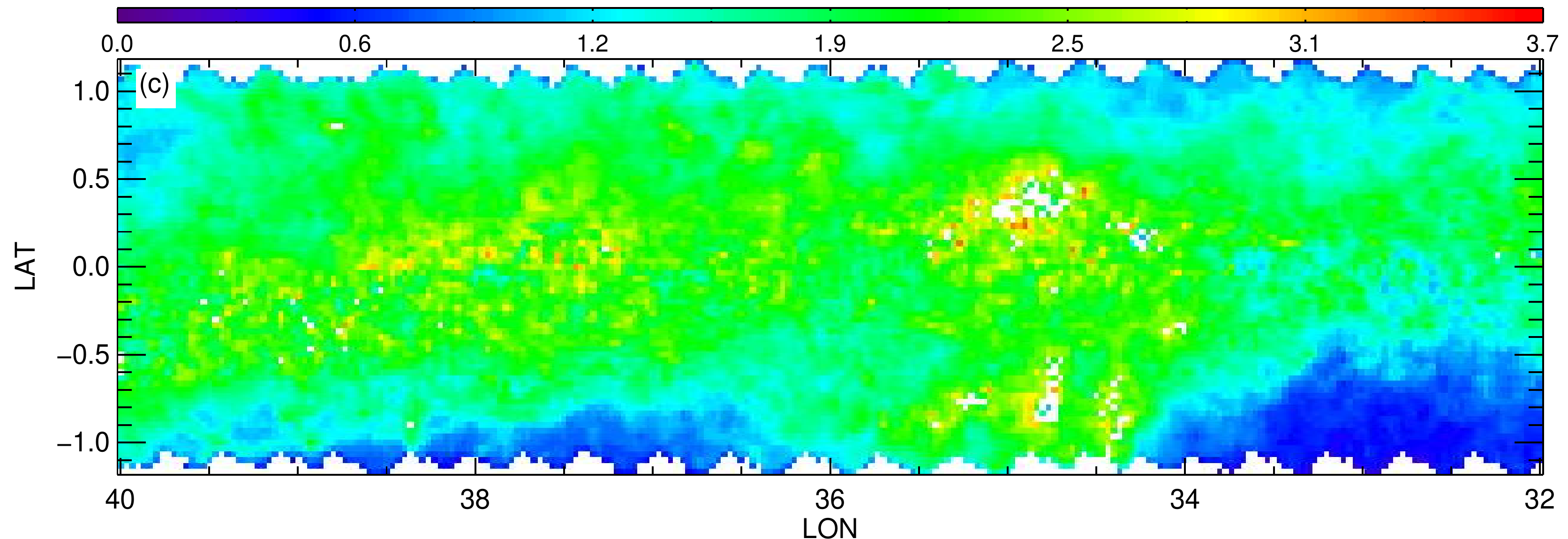} \\
\includegraphics[height=42mm]{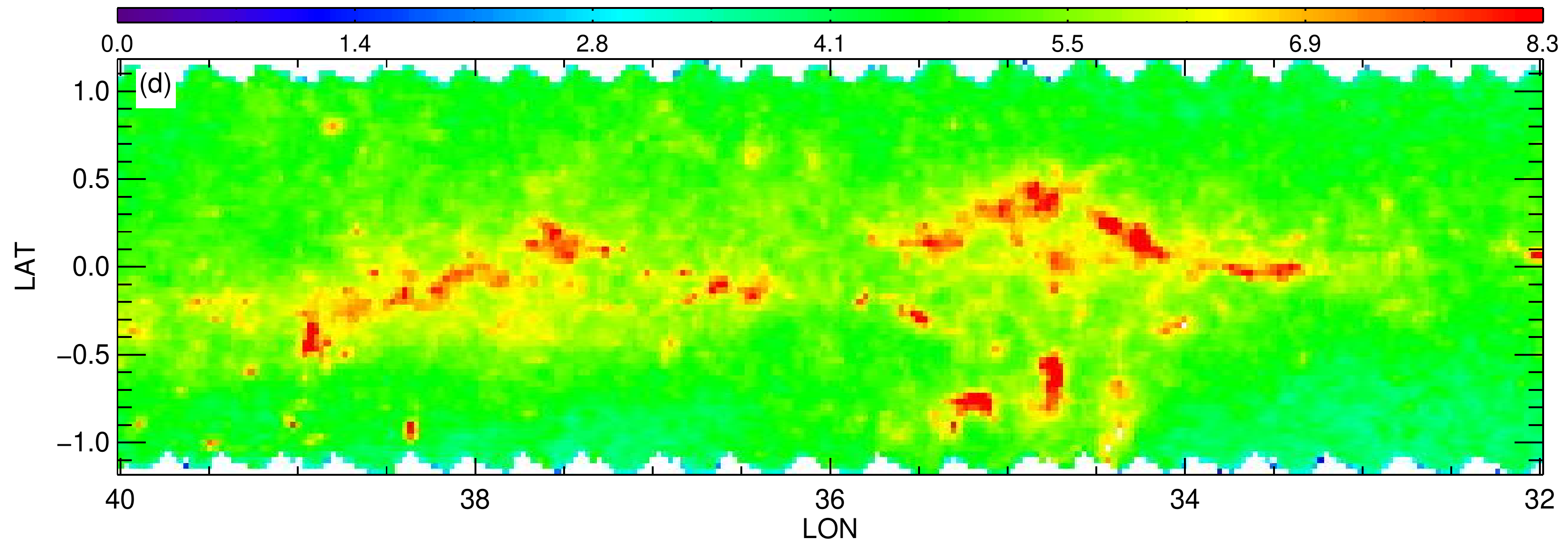} \\
\includegraphics[height=42mm]{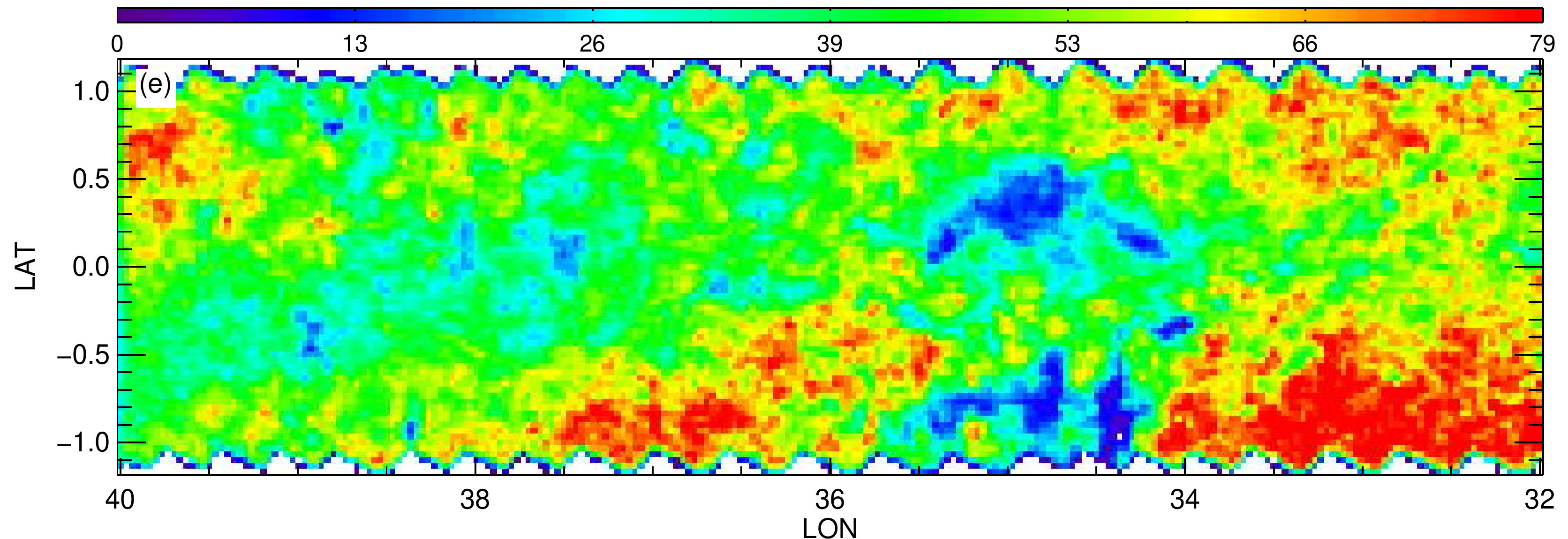}
\end{array}$
\end{center}
\caption{(cont.)}
\label{fig:extmaps3240}
\end{figure*}

\newpage

\addtocounter{figure}{-1}
\begin{figure*}[htpb]
\begin{center}
$\begin{array}{c}
\includegraphics[height=42mm]{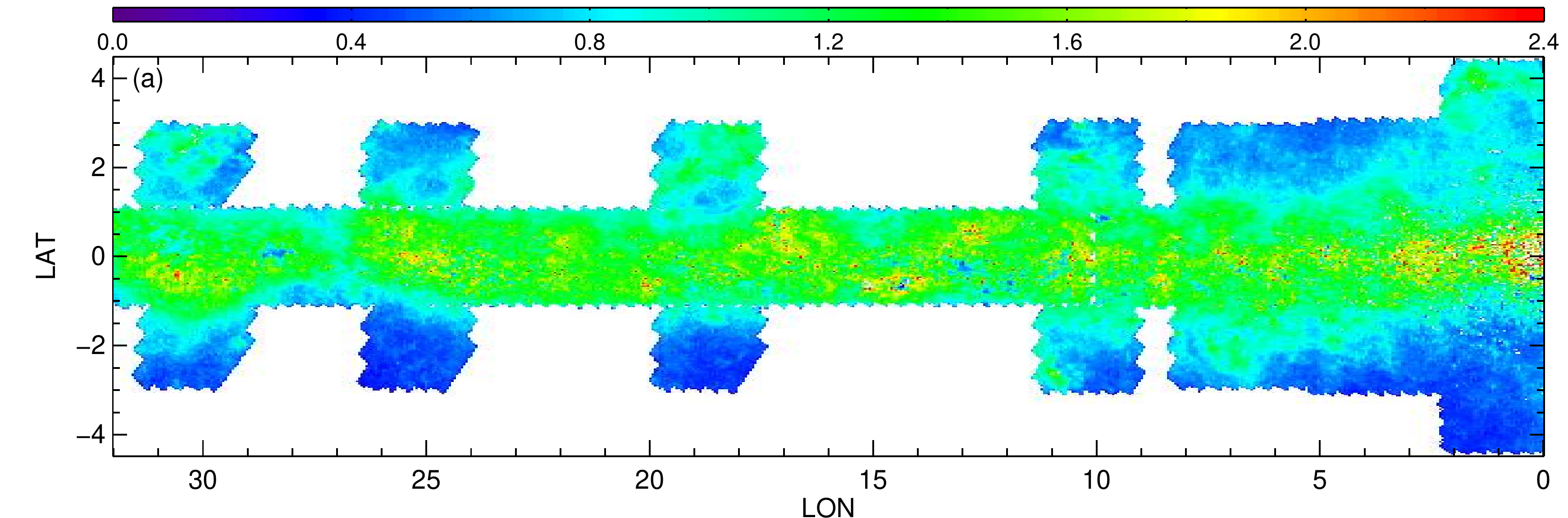} \\
\includegraphics[height=42mm]{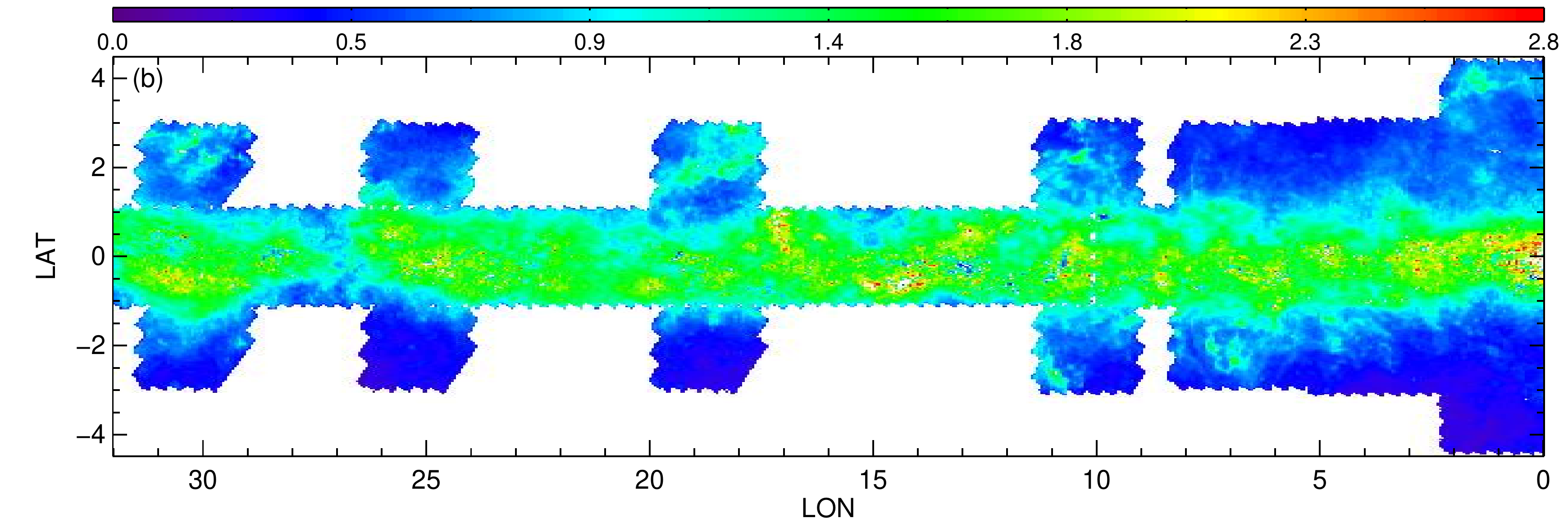} \\
\includegraphics[height=42mm]{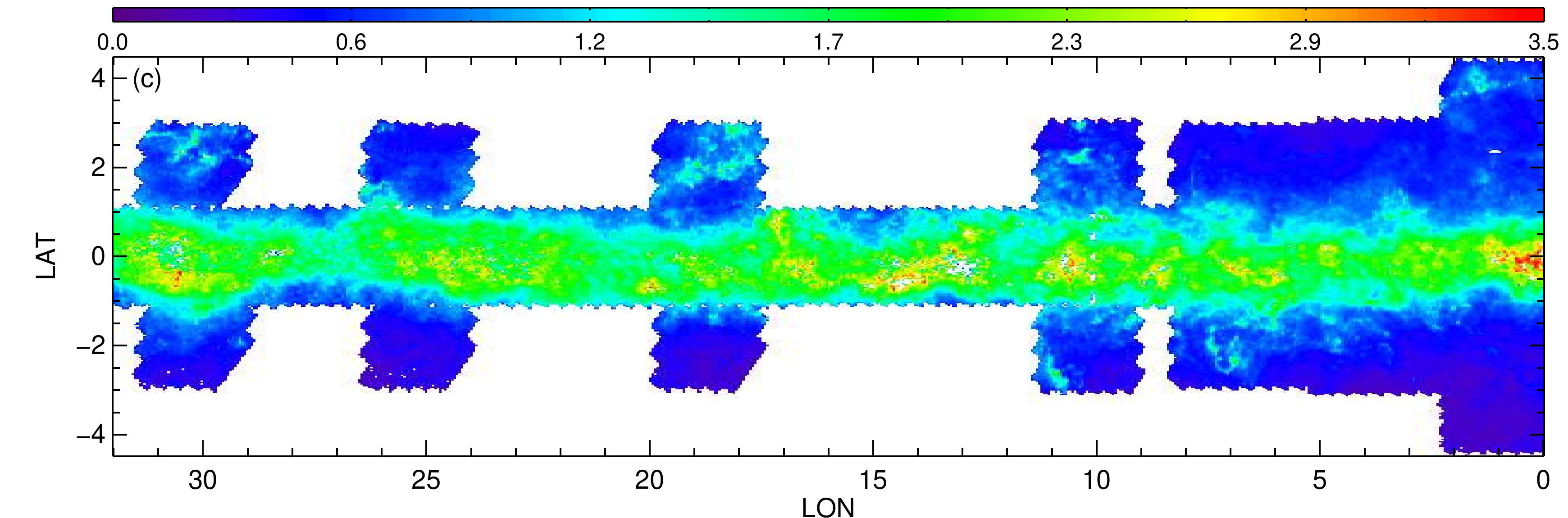} \\
\includegraphics[height=42mm]{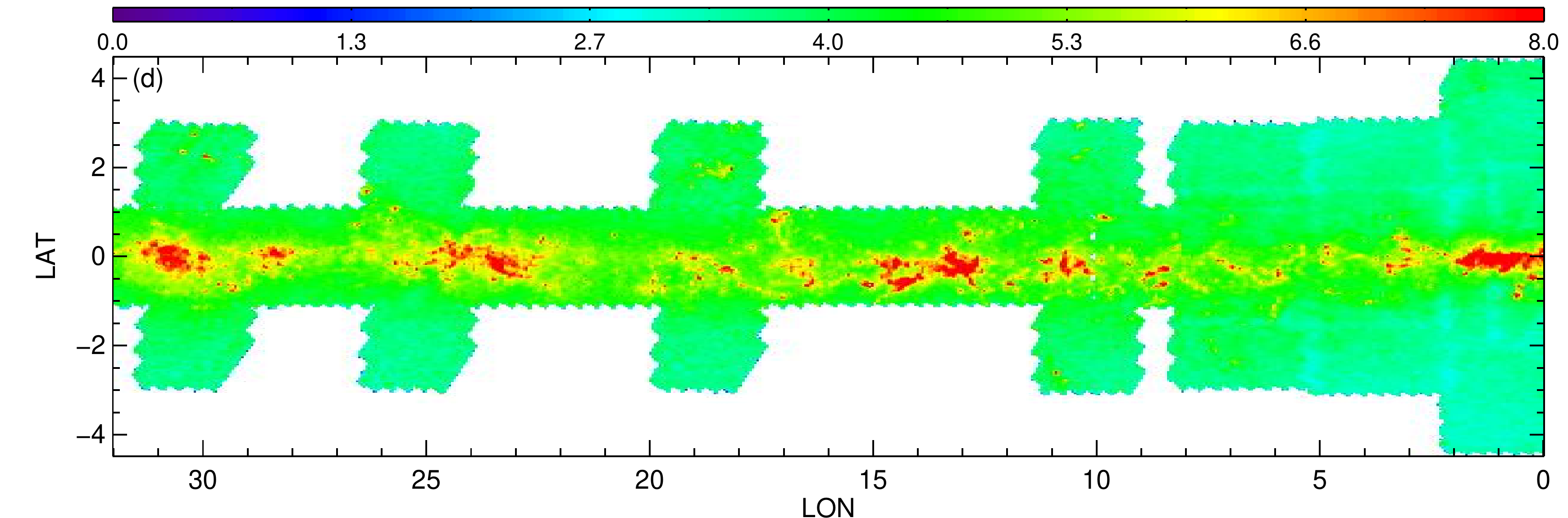} \\
\includegraphics[height=42mm]{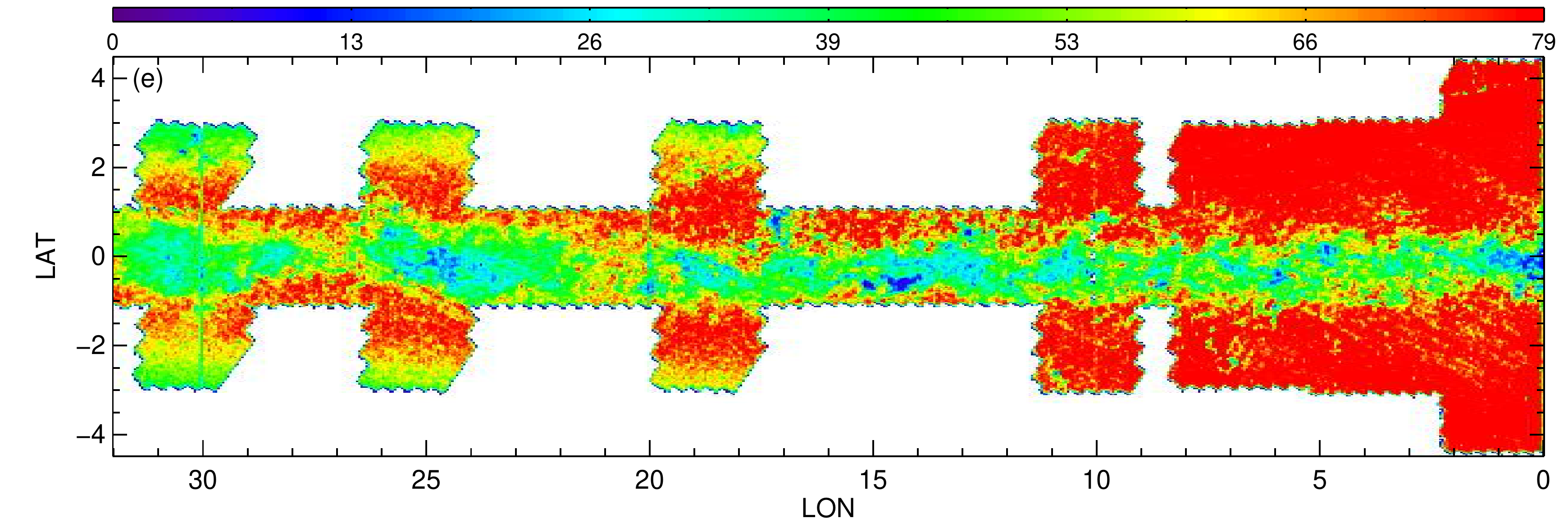}
\end{array}$
\end{center}
\caption{(cont.)}
\label{fig:extmaps032}
\end{figure*}

\newpage

\addtocounter{figure}{-1}
\begin{figure*}[htpb]
\begin{center}
$\begin{array}{c}
\includegraphics[height=42mm]{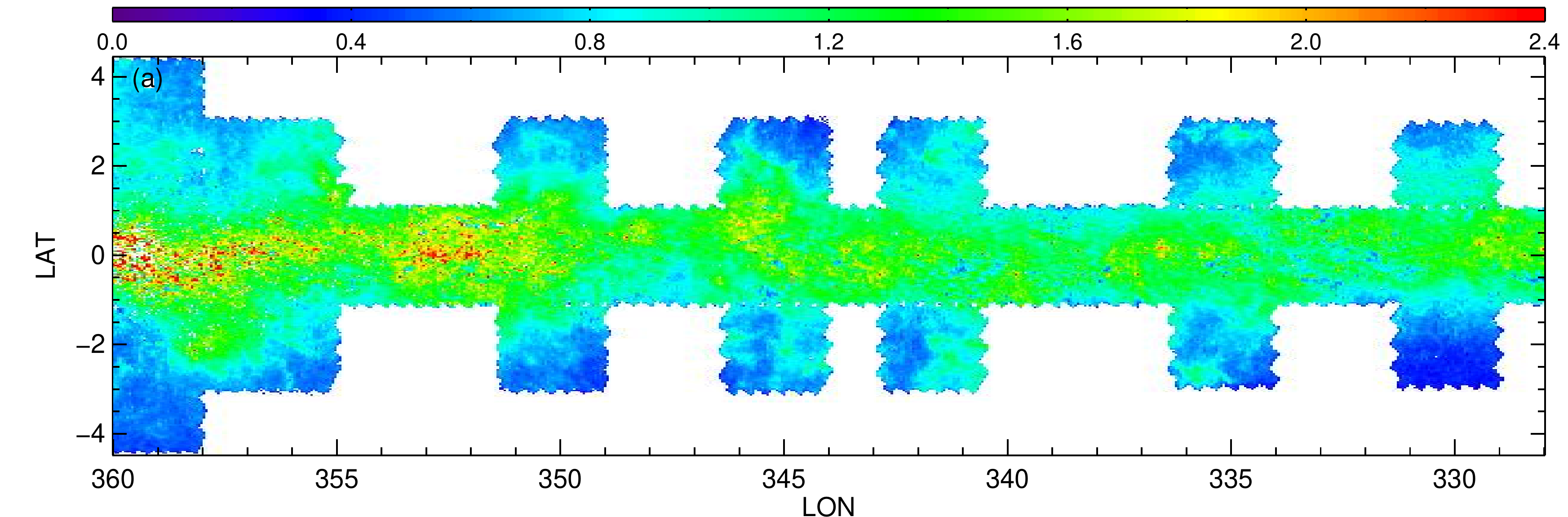} \\
\includegraphics[height=42mm]{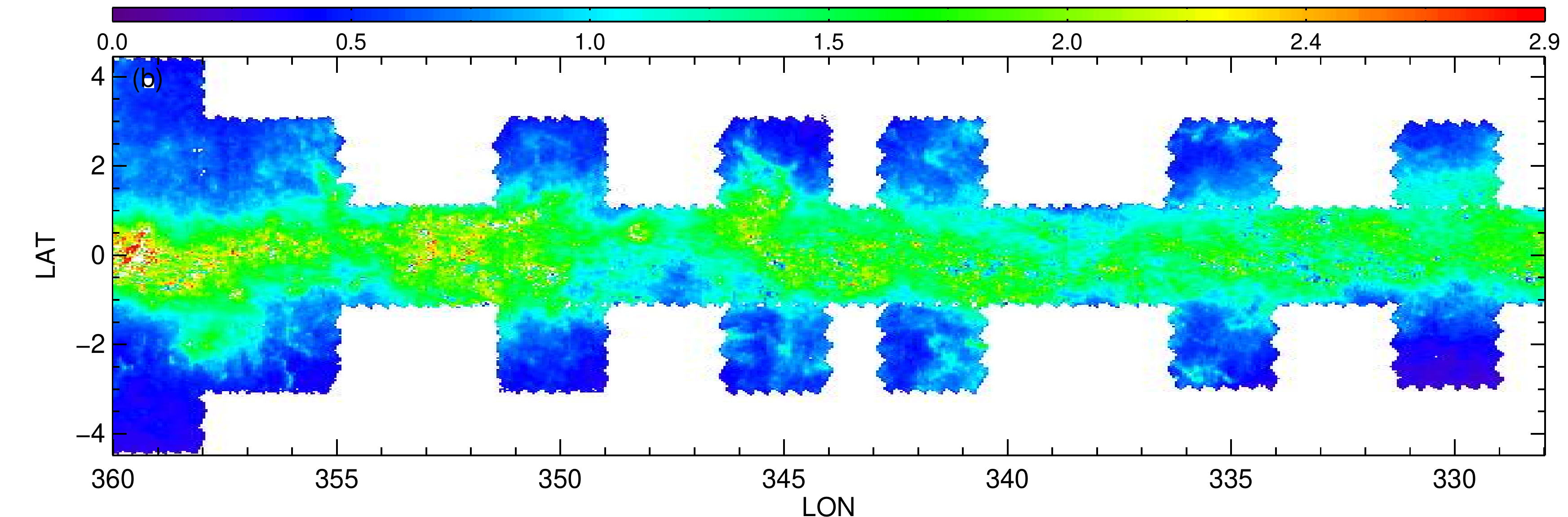} \\
\includegraphics[height=42mm]{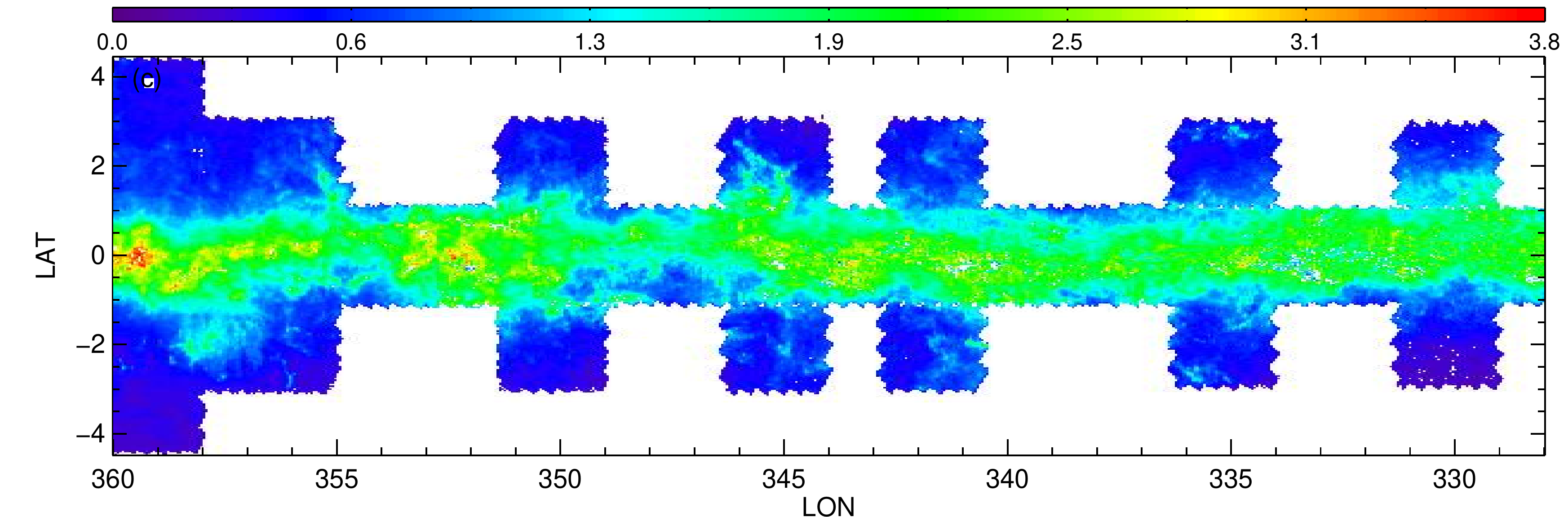} \\
\includegraphics[height=42mm]{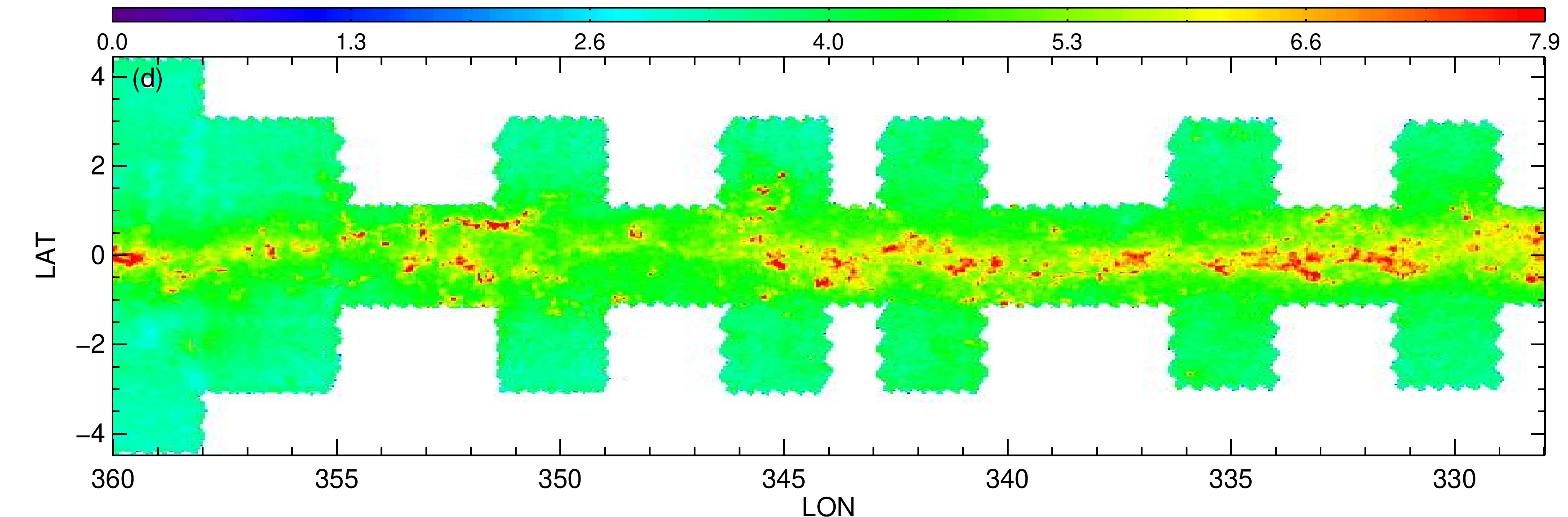} \\
\includegraphics[height=42mm]{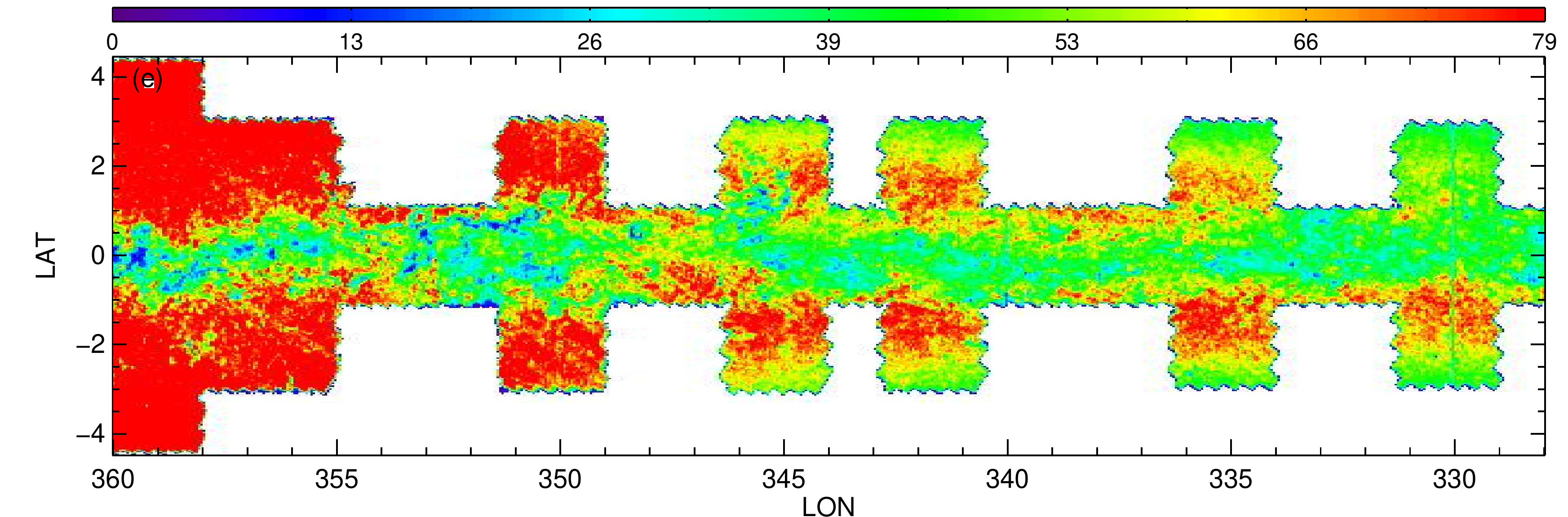}
\end{array}$
\end{center}
\caption{(cont.)}
\label{fig:extmaps328360}
\end{figure*}

\newpage

\addtocounter{figure}{-1}
\begin{figure*}[htpb]
\begin{center}
$\begin{array}{c}
\includegraphics[height=42mm]{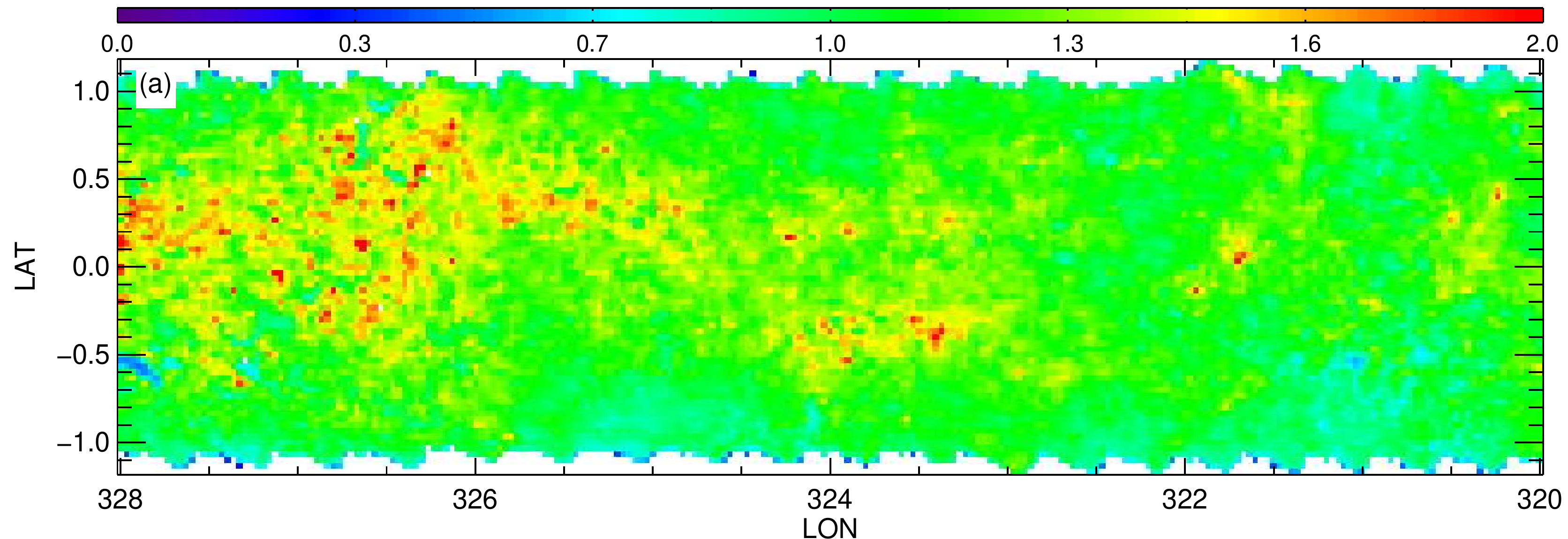} \\
\includegraphics[height=42mm]{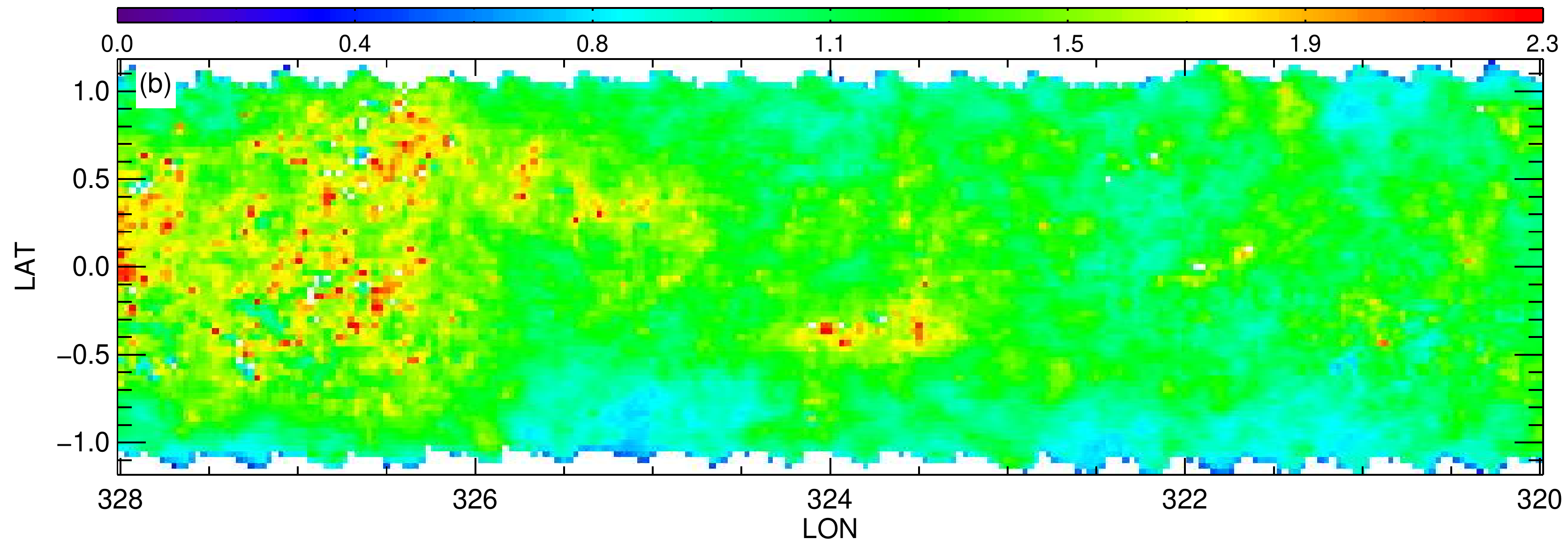} \\
\includegraphics[height=42mm]{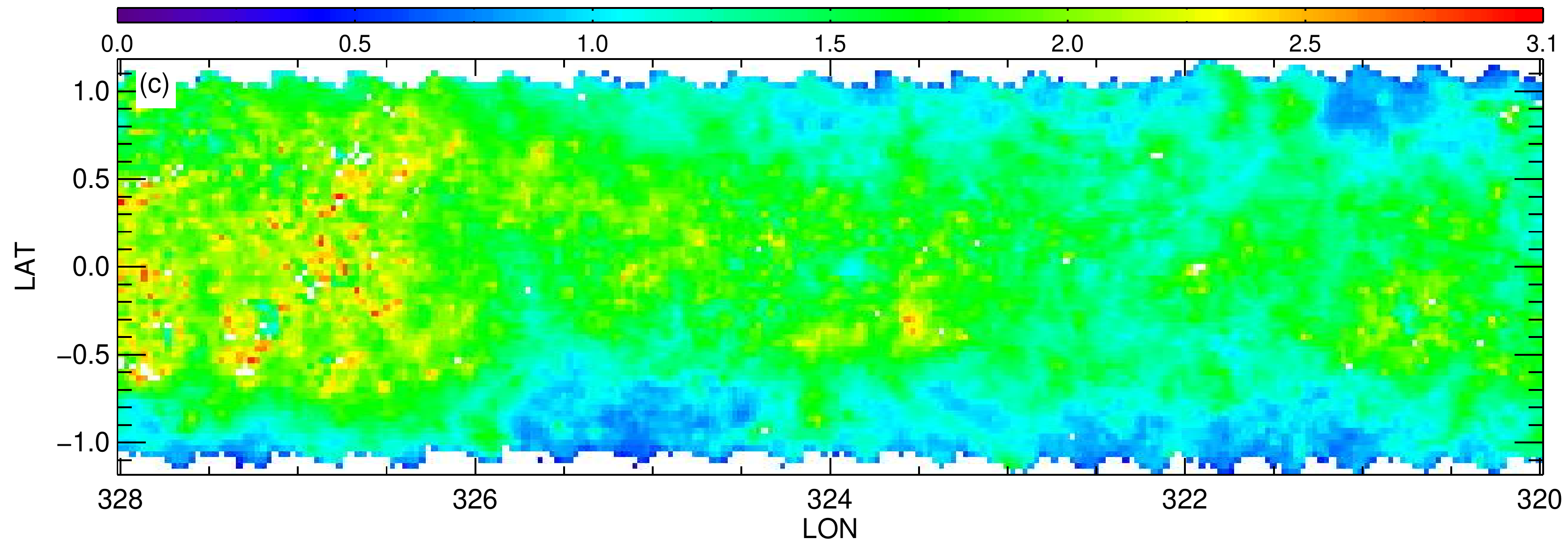} \\
\includegraphics[height=42mm]{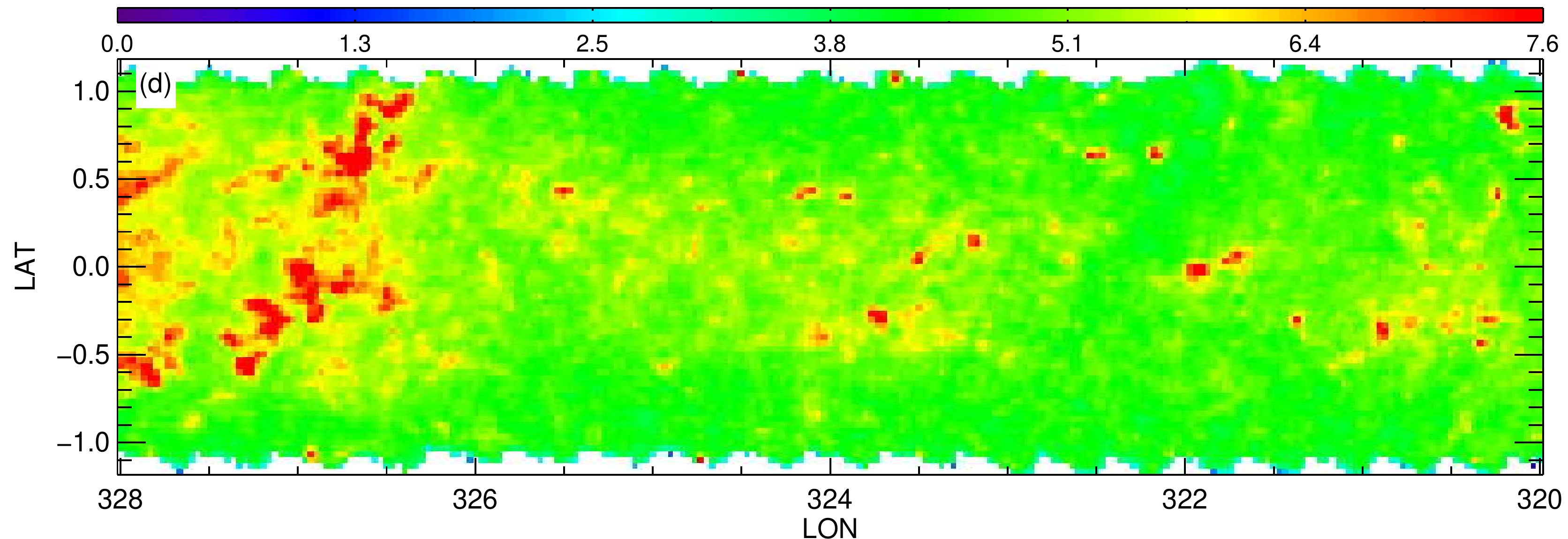} \\
\includegraphics[height=42mm]{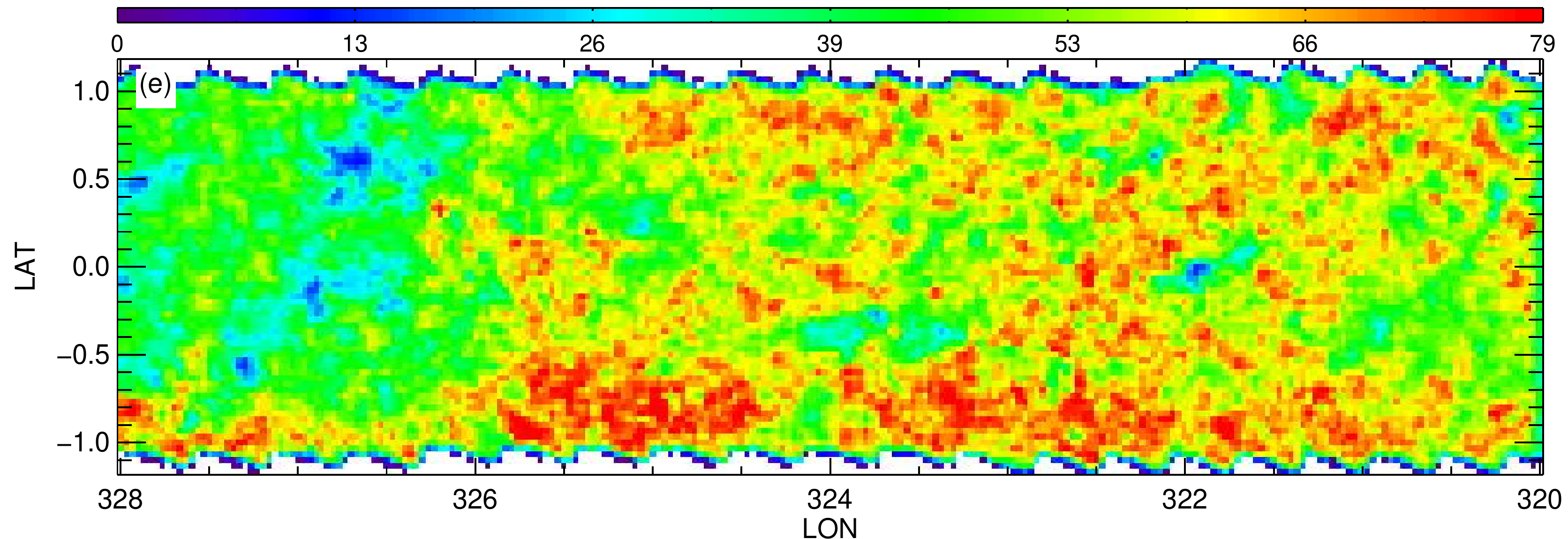}
\end{array}$
\end{center}
\caption{(cont.)}
\label{fig:extmaps320328}
\end{figure*}

\newpage

\addtocounter{figure}{-1}
\begin{figure*}[htpb]
\begin{center}
$\begin{array}{c}
\includegraphics[height=42mm]{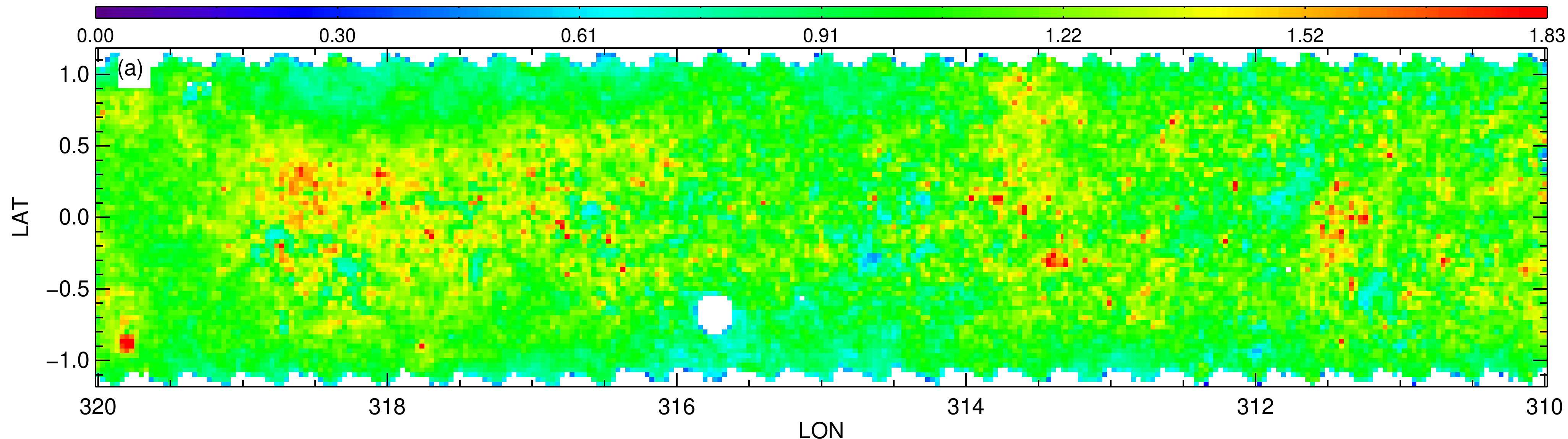} \\
\includegraphics[height=42mm]{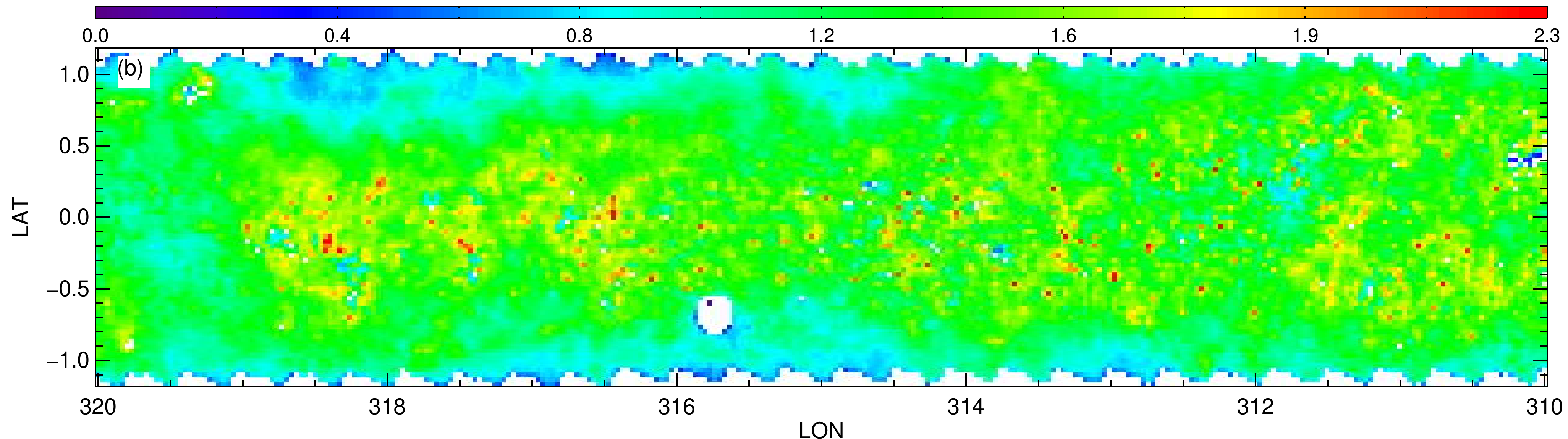} \\
\includegraphics[height=42mm]{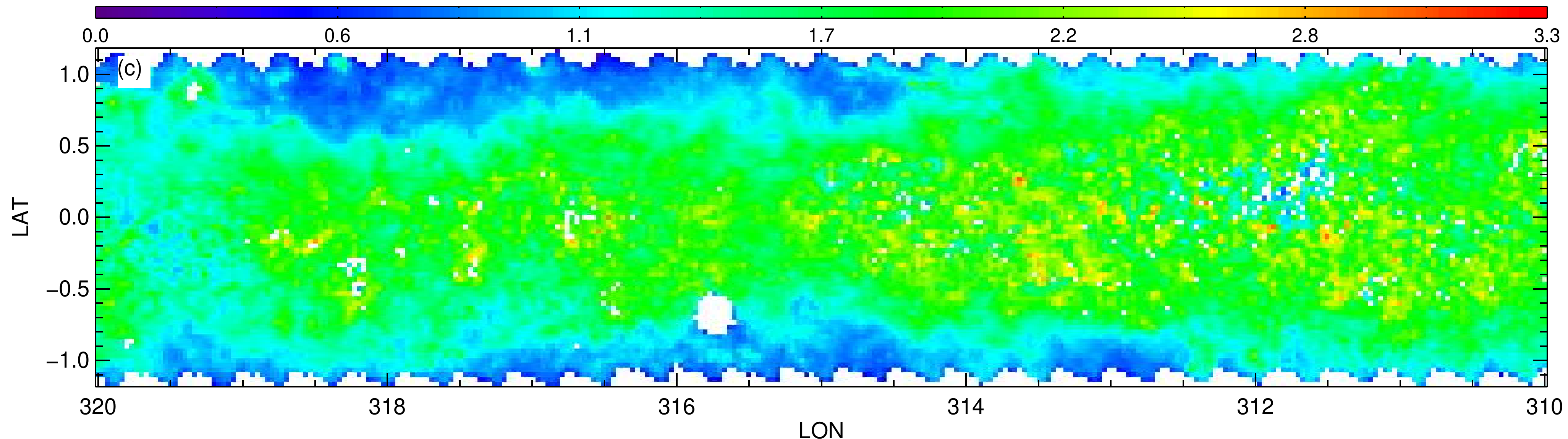} \\
\includegraphics[height=42mm]{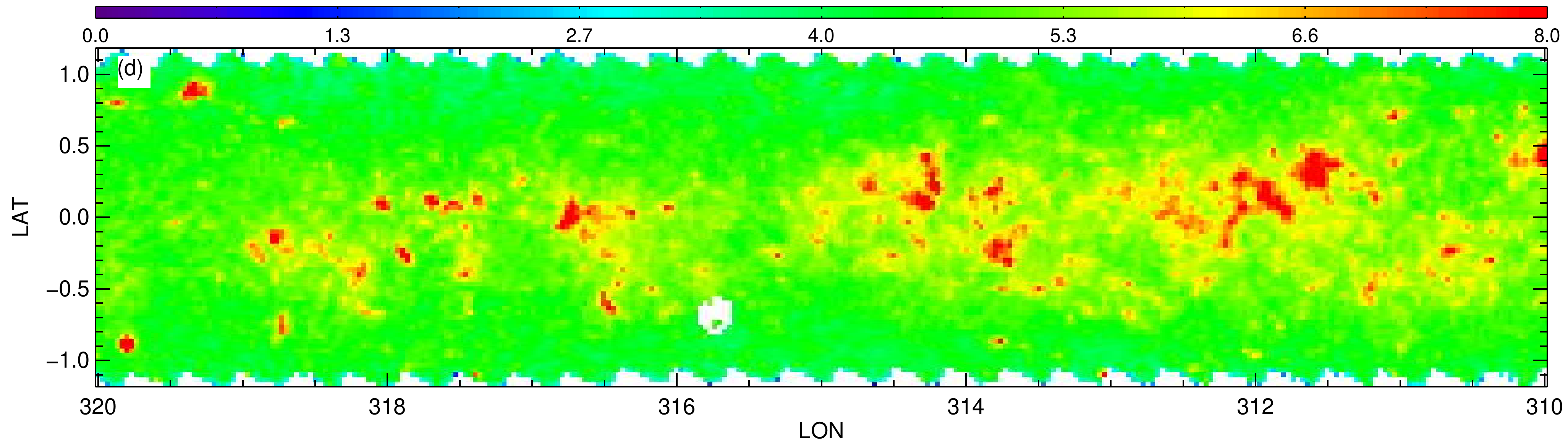} \\
\includegraphics[height=42mm]{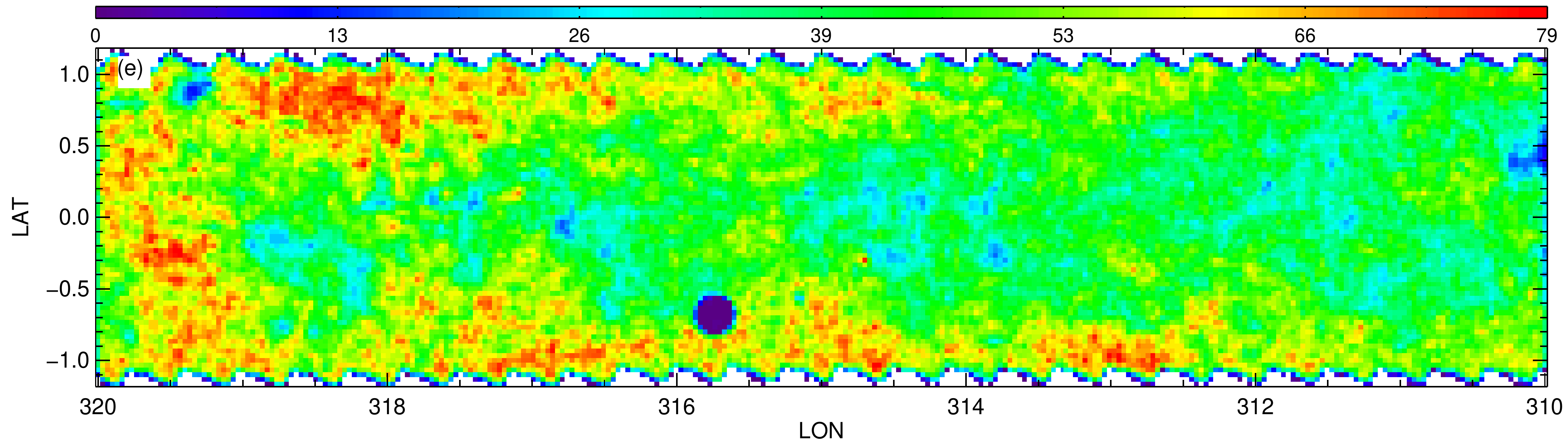}
\end{array}$
\end{center}
\caption{(cont.)}
\label{fig:extmaps310320}
\end{figure*}

\newpage

\addtocounter{figure}{-1}
\begin{figure*}[htpb]
\begin{center}
$\begin{array}{c}
\includegraphics[height=42mm]{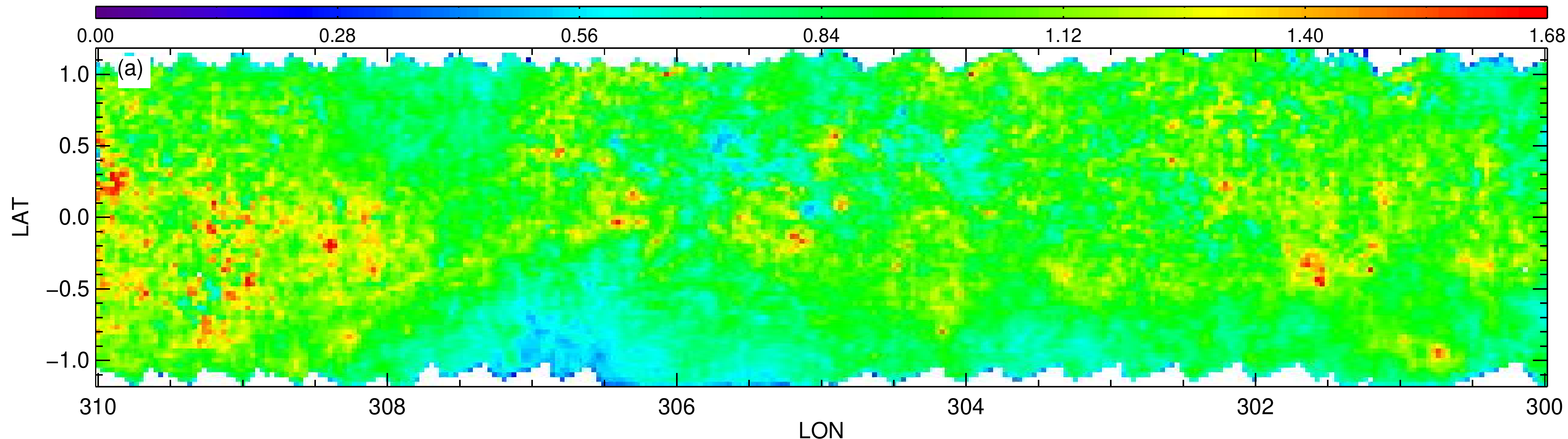} \\
\includegraphics[height=42mm]{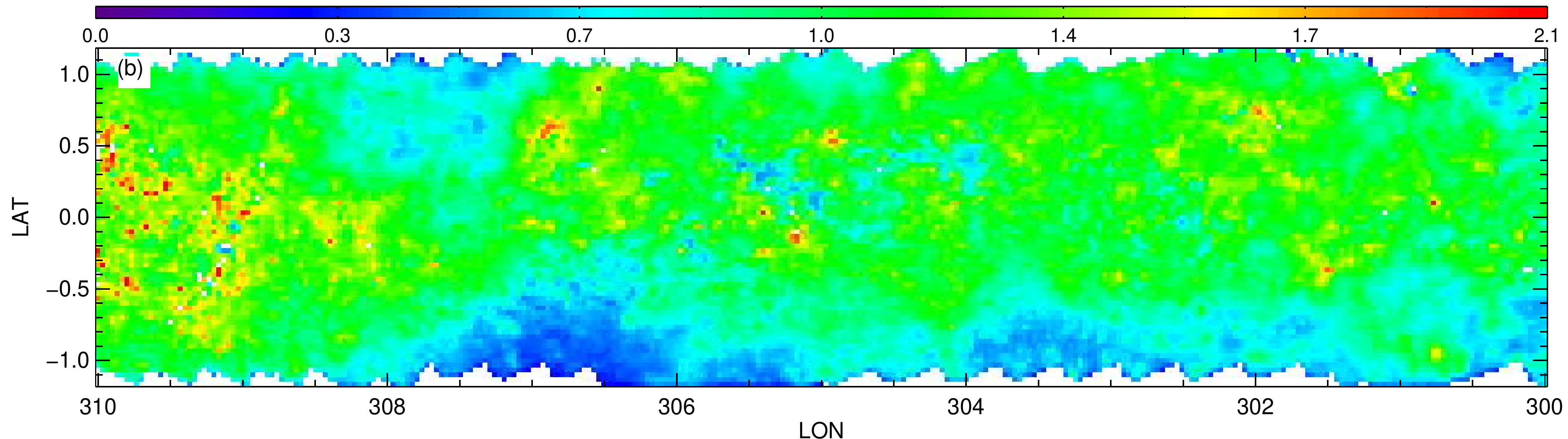} \\
\includegraphics[height=42mm]{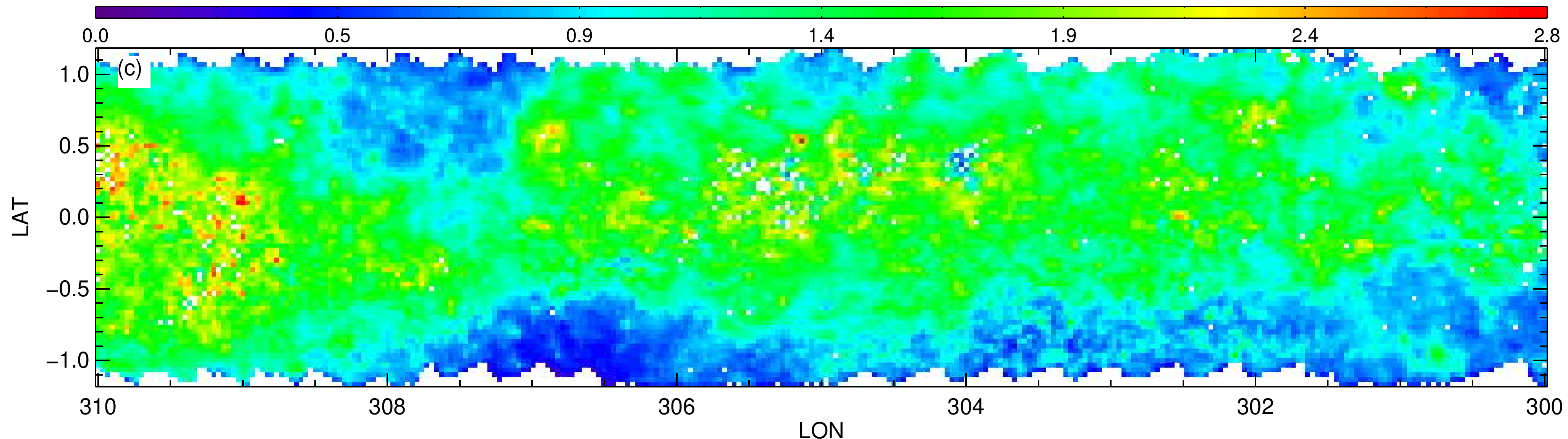} \\
\includegraphics[height=42mm]{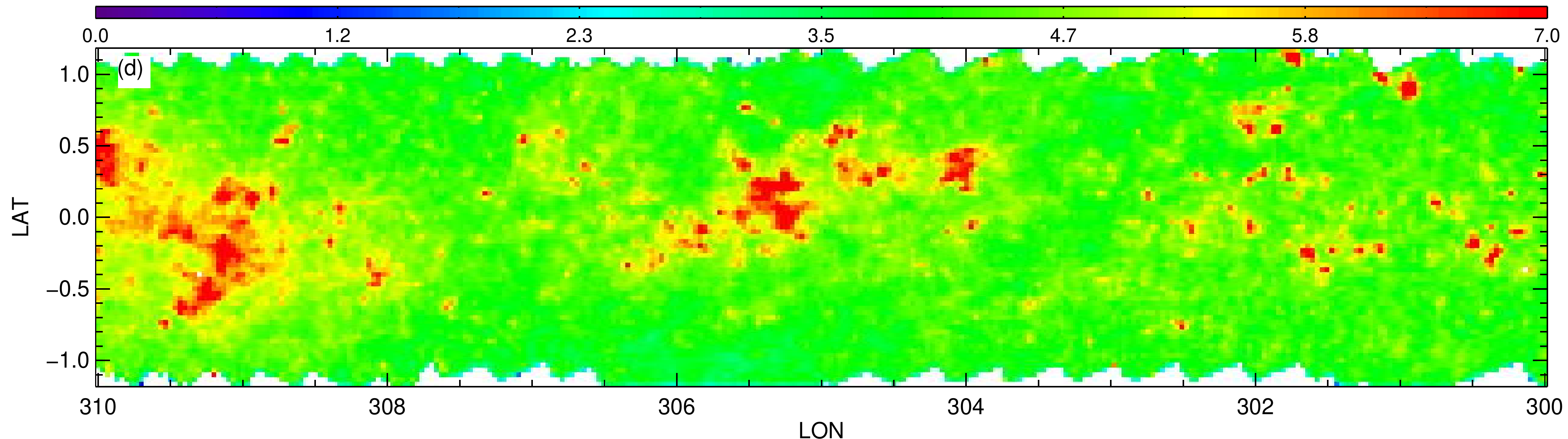}  \\
\includegraphics[height=42mm]{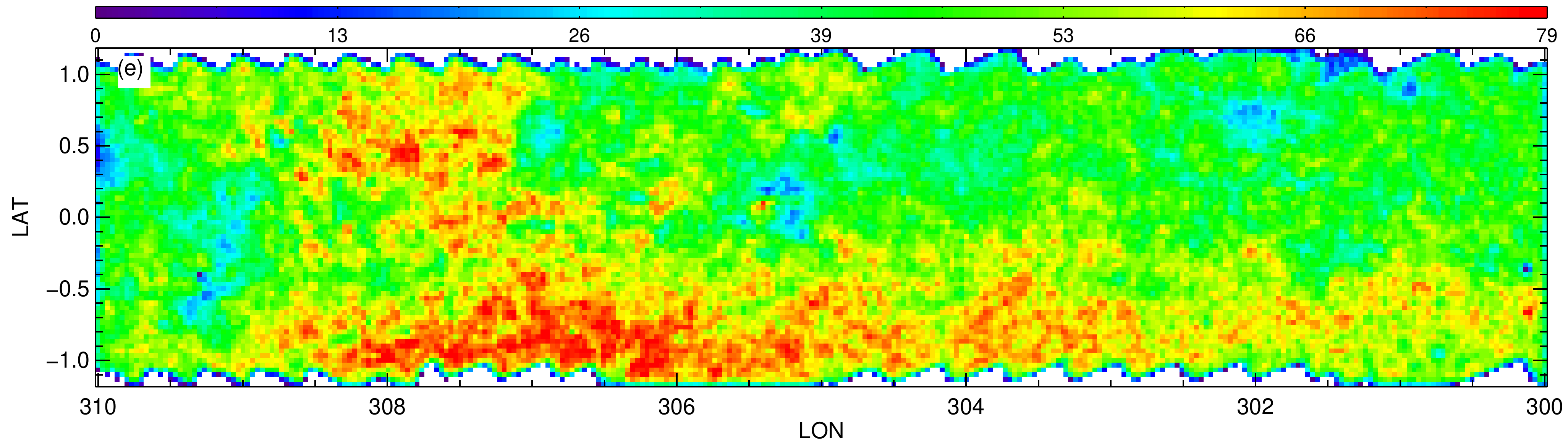}
\end{array}$
\end{center}
\caption{(cont.)}
\label{fig:extmaps300310}
\end{figure*}

\newpage

\addtocounter{figure}{-1}
\begin{figure*}[htpb]
\begin{center}
$\begin{array}{c}
\includegraphics[height=42mm]{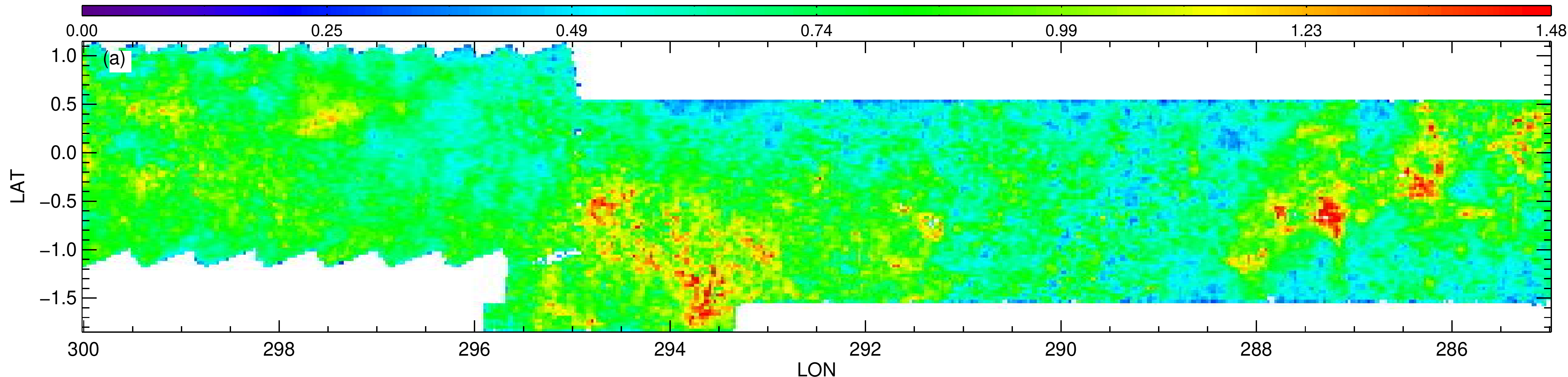} \\
\includegraphics[height=42mm]{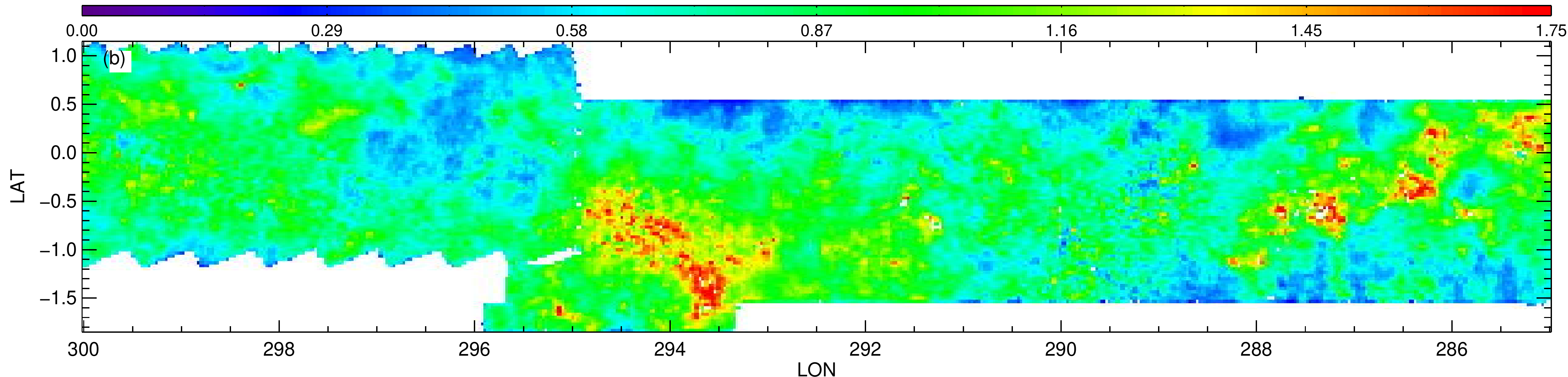} \\
\includegraphics[height=42mm]{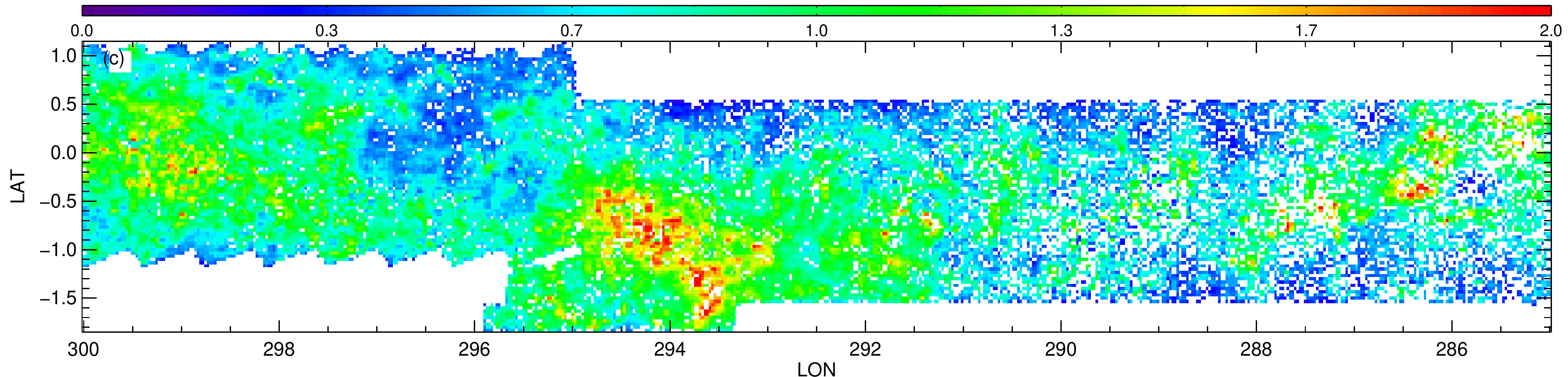} \\
\includegraphics[height=42mm]{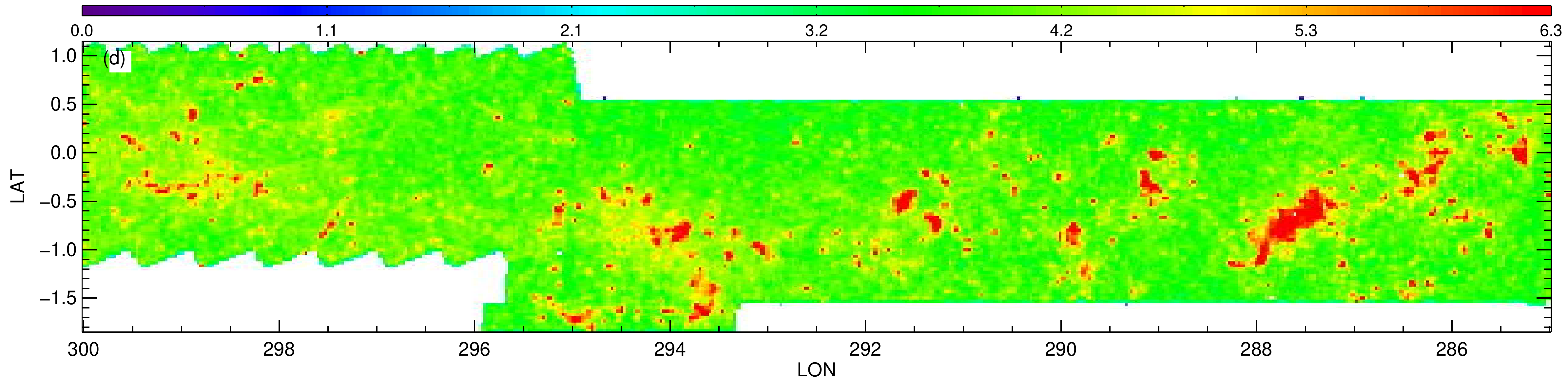} \\
\includegraphics[height=42mm]{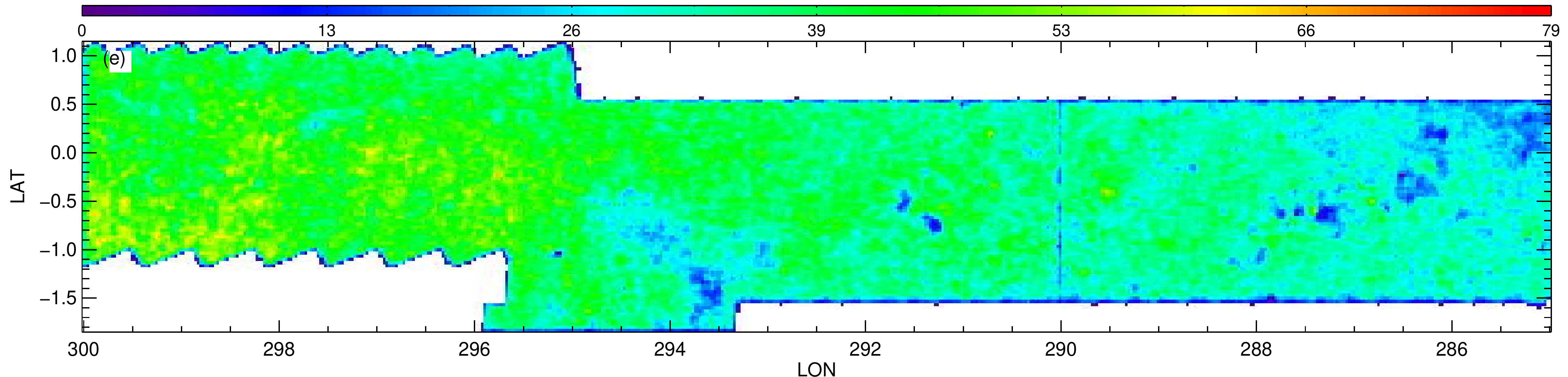}
\end{array}$
\end{center}
\caption{(cont.)}
\label{fig:extmaps285300}
\end{figure*}

\newpage

\addtocounter{figure}{-1}
\begin{figure*}[htpb]
\begin{center}
$\begin{array}{c}
\includegraphics[height=42mm]{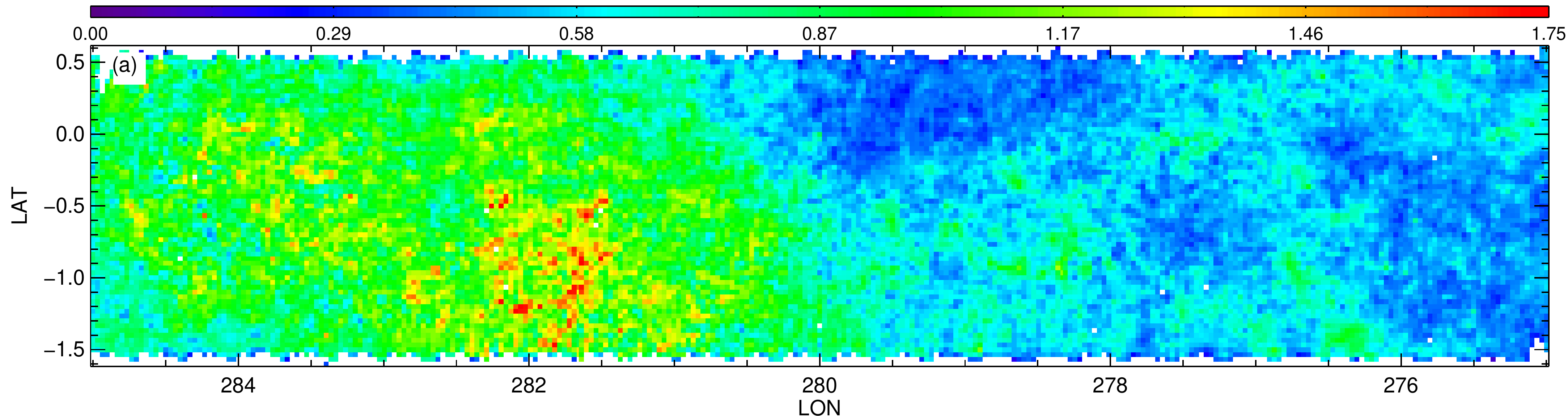} \\
\includegraphics[height=42mm]{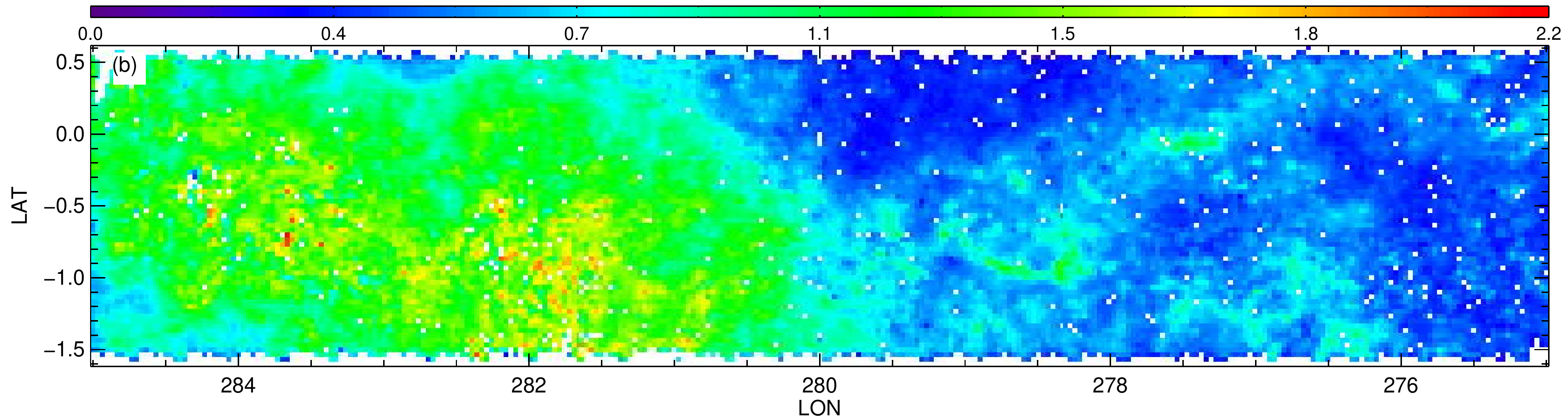} \\
\includegraphics[height=42mm]{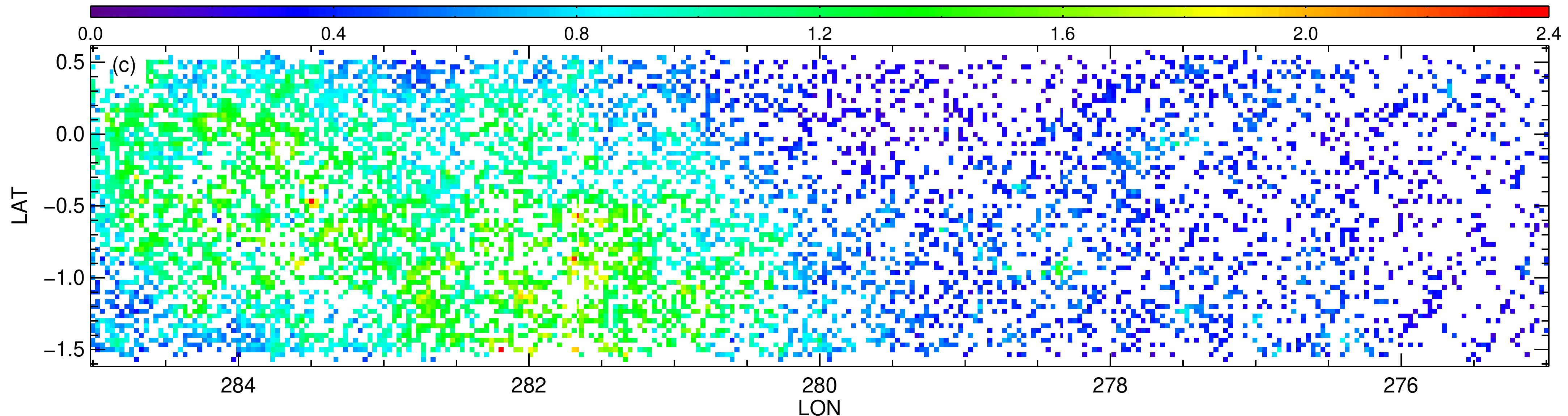} \\
\includegraphics[height=42mm]{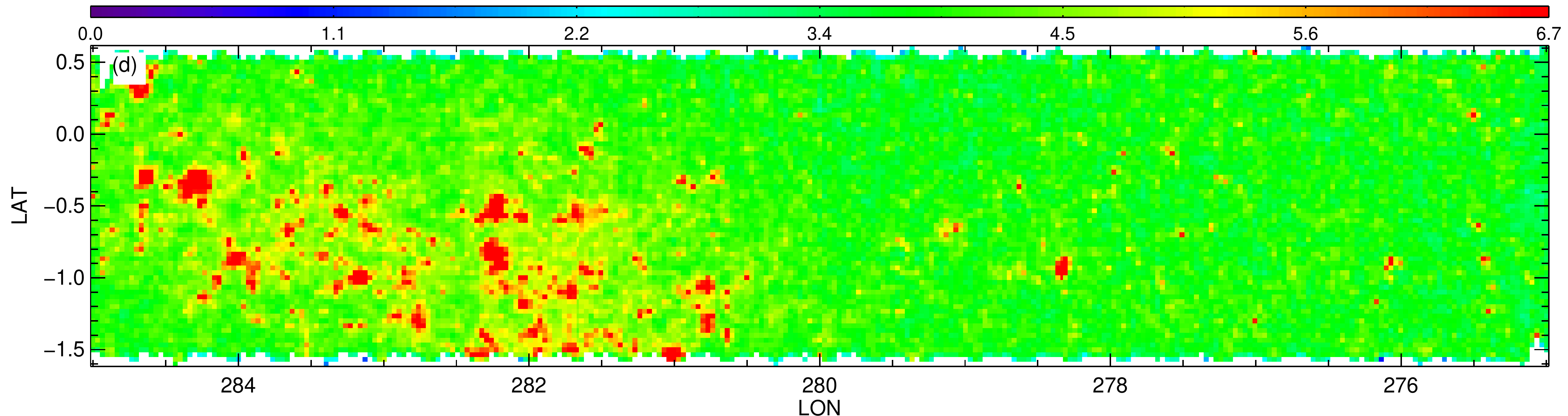} \\
\includegraphics[height=42mm]{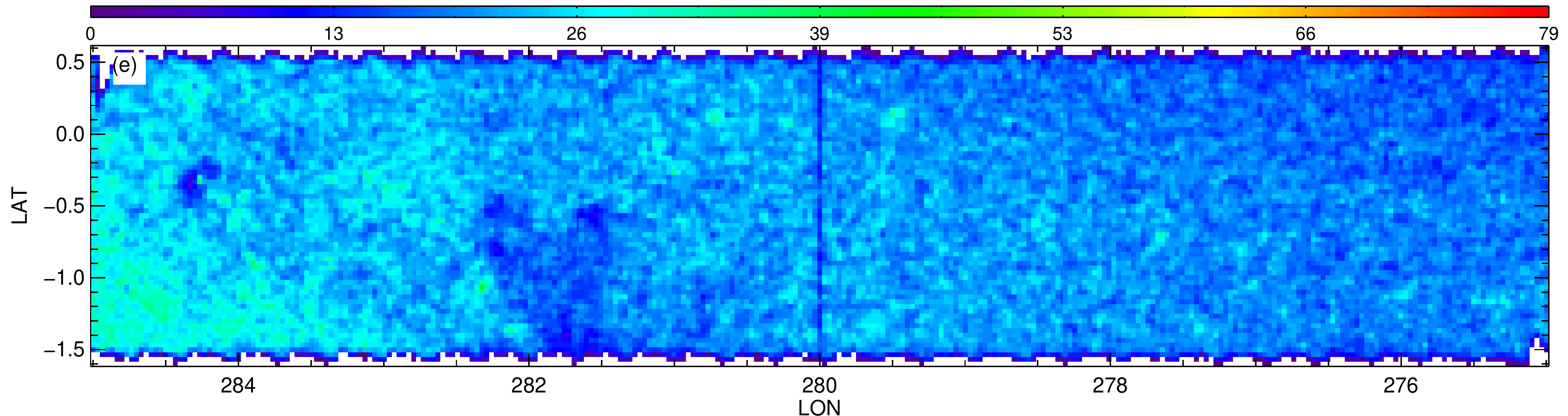}
\end{array}$
\end{center}
\caption{(cont.)}
\label{fig:extmaps275285}
\end{figure*}

\newpage

\addtocounter{figure}{-1}
\begin{figure*}[htpb]
\begin{center}
$\begin{array}{c}
\includegraphics[height=42mm]{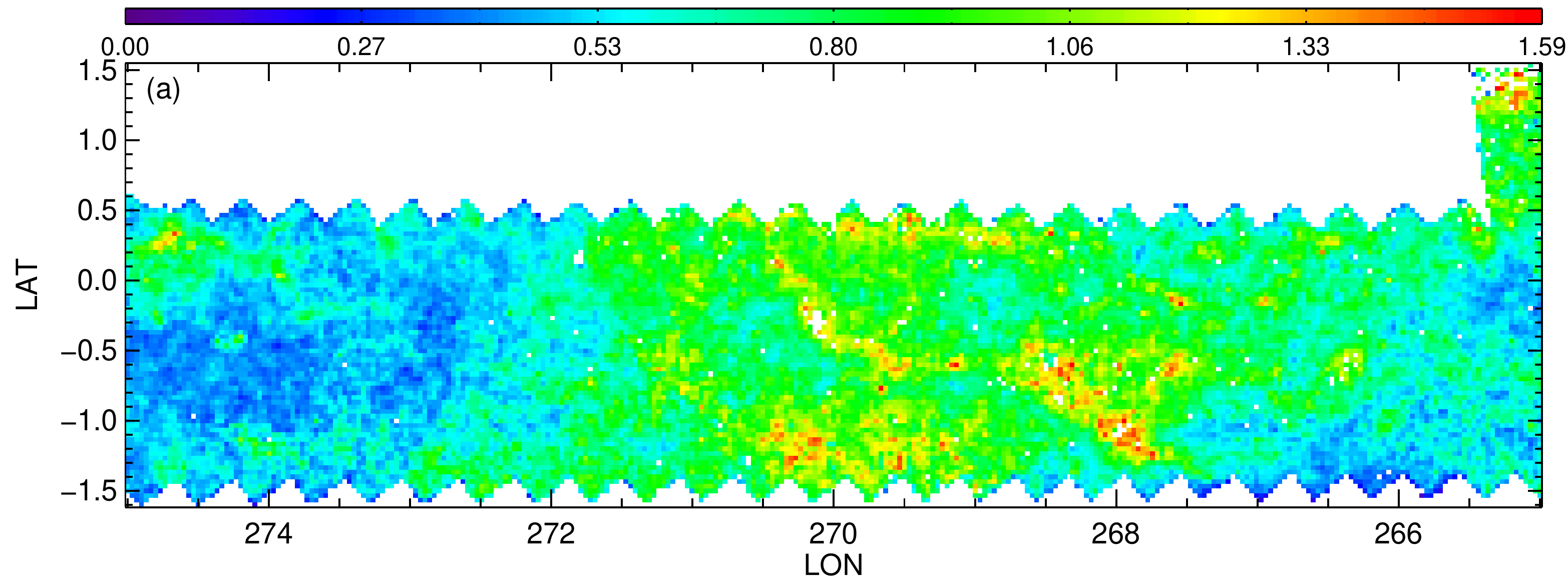} \\
\includegraphics[height=42mm]{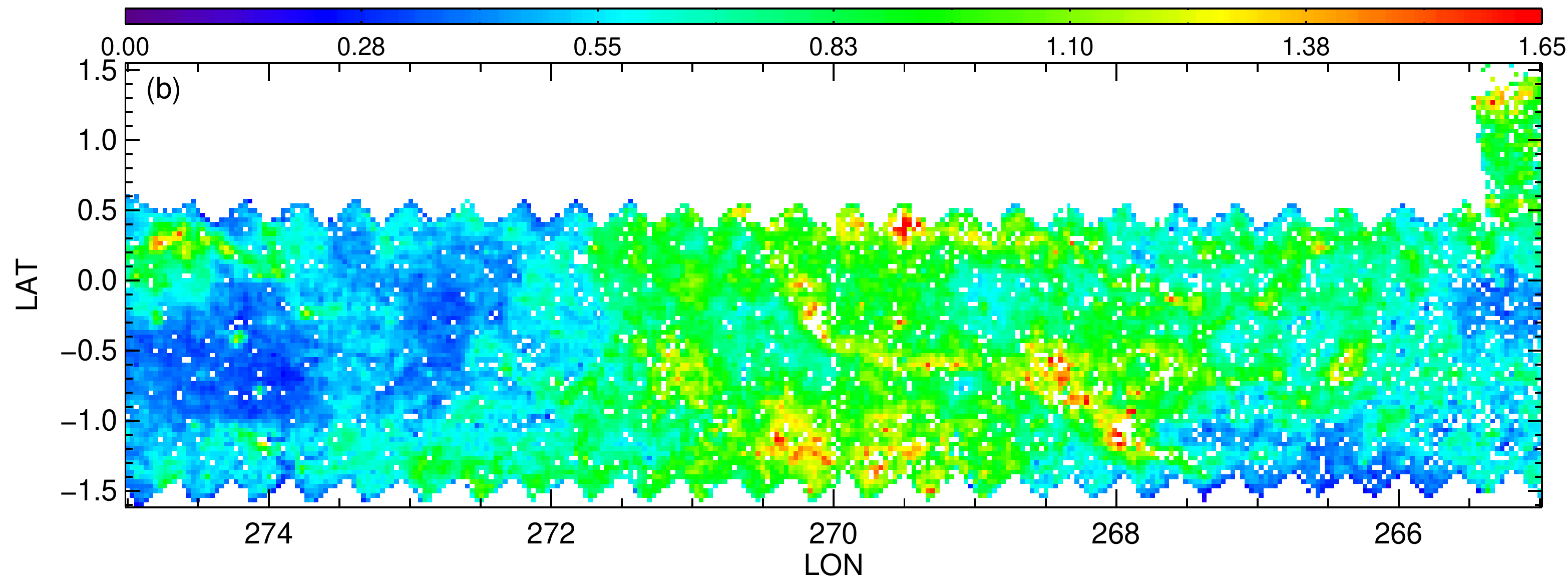} \\
\includegraphics[height=42mm]{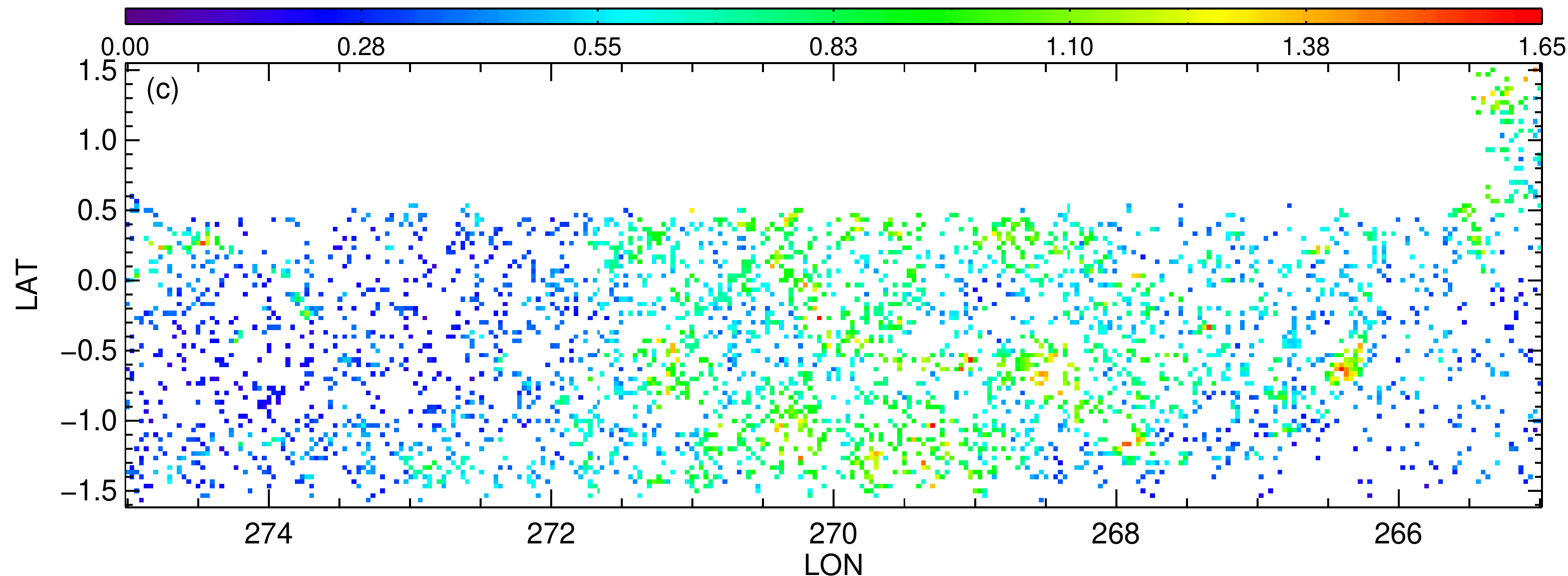} \\
\includegraphics[height=42mm]{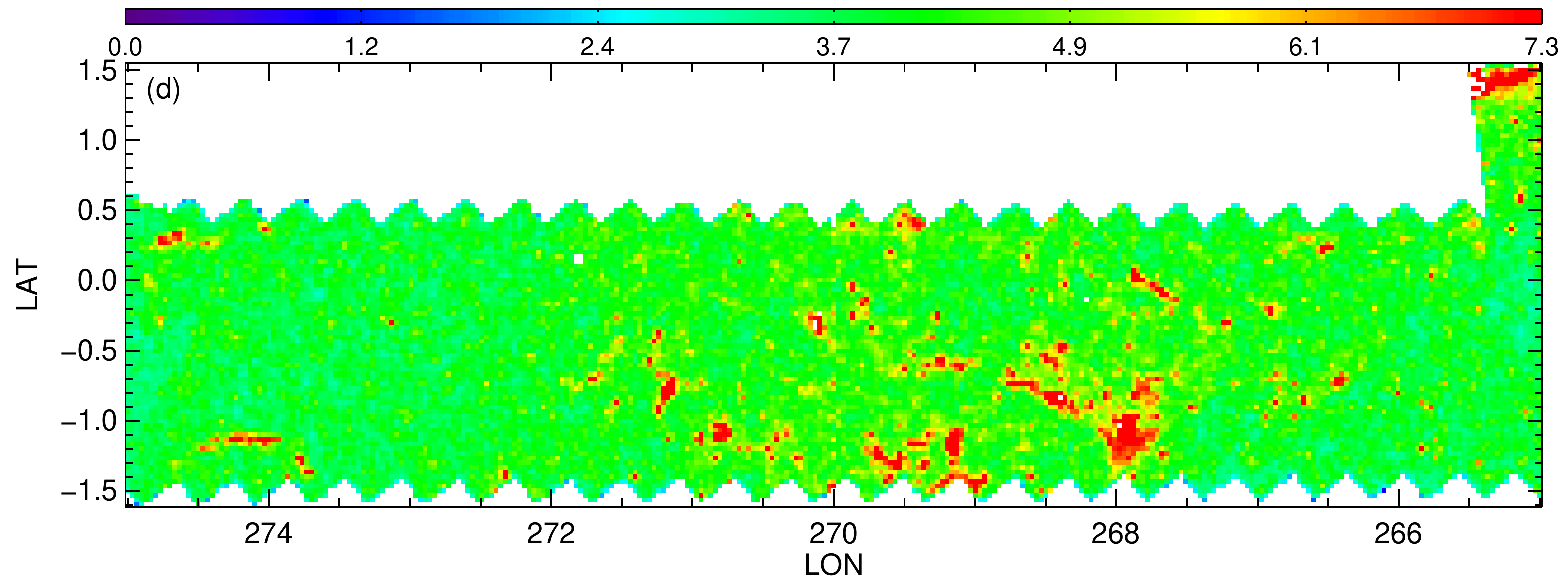} \\
\includegraphics[height=42mm]{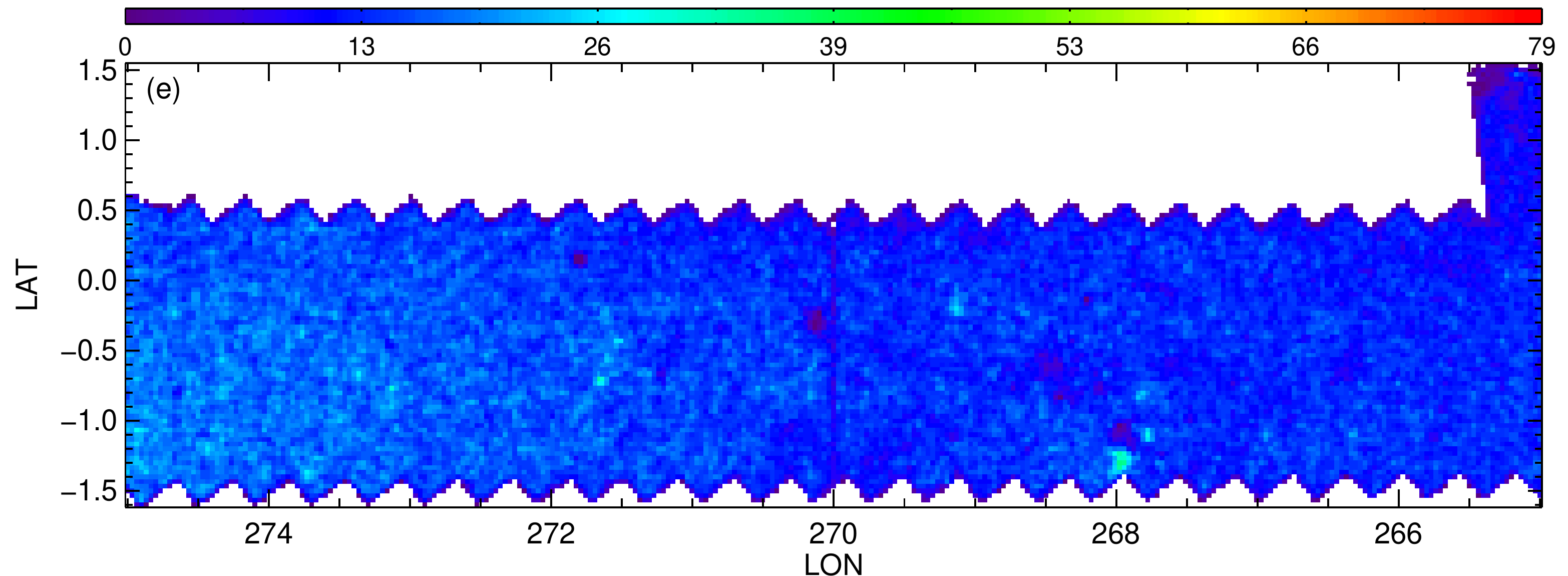}
\end{array}$
\end{center}
\caption{(cont.)}
\label{fig:extmaps265275}
\end{figure*}

\newpage

\addtocounter{figure}{-1}
\begin{figure*}[htpb]
\begin{center}
$\begin{array}{c}
\includegraphics[height=42mm]{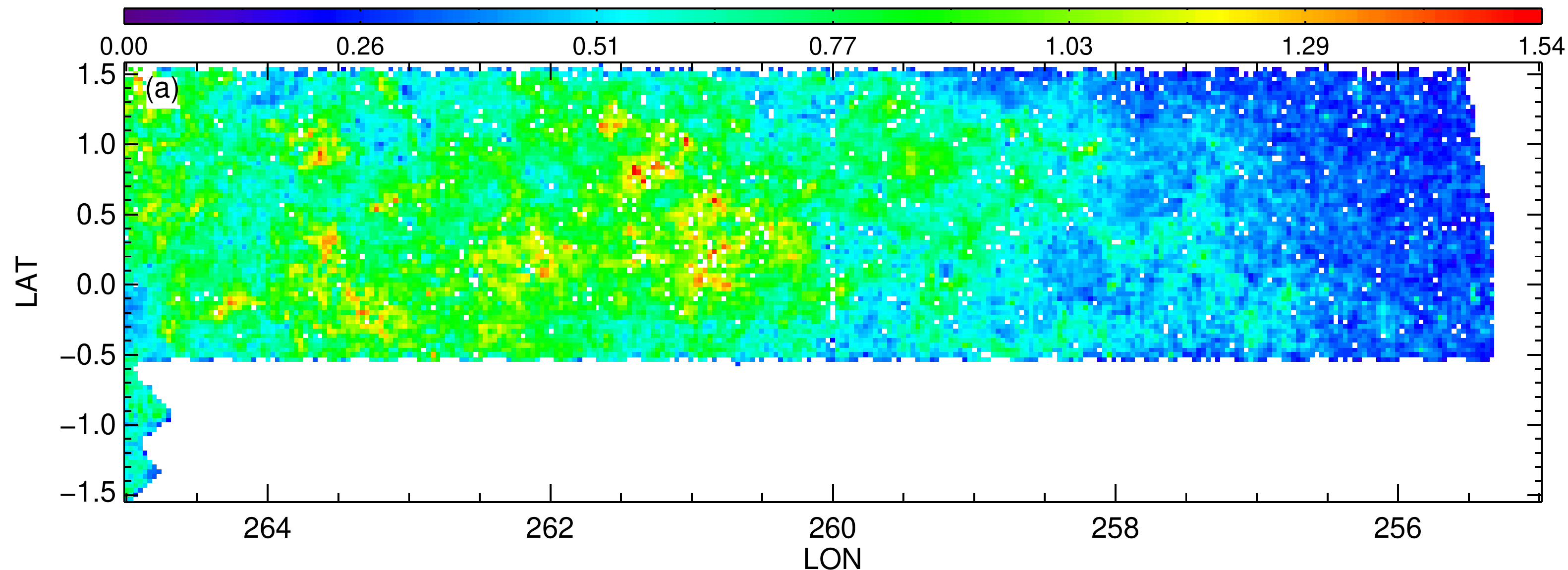} \\
\includegraphics[height=42mm]{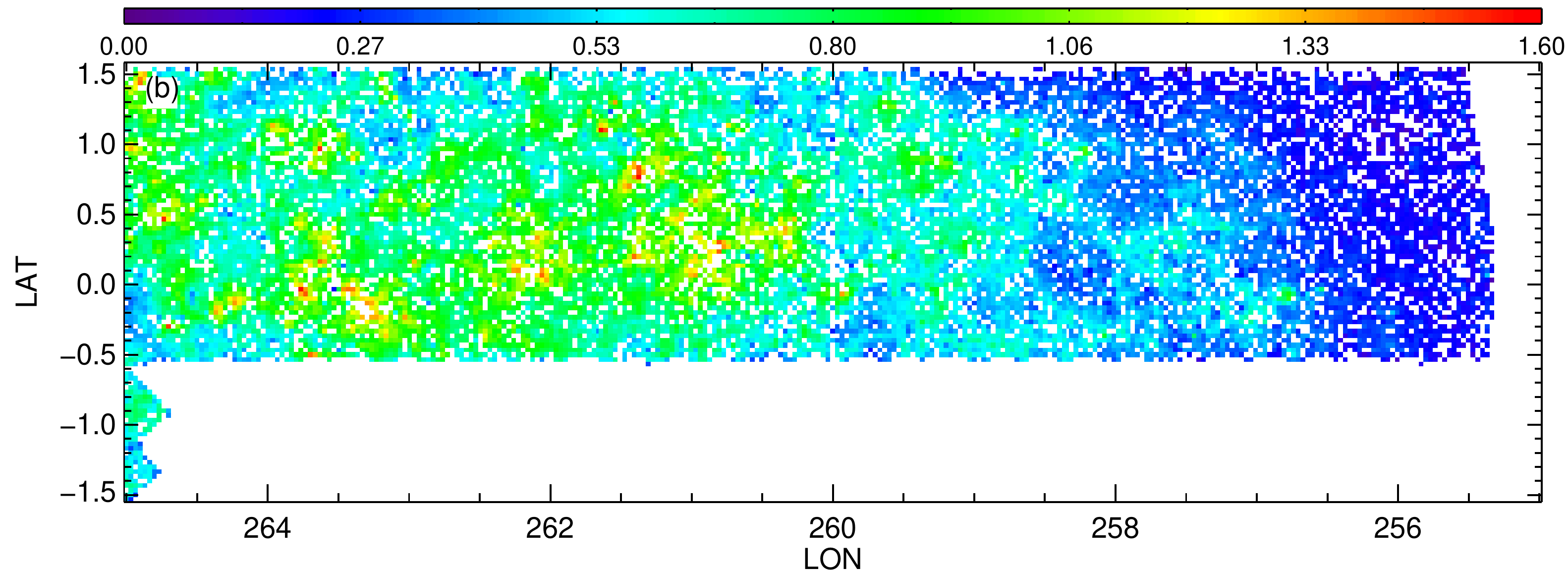} \\
\includegraphics[height=42mm]{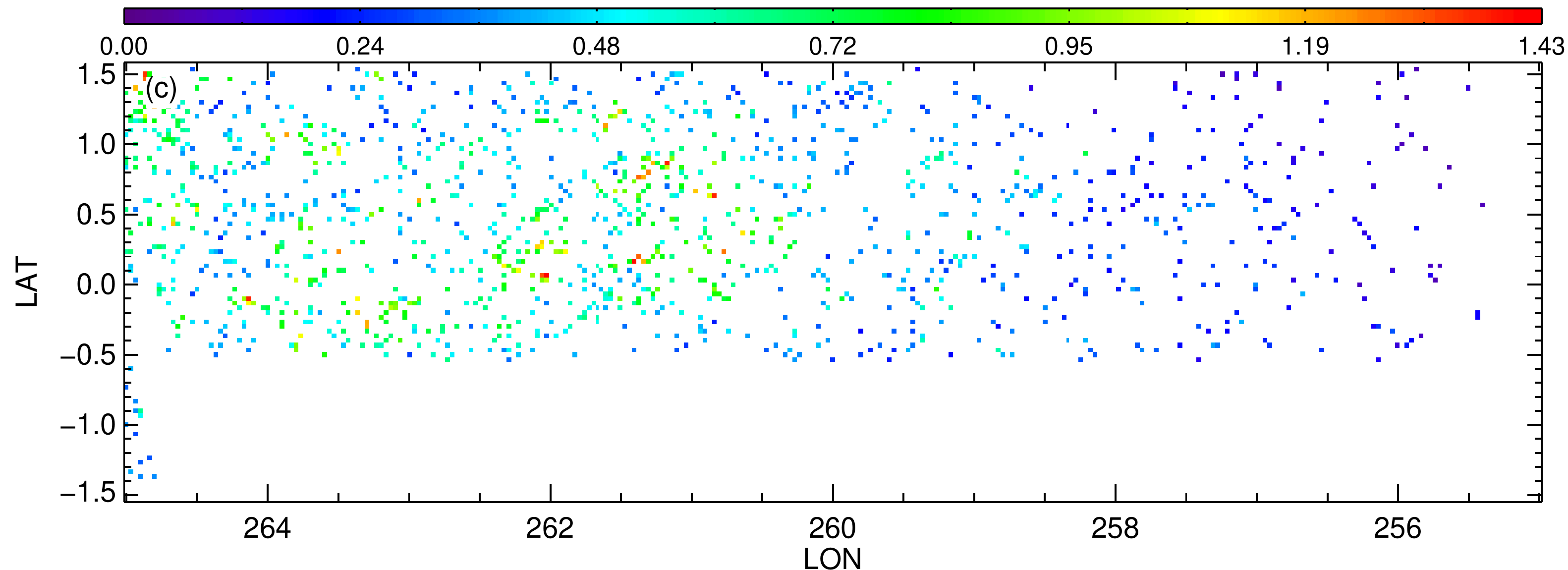} \\
\includegraphics[height=42mm]{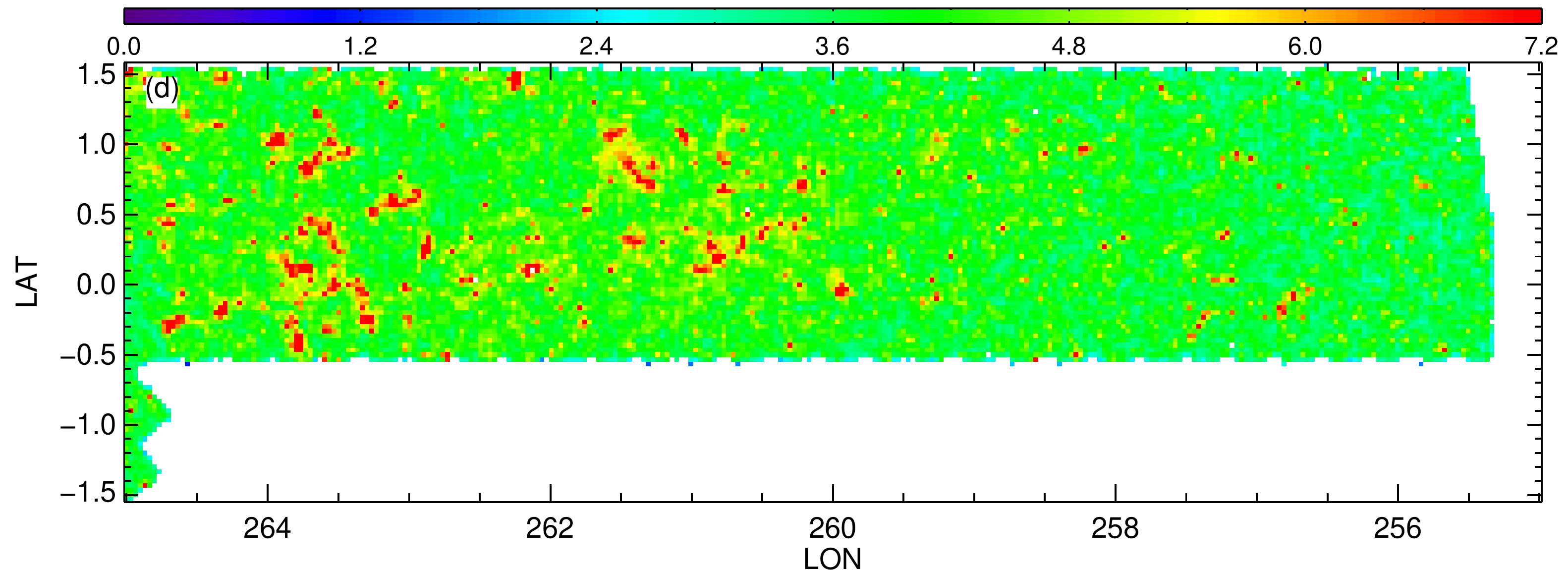} \\
\includegraphics[height=42mm]{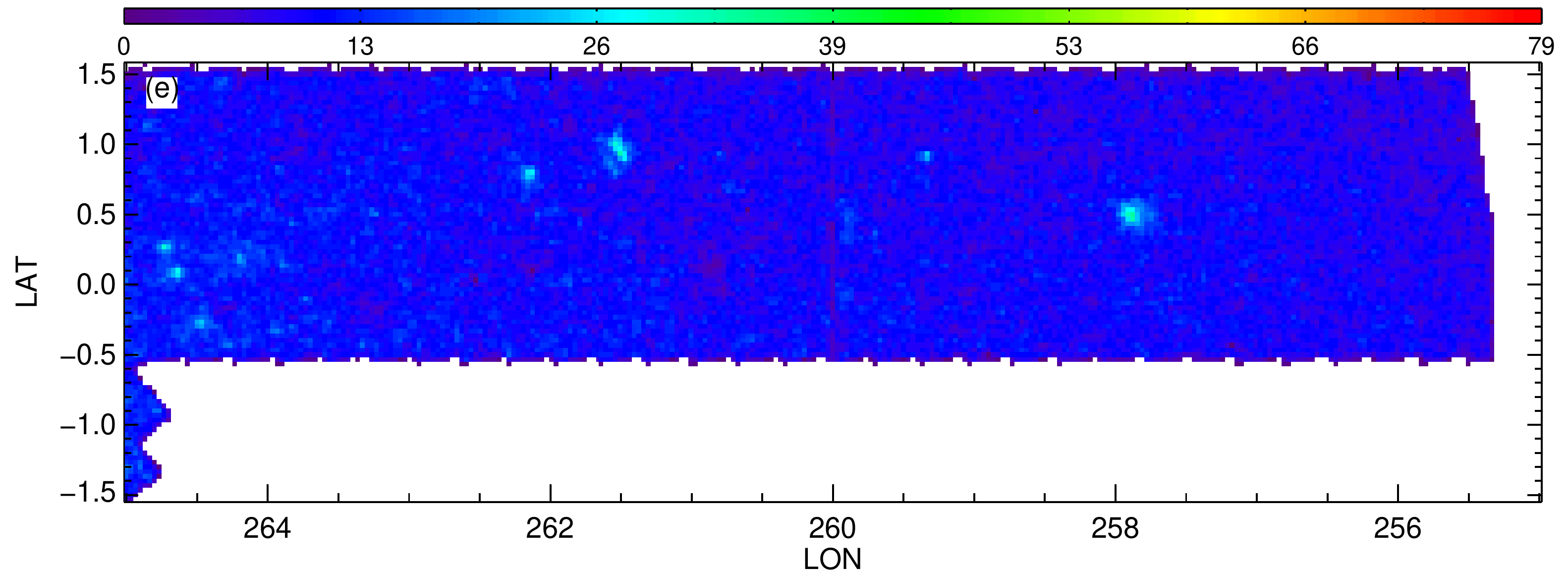}
\end{array}$
\end{center}
\caption{(cont.)}
\label{fig:extmaps255265}
\end{figure*}

\newpage

We note that due to the lower density of stars in the outer Galaxy data, there
are a larger number of ``empty'' pixels in the RGB extinction maps in these directions than in the more populated
inner Galaxy (although the RC and MS extinction maps suffer much less from this deficiency).
Therefore, the RGB maps at these longitudes should be used with caution.  Fortunately, our estimates of the distance
range covered by the RC maps (Section~\ref{sec:dists}) suggest that in these sparser outer Galaxy regions, the farthest
RC stars are very likely to probe extinction to the edge of the Galactic disk; thus, the 90th percentile $A(K_s)$ RC maps
may be used in place of the RGB maps where the latter are unhelpfully incomplete. 

\subsection{Three Dimensional Information}

The derived extinction maps provide coarse {\it three-dimensional} information on the intrinsic interstellar dust distribution. 
Because each of the three stellar subsamples probes different distance ranges (see Section~\ref{sec:dists}),
comparison of the maps for a given section of sky reveals information about the distance, as well as the thickness,
of individual clouds.  The $l$=50--60$^{\circ}$ panels of Figure~\ref{fig:extmaps6065} demonstrate this capability.  
For example, the $\sim$0.5$^\circ$ wide cloud centered near $(l,b)$ $\sim$ (54.8$^\circ$,0.8$^\circ$) in the 
RGB map (Figure~\ref{fig:extmaps6065}c) appears in neither the MS nor the RC map, which indicates that it is a distant feature,
lying beyond the reach of the RC stellar sample ($\sim$8 kpc; Section~\ref{sec:dists}) but in front of the RGB sample.  
Similarly, the filamentary structure near $(l,b)$ $\sim$ (59.0$^\circ$,0.5$^\circ$) is
visible in the RC and RGB maps but not the MS one, which suggests it is closer than the former cloud.
An even closer structure lies at $(l,b)$ $\sim$ (53.7$^\circ$,0.5$^\circ$), as evidenced by the fact that it appears
in the relatively short-range MS map as well as the more distant ones.  In future work, we will be able to refine these
cloud distance measurements after more careful calculation of the stellar distance distribution.

\subsection{Uncertainty Estimates} \label{sec:uncertainties}

Factors contributing to the uncertainty of RJCE extinction
measurements are discussed extensively in Paper I.  In
summary, the uncertainty sources of (1) intrinsic scatter
in the stellar colors, (2) observed photometric uncertainties,
and (3) uncertainty and variation in the extinction law combine
to yield a typical {\it individual} stellar extinction 
uncertainty of $\lesssim$0.11 mag in $K_s$.  The use of
median and 90th percentile statistics in the maps further
reduce the extinction uncertainty per pixel, as described in,
e.g., \citet{Lombardi01} and \citet{Dobashi09}.

\section{Caveats to Usage of the Maps} \label{sec:caveats}

\subsection{Limited Photometry in Highly Reddened Regions} \label{sec:limits}

Apparent in many of the extinction maps in Figure \ref{fig:extmaps6065} are sharp ``holes,'' generally at the
higher-latitude edges of outer disk fields  or at the center of high-extinction regions.  The former are mostly due
to an intrinsic lower stellar density (i.e., a true lack of stars) and could be remedied by reducing the map resolution;
currently, for simplicity, we mask these pixels as white (see Section~\ref{sec:atlas}).

The holes at high extinction, however, are artifacts due to photometric limits in the adopted catalogs (here, the 2MASS
PSC and GLIMPSE/Vela-Carina; see Paper I for discussion of the merged catalog limits), where a particularly dense cloud has a
total extinction sufficient to increase the magnitude of background stars beyond the limit of the catalog.  When this
cloud is relatively nearby, the result is an almost total lack of stars with which to measure the extinction; when the
cloud is farther away, the measured extinction in the affected pixels is dominated by less-reddened foreground stars,
so that the $A(K_s)$ value of the central extinction ``peak'' is reported to be smaller than in the rest of the cloud.

\begin{figure*}[ht!]
\begin{center}
\includegraphics[width=0.8\textwidth]{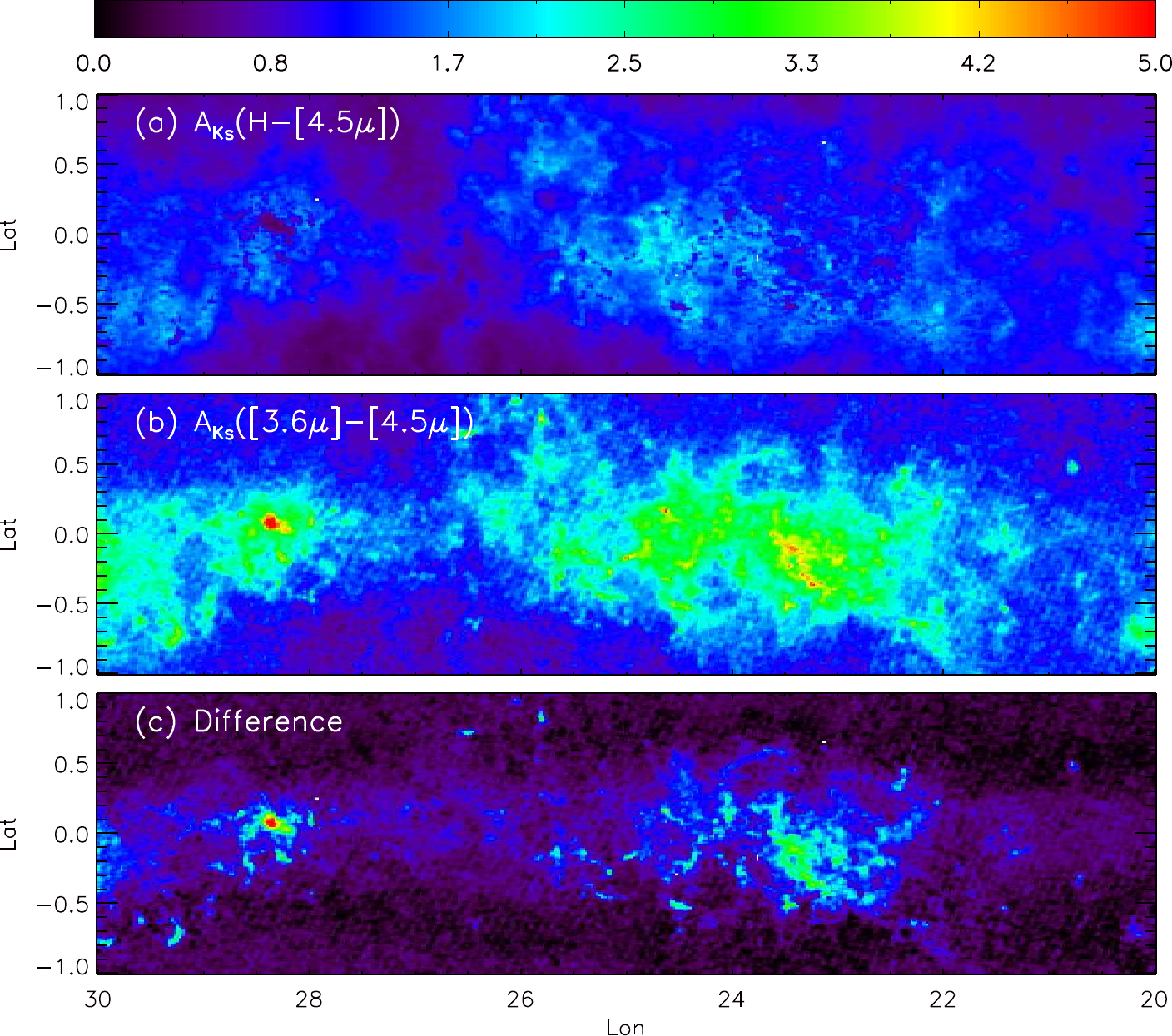}
\end{center}
\caption{Total extinction maps (in units of $A(K_s)$ mags)
made by the RJCE method, but with use of the (a) $(H-[4.5\mu])$
colors and (b) $([3.6\mu]-[4.5\mu])$ colors as the selective extinction measure (where we 
have adopted $([3.6\mu]-[4.5\mu])_0$ = $-0.1$ as a baseline color).  
In both cases, the stars used 
are those satisfying a 0.5 mag
photometric uncertainty limit in only the two filters used 
in the color-excess measurement.  Panel (c) shows the difference between the 
maps in panels (a) and (b).  The few white pixels indicate bins containing $<$2 stars.  }
\label{fig:rjce-diff}
\end{figure*}

Automatically identifying which ``holes'' are due to which cause(s) is a nontrivial task.  For the purposes of mapping
extinction, there are three approaches to dealing with the second problem (i.e., artificially low $A(K_s)$ values due
to the presence of a nearby dense cloud): (1) acquire deeper photometry, (2) ignore (identify and flag) the data in
such pixels, or (3) use photometric bands with even lower dust sensitivity in an attempt to ``see'' beyond the cloud.
Option (1) will become increasingly feasible as large ongoing and future IR surveys release their data, but until
then, deeper IR photometry is limited to much smaller patches of sky (mostly targeted at dark clouds or the central
Galactic regions), not amenable to large-scale or diffuse extinction mapping.  We have chosen option (2) for the
extinction maps presented here, but we enable option (3) by including ([3.6$\mu$]$-$[4.5$\mu$]) extinction maps for
possible use in high extinction regions.

Our analysis of catalog magnitude limits in Paper I suggests that the use of MIR data alone (i.e., not requiring a
{\it JHK$_s$} detection) holds great promise for detecting stars behind extremely dense clouds.  The extinction at
3.6$\mu$m and 4.5$\mu$m is only $\sim$0.04$\times A(V)$ (Indebetouw et al.\ 2005, Zasowski et al.\ 2009), compared to
$\sim$0.2$\times A(V)$ for the $H$ band (Cardelli et al.\ 1989).  Figure~\ref{fig:rjce-diff} demonstrates that by using
IRAC's [3.6$\mu$] and [4.5$\mu$] bands alone one can trace not only the extinction peaks appearing as ``holes'' in the
primary RJCE $(H-[4.5\mu])$ map but also an overall higher level of extinction.  This is because these redder MIR bands
are less susceptible to dust reddening than the NIR bands and so probe farther, while detections in $JHK_s$, the lack of which
is the dominant cause for highly-reddened stars to be extinguished out of the band-merged 2MASS+{\it Spitzer} catalog, are
not required.  However, as discussed in greater depth in Paper I, a number of factors --- including the greater intrinsic
scatter in $([3.6\mu]-[4.5\mu])_0$, the increased photometric uncertainties at longer wavelengths, and the inability to
distinguish stellar type (and hence identify the most distant extinction probes) because all stars have essentially the
same ([3.6$\mu$]--[4.5$\mu$]) color --- render IRAC-only colors less suited for producing dereddened NIR CMDs or
three-dimensional dust maps.  Nevertheless, in our publicly-available maps (Section~\ref{sec:atlas}), we include those
derived using $E([3.6\mu]-[4.5\mu])$ as an option for users interested in estimates of the {\it total} line-of-sight extinction.

\begin{figure*}[ht!]
\begin{center}
\includegraphics[width=0.8\textwidth]{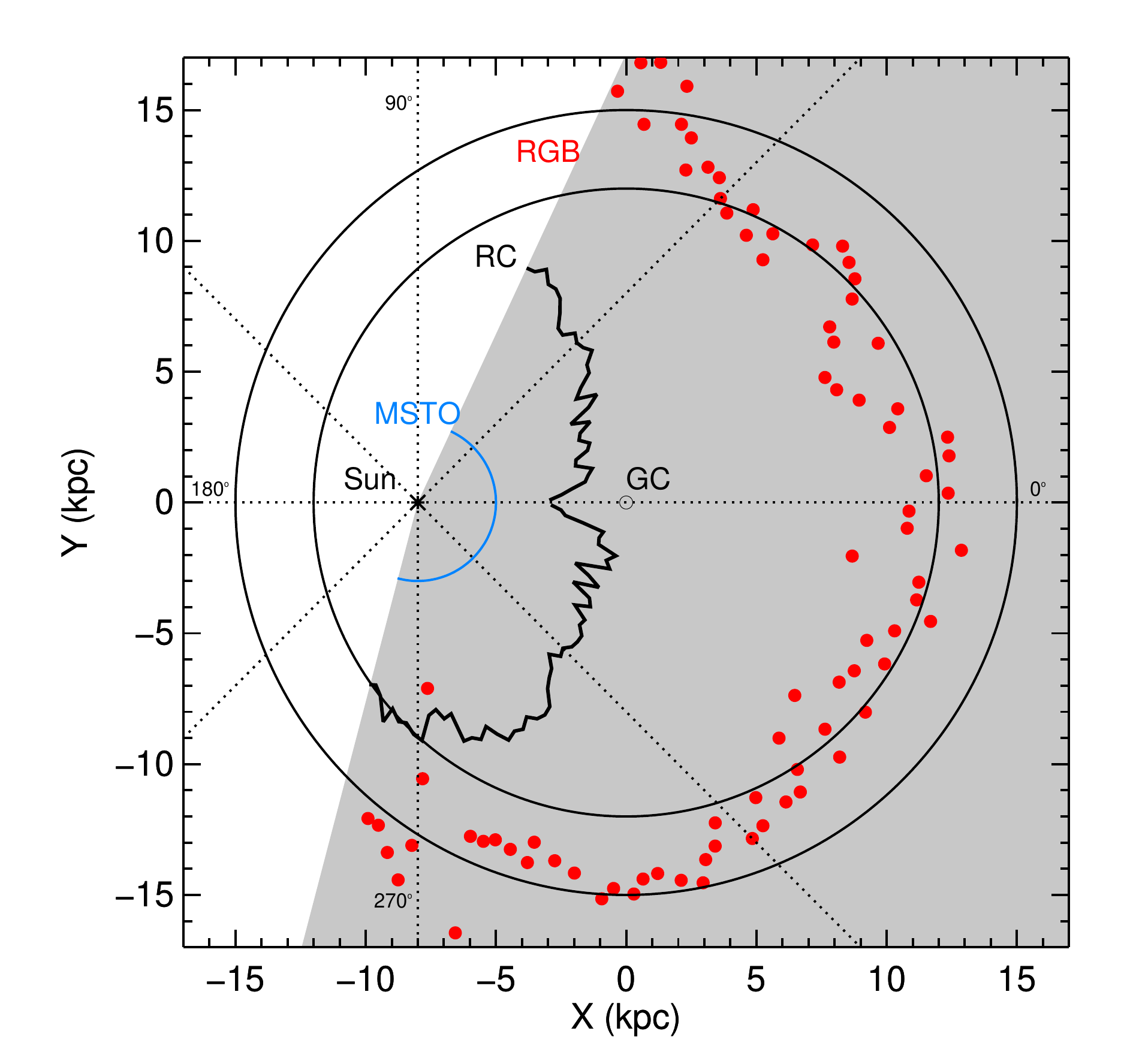}
\end{center}
\caption{Distances of the stellar tracer populations used for the extinction maps as a function of Galactic longitude.
Red filled circles indicate the maximum distance probed by the red giant branch as determined
from isochrone fits to the lower edge of the dereddened RGB (\S \ref{sec:dists}).
Median RC distances (in 2$^{\circ}$ bins) are shown by the black line, 
while the typical main sequence distance of $\sim$3 kpc is shown in blue.
The solid circles indicate Galactocentric radii of 12 and 15 kpc.  The shaded region indicates the 
longitude range covered by our extinction maps.}
\label{fig:tracerdist}
\end{figure*}

\subsection{Distance Limits} \label{sec:dists}

Obviously, the catalog magnitude limits discussed above (Section~\ref{sec:limits}) translate to distance limits in
the maps and define the typical depth to which each map probes.  Characterization of these
limits is crucial to understanding how to interpret the extinction data contained in the maps.  This characterization
was done in two ways.

First, we analyzed the drop in stellar density at the faint edge of the RGB in dereddened CMDs throughout the GLIMPSE
and Vela-Carina survey coverage.  In summary, we compared the faint edge of the RGB to a suite of stellar isochrones
spanning
$-1.3$ $\leq$ [Fe/H] $\leq$ $+1.7$ dex and $9.0$ $\leq$ log(Age) $\leq$ $10.05$ yrs
(Girardi et al.\ 2002), leaving distance
as a free parameter.  This provided us with a rough estimate of the distance of the farthest stars in each CMD, as well
as an estimate of how likely those stars were to be at the magnitude limit of the catalog (as opposed to representing
a genuine drop in stellar density, such as at the edge of the Galactic disk).  The procedure used here is very similar
to the one described in Paper I with the main
difference that the disk edge (at a given color bin) was determined by finding the magnitude for which 90 percent of all stars in that
bin are brighter.  This updated method was found to produce reliable fits to the faint edge of the RGB and
the results are presented in Figure~\ref{fig:tracerdist}.
The filled red circles indicate
the derived Galactic positions for the most distant stars in the $\Delta l = 2^\circ$ wide fields that were fit
reasonably well with isochrones.  Figure~\ref{fig:tracerdist} and our comparison of the CMDs to the catalog magnitude
limits are consistent in suggesting that
our extinction maps from the most distant
tracers (i.e., those made using the mean and 90th percentile of RGB extinction;
see Section~\ref{sec:atlas}) likely measure the true total Galactic extinction. 

\begin{itemize}

\item {\em MSTO:} We selected all TRILEGAL model stars in a test field at ($l$,$b$)=($30^{\circ},0^{\circ}$)
in the CMD region $0.0<(J-K_s)_0<0.4$ and $12.5<K_{s,0}<13.5$,
where a ``clump'' of MSTO stars is seen at these Galactic coordinates.
The median distance of these simulated stars is $\sim$2 kpc and
the 90th percentile is $\sim$4 kpc.  Therefore, the MSTO distances are roughly 2--4 kpc.

\item {\em RC:}  Because red clump stars are standard candles (almost independent of [Fe/H]; Salaris \& Girardi 2002) and
their absolute magnitude is known fairly accurately ($M_{\rm Ks} = -1.54\pm0.04$; Groenewegen 2008), we were able to calculate
distances directly for our RC sample at all longitudes.  Stars in the RC color range ($0.55<(J-K_s)_0<0.85$) were selected and binned
in 3$^\prime$ pixels.  Then, for each pixel, the RC star with the 90th percentile extinction was selected and its distance
was calculated using its dereddened $K_s$ magnitude and $M_{\rm Ks} = -1.54$.  Finally, the median distance was calculated
in longitude bins of 2$^{\circ}$; these are shown as a solid, jagged, black line in Figure \ref{fig:tracerdist}.  These distances are
roughly 6--10 kpc with an average distance of 8.4 kpc.

\item {\em RGB:} Both a TRILEGAL model and isochrones were used to estimate the RGB distances in a test field
at ($l$,$b$)=($30^{\circ},0^{\circ}$).  Sources with $0.85<(J-K_s)_0<1.4$ and $K_{s,0}<13.5$ were selected from the
TRILEGAL model as RGB stars (only four red dwarfs contaminated this sample).  The median distance for these sources
is $\sim$15 kpc and the 90th percentile is $\sim$19 kpc.
In addition, a 5 Gyr, solar metallicity Padova isochrone was used to obtain distances for observed
stars with $29^{\circ}<l<31^{\circ}$, $-1^{\circ}<b<+1^{\circ}$ and RGB colors of $0.85<(J-K_s)_0<1.25$ (since the isochrone
terminates at 1.25).  Distances were derived assuming these were RGB stars.  The median isochrone distance
is $\sim$10 kpc, the 90th percentile distance is $\sim$24 kpc, and the median value in a 90th percentile binned
distance map (with 3$^\prime$ pixels) is 22.9 kpc.  In comparison, the ``CMD edge'' distances described above
are on average 15 kpc. Therefore, the RGB distances are roughly 15--24 kpc.

\item {\em ALL:} Since the vast majority of the stars in the data set are red clump stars, the distance limits for
maps that make use of all stars are likely to be similar to the limits for RC stars.

\end{itemize}

The reader is referred to Figure \ref{fig:tracerdist}, which shows distances of the stellar tracers.
The RGB maps probe to the edge of the Galactic disk and can be 
considered close to ``total line-of-sight extinction'' maps in nearly all directions covered by the maps.

\section{Public Access to the Maps} \label{sec:publicmaps}

To encourage usage, additional comparisons, and feedback from the astronomical community, 
we have made public 515 deg$^2$ of midplane and bulge extinction maps using the RJCE
method outlined in Paper I and the present paper --- i.e., using the ($H-[4.5\mu]$) color with $H$-band photometry 
from 2MASS and [4.5$\mu$] data from {\it Spitzer}-IRAC.
In addition, we provide $E([3.6\mu]-[4.5\mu])$-derived extinction maps for ``all stars'' which
are useful in heavily-extincted regions, and 2MASS starcount maps that are roughly anti-correlated with extinction.
The regions included in this public release are those covered by the GLIMPSE-I, -II, -3D, and Vela-Carina surveys,
spanning 255.3 $\leq$ $l$ $\leq$ 65.4$^\circ$ and, in general, for $|b|\leq1^{\circ}-1.5^{\circ}$ (the Vela-Carina data
is centered slightly off the midplane to track the highest extinction regions) but up to
$b\sim4^{\circ}$ for certain longitudes (for GLIMPSE-II and -3D).
These maps are available
at \url{http://www.astro.virginia.edu/rjce/} as quick-look GIF images and downloadable FITS files along with
querying software written in IDL.

Included on the website are instructions for the download and use of the maps and software, 
as well as notes regarding the known caveats.  We welcome any feedback from
the community that may help improve these tools, and we plan to post refined and improved maps 
as our methodology evolves.  

\section{Summary and Future Work} \label{sec:conc}

We have developed a method for measuring line-of-sight extinction that combines
near- and mid-infrared photometry (from 2MASS + {\it Spitzer}-IRAC) to calculate stellar reddening in a process that is
(i) robust
to variations in intrinsic stellar type {\it and} (ii) recovers the intrinsic stellar type.
This Rayleigh-Jeans Color Excess (RJCE) technique represents a major improvement on previous color-excess measurements
because it includes reddening corrections of temperature-sensitive near-infrared colors
which allows us to create dereddened CMDs (extremely useful for
stellar population studies) and select reasonably pure samples of stars of specific stellar type or
luminosity class.
In Paper I of this series, we described the RJCE method in great detail.  This installment presents a large,
publicly-available atlas of two-dimensional, high angular resolution extinction maps created from RJCE-derived extinction values.

We take advantage of RJCE's ability to discriminate stellar spectral type to produce extinction maps generated independently
from three groups of stars dominated by distinct stellar populations (main sequence [MS] dwarfs, red clump [RC] giants,
and red giant branch [RGB] stars), each of which probes different heliocentric distance ranges and thus traces the
net dust column to different distances.  Comparison of these maps for any given section of the Galactic midplane
provides rough {\it three}-dimensional information on the intrinsic dust distribution and allows clouds and cloud complexes
to be isolated to specific distance ranges.
We present a total of ten different types of extinction maps (in units of $A[K_s]$), each encompassing
$\sim$515 deg$^2$ of the dustiest regions of the Galactic midplane, using data from the 2MASS-PSC merged with
{\it Spitzer}-IRAC photometry from the GLIMPSE-I/-II/-3D and Vela-Carina surveys.  For each of
the three stellar tracer subsamples (MS, RC, and RGB) and ``all stars'', we provide maps giving
two different statistical estimates of the $(H-[4.5\mu])$-derived extinction: the median and 90th percentile
$A(K_s)$ in each spatial pixel.
The three 90th percentile maps probably provide the most reliable estimate of the total integrated extinction
affecting each subsample population to the maximum distance probed.
Additionally, we provide median and 90th percentile $([3.6\mu]-[4.5\mu])$-derived extinction maps of ``all stars'',
which provide reasonable extinction estimates for very dusty regions where the other maps fail due to lack of stars.

Section \ref{sec:caveats} provides caveats to the applicability of our maps as well as distance limits for
the different tracer populations.
We find that our RGB 90th percentile maps
very likely represent the total Galactic extinction in near all directions.
Our 90th percentile RC maps
probe to an average distance of $\sim$8 kpc, while the MS maps are limited to 2--4 kpc.

The possibility of using MIR-only colors to estimate extinction where NIR-MIR colors are unavailable (due to the
increased extinction in the NIR bands or the lower resolution of 2MASS) is also explored.  
Though MIR-only colors do not carry enough information to allow stellar type selection (and thus reliable
distance estimates), we find that they can provide extinction estimates in the most highly extinguished portions of our
maps (such as cloud cores and regions near the Galactic center).  Thus, we include in our public atlas extinction
maps derived from MIR-only reddening to serve as strong lower limits to the total extinction;
this is most useful in the inner Galaxy, where even the RGB maps from typical RJCE colors are insufficient to probe
to the far edge of the Galactic disk.

An extensive program for application and improvement of the RJCE methodology is anticipated.
We are working to improve the RJCE extinction estimates using a more refined treatment of stellar properties, and
the ongoing APOGEE survey (Majewski et al.\ 2010) will allow us to compare RJCE-derived stellar types to
spectroscopically-determined ones.  We are evaluating the benefit of using optimized ``hybrid''-color schemes 
(i.e., merging extinction estimated using long-baseline NIR-MIR colors and MIR-only colors) where high extinction
or confusion limits the depth of the NIR photometry.  Our ultimate goal is to apply this dereddening
methodology to ascertain the three-dimensional distribution of both dust and stars within the Galactic 
midplane.  We have already taken advantage of our extensive data set to characterize the wavelength dependence
of IR extinction throughout the midplane (Zasowski et al.\ 2009), and work is under way
to use the new, reliably-cleaned CMDs to explore heavily-reddened or otherwise elusive Galactic structures, 
such as the central bar(s) and the outer spiral arms and the edges of the disk.

\begin{acknowledgments}
We thank R.~Indebetouw and M.~F.~Skrutskie for helpful discussions, and M.~Meade and B.~Babler for their efforts to
reduce the Vela-Carina dataset with the GLIMPSE pipeline.
We thank the anonymous referee for useful comments and suggestions.
SRM appreciates the hospitality of the Observatories of the Carnegie Institution of Washington for hosting a sabbatical
visit during which this project was conceived.
We acknowledge support from NSF grants AST-0307851
and AST-0807845, and funding from NASA {\it Spitzer} grants 1276756 and 1316912.
DLN has been supported by an ARCS Scholarship, a VSGC Graduate Research Fellowship, and an SDSS-III APOGEE postdoc, and
GZ has been supported by a VSGC Graduate Research Fellowship, a NASA Earth \& Space Science Fellowship, and a UVa
Dissertation Acceleration Fellowship.  This work is based in part on observations made with the {\it Spitzer} Space
Telescope, and has made use of the NASA~/~IPAC Infrared Science Archive, 
which are operated by the Jet Propulsion Laboratory, California Institute of Technology under a contract with NASA. 
We acknowledge use of data products from the Two Micron All Sky Survey, a joint project of the University of Massachusetts and the 
Infrared Processing and Analysis Center~/~California Institute of Technology, funded by NASA and the NSF.  
\end{acknowledgments}


\begin{thebibliography}{}

\bibitem[Alves et al.(1998)]{1998ApJ...506..292A} Alves, J., Lada, C.~J., 
Lada, E.~A., Kenyon, S.~J., \& Phelps, R.\ 1998, \apj, 506, 292 

\bibitem[Am{\^o}res 
\& L{\'e}pine(2005)]{2005AJ....130..659A} Am{\^o}res, E.~B., \& L{\'e}pine, J.~R.~D.\ 2005, \aj, 130, 659 

\bibitem[Am{\^o}res 
\& L{\'e}pine(2007)]{2007AJ....133.1519A} Am{\^o}res, E.~B., \& L{\'e}pine, J.~R.~D.\ 2007, \aj, 133, 1519 

\bibitem[Arce \& Goodman(1999a)]{AG99a} Arce, H., Goodman, A. 1999a, ApJLetters, 512, L135 

\bibitem[Barnard et al.(1927)]{1927QB819.B3.......} Barnard, E.~E., Frost, E.~B., 
\& Calvert, M.~R.\ 1927, A Photographic Atlas of Selected Regions of the Milky Way,
[Washington] Carnegie institution of Washington, 1927.,  

\bibitem[Benjamin et al.(2003)]{2003PASP..115..953B} Benjamin, R.~A., et al.\ 2003, \pasp, 115, 953 

\bibitem[Bohlin(1975)]{1975ApJ...200..402B} Bohlin, R.~C.\ 1975, \apj, 200, 402 

\bibitem[Bok(1956)]{1956AJ.....61..309B} Bok, B.~J.\ 1956, \aj, 61, 309 

\bibitem[Burstein 
\& Heiles(1978)]{1978ApJ...225...40B} Burstein, D., \& Heiles, C.\ 1978, \apj, 225, 40 

\bibitem[Burstein 
\& Heiles(1982)]{1982AJ.....87.1165B} Burstein, D., \& Heiles, C.\ 1982, \aj, 87, 1165 

\bibitem[Burstein et al.(1988)]{1988ApJ...328..440B} Burstein, D., Bertola, 
F., Buson, L.~M., Faber, S.~M., \& Lauer, T.~R.\ 1988, \apj, 328, 440 

\bibitem[Burstein(2003)]{2003AJ....126.1849B} Burstein, D.\ 2003, \aj, 126, 1849 

\bibitem[Cambr{\'e}sy et 
al.(2005)]{2005A&A...435..131C} Cambr{\'e}sy, L., Jarrett, T.~H., \& Beichman, C.~A.\ 2005, \aap, 435, 131 

\bibitem[Cardelli, Clayton \& Mathis]{CCM} Cardelli, J., Clayton, G., Mathis, J. 1989, ApJ,
345, 245

\bibitem[Chen et 
al.(1999)]{1999A&A...352..459C} Chen, B., Figueras, F., Torra, J., Jordi, C., Luri, X., \& Galad{\'{\i}}-Enr{\'{\i}}quez, D.\ 1999, \aap, 352, 459 

\bibitem[Choloniewski 
\& Valentijn(2003)]{Choloniewski03} Choloniewski, J., \& Valentijn, E.~A.\ 2003, Acta Astronomica, 53, 265 

\bibitem[Churchwell et al.(2009)]{Churchwell09} Churchwell, E., 
Babler, B.~L., Meade, M.~R., et al.\ 2009, \pasp, 121, 213 

\bibitem[de Vaucouleurs 
\& Buta(1983)]{1983AJ.....88..939D} de Vaucouleurs, G., \& Buta, R.\ 1983, \aj, 88, 939 

\bibitem[de Vaucouleurs et al.(1976)]{1976RC2...C......0D} de Vaucouleurs, 
G., de Vaucouleurs, A., 
\& Corwin, J.~R.\ 1976, Second reference catalogue of bright galaxies, 1976, Austin: University of Texas Press., 0 

\bibitem[Dobashi et al.(2009)]{Dobashi09} Dobashi, K., Bernard, 
J.-P., Kawamura, A., et al.\ 2009, \aj, 137, 5099 

\bibitem[Dutra et al.(2003a)]{dutra} Dutra, C., Santiago, B., Bica, E., Barbuy, B. 
2003a, MNRAS, 338, 253

\bibitem[Dutra et 
al.(2003b)]{2003A&A...408..287D} Dutra, C.~M., Ahumada, A.~V., Clari{\'a}, J.~J., Bica, E., \& Barbuy, B.\ 2003b, \aap, 408, 287 

\bibitem[Fabbri et al.(1986)]{1986mgm..conf..789F} Fabbri, R., Guidi, I., Natale, V., 
\& Ventura, G.\ 1986, Fourth Marcel Grossmann Meeting on General Relativity, 789 

\bibitem[Faber et al.(1989)]{1989ApJS...69..763F} Faber, S.~M., Wegner, G., 
Burstein, D., Davies, R.~L., Dressler, A., Lynden-Bell, D., 
\& Terlevich, R.~J.\ 1989, \apjs, 69, 763 

\bibitem[Fong et al.(1987)]{1987MNRAS.224.1059F} Fong, R., Jones, L.~R., 
Shanks, T., Stevenson, P.~R.~F., \& Strong, A.~W.\ 1987, \mnras, 224, 1059 

\bibitem[Froebrich et al.(2005)]{2005A&A...432L..67F} Froebrich, D., Ray, T.~P., Murphy, G.~C., \& Scholz, A.\ 2005, \aap, 432, L67 

\bibitem[Girardi et 
al.(2005)]{2005A&A...436..895G} Girardi, L., Groenewegen, M.~A.~T., 
Hatziminaoglou, E., \& da Costa, L.\ 2005, \aap, 436, 895 

\bibitem[Girardi et al.(2002)]{2002A&A...391..195G} Girardi, L., et al. 2002, A\&A, 391, 195

\bibitem[Groenewegen (2008)]{Groenewegen08} Groenewegen, M.~A.~T.\ 2008, \aap, 488, 935 

\bibitem[Gosling et al.(2009)]{2009MNRAS.394.2247G} Gosling, A.~J., 
Bandyopadhyay, R.~M., \& Blundell, K.~M.\ 2009, \mnras, 394, 2247 

\bibitem[Heiles(1976)]{1976ApJ...204..379H} Heiles, C.\ 1976, \apj, 204, 379 

\bibitem[Herschel(1785)]{herchel} Herschel, W. 1785, Philosophical Transactions, 75, 213

\bibitem[Hilditch et al.(1983)]{1983MNRAS.204..241H} Hilditch, R.~W., Hill, 
G., \& Barnes, J.~V.\ 1983, \mnras, 204, 241 

\bibitem[Holmberg(1974)]{1974A&A....35..121H} Holmberg, E.~B.\ 1974, \aap, 35, 121 

\bibitem[Indebetouw et al.(2005)]{I05} Indebetouw, R., et al. 2005, ApJ, 619, 931

\bibitem[Ivans et al.(1999)]{1999AJ....118.1273I} Ivans, I.~I., Sneden, C., 
Kraft, R.~P., Suntzeff, N.~B., Smith, V.~V., Langer, G.~E., 
\& Fulbright, J.~P.\ 1999, \aj, 118, 1273

\bibitem[Jenkins \& Savage(1974)]{1974ApJ...187..243J} Jenkins, E.~B., \& Savage, B.~D.\ 1974, \apj, 187, 243 

\bibitem[Jones et al.(1984)]{1984ApJ...282..675J} Jones, T.~J., Hyland, 
A.~R., \& Bailey, J.\ 1984, \apj, 282, 675 

\bibitem[Jones et al.(1980)]{1980ApJ...242..132J} Jones, T.~J., Hyland, 
A.~R., Robinson, G., Smith, R., \& Thomas, J.\ 1980, \apj, 242, 132 

\bibitem[Knapp \& Kerr(1974)]{1974A&A....35..361K} Knapp, G.~R., \& Kerr, F.~J.\ 1974, \aap, 35, 361 

\bibitem[Knude(1996)]{1996A&A...306..108K} Knude, J.\ 1996, \aap, 306, 108 

\bibitem[Kron \& Guetter(1973)]{1973PASP...85..534K} Kron, G.~E., \& Guetter, H.~H.\ 1973, \pasp, 85, 534 

\bibitem[Lada et al.(1994)]{1994ApJ...429..694L} Lada, C.~J., Lada, E.~A., 
Clemens, D.~P., \& Bally, J.\ 1994, \apj, 429, 694 

\bibitem[Lada et al.(2009)]{2009ApJ...703...52L} Lada, C.~J., Lombardi, M., 
\& Alves, J.~F.\ 2009, \apj, 703, 52 

\bibitem[Lilley(1955)]{1955ApJ...121..559L} Lilley, A.~E.\ 1955, \apj, 121, 
559 

\bibitem[Lombardi \& Alves(2001)]{Lombardi01} Lombardi, M., \& Alves, J.\ 2001, \aap, 377, 1023 

\bibitem[Lombardi(2009)]{Lombardi09} Lombardi, M.\ 2009, \aap, 493, 735 

\bibitem[Lombardi et al.(2010)]{Lombardi10} Lombardi, M., Lada, C.~J., \& Alves, J.\ 2010, \aap, 512, A67 

\bibitem[Lucas et al.(2008)]{2008MNRAS.391..136L} Lucas, P.~W., et al.\
2008, MNRAS, 391, 136

\bibitem[Majewski et al.(2010)]{2010IAUS..265..480M} Majewski, S.~R., 
Wilson, J.~C., Hearty, F., Schiavon, R.~R., 
\& Skrutskie, M.~F.\ 2010, IAU Symposium, 265, 480 

\bibitem[Majewski et al.(2007)]{Majewski07} Majewski, S., Babler, 
B., Churchwell, E., et al.\ 2007, Spitzer Proposal ID \#40791, 40791 

\bibitem[Majewski et al.(2011)]{Majewski11} Majewski, S.~R., 
Zasowski, G., \& Nidever, D.~L.\ 2011, \apj, 739, 25 

\bibitem[Noonan(1971)]{1971AJ.....76..190N} Noonan, T.~W.\ 1971, \aj, 76, 190 

\bibitem[Peek 
\& Graves(2010)]{2010ApJ...719..415P} Peek, J.~E.~G., \& Graves, G.~J.\ 2010, \apj, 719, 415 

\bibitem[Peterson(1970)]{1970AJ.....75..695P} Peterson, B.~A.\ 1970, \aj, 75, 695 

\bibitem[Rocha-Pinto et al.(2004)]{2004ApJ...615..732R} Rocha-Pinto, H.~J., 
Majewski, S.~R., Skrutskie, M.~F., Crane, J.~D., 
\& Patterson, R.~J.\ 2004, \apj, 615, 732 

\bibitem[Salaris \& Girardi(2002)]{2002MNRAS.337..332S} 
Salaris, M., Girardi, L. 2002, MNRAS, 337, 332 

\bibitem[Sandage(1973)]{1973ApJ...183..711S} Sandage, A.\ 1973, \apj, 183, 711 

\bibitem[Savage 
\& Jenkins(1972)]{1972ApJ...172..491S} Savage, B.~D., \& Jenkins, E.~B.\ 1972, \apj, 172, 491 

\bibitem[Savage et al.(1977)]{1977ApJ...216..291S} Savage, B.~D., Bohlin, 
R.~C., Drake, J.~F., \& Budich, W.\ 1977, \apj, 216, 291 

\bibitem[Schlegel, Finkbeiner \& Davis(1998)]{SFD}
Schlegel, D., Finkbeiner, D., Davis, M. 1998, ApJ, 500, 525 (SFD)

\bibitem[Seki(1973)]{1973A&A....28..207S} Seki, M.\ 1973, \aap, 28, 207 

\bibitem[Skrutskie et al.(2006)]{2006AJ....131.1163S} Skrutskie, M.~F., et 
al.\ 2006, \aj, 131, 1163 

\bibitem[Smith(1987)]{1987MNRAS.227..943S} Smith, R.~G.\ 1987, \mnras, 227, 943 

\bibitem[Stanek(1998a)]{1998astro.ph..2093S} Stanek, K.~Z.\ 1998a, 
arXiv:astro-ph/9802093 

\bibitem[Stanek(1998b)]{1998astro.ph..2307S} Stanek, K.~Z.\ 1998b, 
arXiv:astro-ph/9802307 

\bibitem[Struve(1847)]{struve} Struve, F. G. W. 1847, {\it Etudes d'astronomie stellaire sur la voie lacte{\'e} et sur 
la distance des {\'e}toiles fixes}, (St. Petersbourg: Imprimerie de L'Acad{\'e}mie Imp{\'e}riale des Sciences)

\bibitem[Sturch(1969)]{1969AJ.....74...82S} Sturch, C.\ 1969, \aj, 74, 82 

\bibitem[Trumpler(1930)]{1930LicOB..14..154T} 
Trumpler, R. 1930, Lick Obs. Bull., 14, 154

\bibitem[von Braun 
\& Mateo(2001)]{2001AJ....121.1522V} von Braun, K., \& Mateo, M.\ 2001, \aj, 121, 1522 

\bibitem[Wolf(1923)]{1923AN....219..109W} Wolf, M.\ 1923, Astronomische 
Nachrichten, 219, 109 

\bibitem[Yasuda et al.(2007)]{2007AJ....134..698Y} Yasuda, N., Fukugita, 
M., \& Schneider, D.~P.\ 2007, \aj, 134, 698

\bibitem[Zasowski et al.(2009)]{2009ApJ...707..510Z} Zasowski, G., et al.\ 
2009, \apj, 707, 510 

\end{thebibliography}
\end{document}